\documentclass[10pt, aps, pra, superscriptaddress, email, longbibliography, nofootinbib, noeprint]{revtex4-2}

\usepackage[utf8]{inputenc}
\usepackage[english]{babel}

\usepackage[T1]{fontenc}
\usepackage{dsfont}
\usepackage{siunitx}

\usepackage{mathtools}
\usepackage{amssymb}
\usepackage{braket}
\usepackage{bbm}

\usepackage{graphicx}
    \graphicspath{{./images/}}
\usepackage{xcolor}

\usepackage{array}
\usepackage{multirow}
\usepackage{dcolumn}

\usepackage{enumitem}

\usepackage[hidelinks]{hyperref}
    \hypersetup{
        colorlinks=true,
        linkcolor=blue,
        filecolor=blue,      
        urlcolor=blue,
        citecolor=blue
    }

\definecolor{white}{rgb}{1.0, 1.0, 1.0}
\definecolor{black}{rgb}{0.0, 0.0, 0.0}
\definecolor{red}{rgb}{0.8, 0.2, 0.2}
\definecolor{blue}{rgb}{0.0, 0.3, 0.7}
\definecolor{green}{rgb}{0.2, 0.7, 0.2}
\definecolor{yellow}{rgb}{1.0, 0.9, 0.2}
\definecolor{purple}{rgb}{0.6, 0.0, 0.6}
\definecolor{orange}{rgb}{1.0, 0.6, 0.0}

\renewcommand{\d}{\mathrm{d}} 
\newcommand{\e}{\mathrm{e}} 
\renewcommand{\i}{\mathrm{i}} 

\newcommand{\h}[1]{\hat{#1}}

\newcommand{\qo}[1]{#1} 
\renewcommand{\bm}[1]{\mathbf{#1}} 

\newcommand{\ab}[1]{|#1|} 

\newcommand{\ip}[2]{\braket{#1|#2}} 
\newcommand{\op}[2]{\ket{#1}\!\!\bra{#2}} 
\newcommand{\me}[3]{\braket{#1|#2|#3}} 

\newcommand{\magn}[1]{\!\times\!10^{#1}} 

\renewcommand{\r}{r} 
    \newcommand{\rv}{\mathbf{\r}} 
    \newcommand{\ro}{\qo{\r}} 

\newcommand{\R}{R} 
    \newcommand{\Rv}{\mathbf{\R}} 
    
\newcommand{\nv}{\mathbf{n}} 

\begin{document}


\title{Two-qubit gate protocols with microwave-dressed Rydberg ions in a linear Paul trap}

\author{Joseph W. P. Wilkinson}
\email{joseph.wilkinson@uni-tuebingen.de}
\affiliation{Institut f\"{u}r Theoretische Physik and Center for Integrated Quantum Science and Technology, Universit\"{a}t T\"{u}bingen, Auf der Morgenstelle 14, 72076 T\"{u}bingen, Germany}

\author{Katrin Bolsmann}
\affiliation{Institut f\"{u}r Quanteninformation, RWTH Aachen University, Otto-Blumenthal-Stra\ss{}e 20, 52074 Aachen, Germany}
\affiliation{Institut f\"{u}r Theoretische Nanoelektronik, Forschungszentrum J\"{u}lich, Wilhelm-Johnen-Stra\ss{}e, 52428 J\"{u}lich, Germany}

\author{Thiago L. M. Guedes}
\affiliation{Institut f\"{u}r Quanteninformation, RWTH Aachen University, Otto-Blumenthal-Stra\ss{}e 20, 52074 Aachen, Germany}
\affiliation{Institut f\"{u}r Theoretische Nanoelektronik, Forschungszentrum J\"{u}lich, Wilhelm-Johnen-Stra\ss{}e, 52428 J\"{u}lich, Germany}

\author{Markus M\"{u}ller}
\affiliation{Institut f\"{u}r Quanteninformation, RWTH Aachen University, Otto-Blumenthal-Stra\ss{}e 20, 52074 Aachen, Germany}
\affiliation{Institut f\"{u}r Theoretische Nanoelektronik, Forschungszentrum J\"{u}lich, Wilhelm-Johnen-Stra\ss{}e, 52428 J\"{u}lich, Germany}

\author{Igor Lesanovsky}
\affiliation{Institut f\"{u}r Theoretische Physik and Center for Integrated Quantum Science and Technology, Universit\"{a}t T\"{u}bingen, Auf der Morgenstelle 14, 72076 T\"{u}bingen, Germany}
\affiliation{School of Physics and Astronomy and Centre for the Mathematics and Theoretical Physics of Quantum Non-Equilibrium Systems, The University of Nottingham, The University of, Nottingham, NG7 2RD, United Kingdom}

\date{\today}


\begin{abstract}
    Ultracold trapped atomic ions excited into highly energetic Rydberg states constitute a promising platform for scalable quantum information processing.
    Elementary building blocks for such tasks are high-fidelity and sufficiently fast entangling two-qubit gates, which can be achieved via strong dipole-dipole interactions between microwave-dressed Rydberg ions, as recently demonstrated in a breakthrough experiment~\cite{zhang2020}.
    We theoretically investigate the performance of three protocols leading to controlled-phase gate operations.
    Starting from a microscopic description of Rydberg ions in a linear Paul trap, we derive an effective Hamiltonian that faithfully captures the essential dynamics underlying the gate protocols.
    We then use an optimization scheme to fine-tune experimentally controllable parameters like laser detuning and Rabi frequency to yield maximal gate fidelity under each studied protocol.
    We show how non-adiabatic transitions resulting from fast laser driving relative to the characteristic time scales of the system detrimentally affect the fidelity.
    Despite this, we demonstrate that in the realistic scenario of Rydberg ions with finite radiative lifetimes, optimizing the best found gate protocol enables achievement of fidelities as high as $99.25\,\%$ for a gate time of $0.2\,\unit{\micro\second}$.
    This considerably undercuts entangling gate durations between ground-state ions, for which gate times are typically limited by the comparably slower time scales of vibrational modes.
    Overall, this places trapped Rydberg ions into the regime where fast high-accuracy quantum computing and eventually quantum error correction become possible.
\end{abstract}


\maketitle

\tableofcontents

\addtocontents{toc}{\protect\setcounter{tocdepth}{1}}

\pagenumbering{arabic}


\section{Introduction}\label{sec:introduction}

Trapped ions are among the most promising platforms for quantum technologies, such as quantum computers~\cite{cirac1995, cirac2000, haffner2008, schindler2013, ballance2016, bruzewicz2019, schmidtkaler2003, blatt2008, duan2010, bermudez2017, moses2023}, quantum simulators~\cite{porras2004, kim2010, barreiro2011, friedenauer2008, blatt2012, georgescu2014, monroe2021}, and quantum sensors~\cite{baumgart2016, degen2017, hempel2013, kotler2011, campbell2017, carney2021, gilmore2021}.
They have been successfully used to actualize high-fidelity one- and two-qubit quantum gates~\cite{ballance2016, gaebler2016, benhelm2008}, even enabling the implementation of fault-tolerant protocols~\cite{ryananderson2021, heussen2023, postler2024, pogorelov2024, paetznick2024, reichardt2024a, berthusen2024, wang2024}.
Additionally, gate times on the scale of $\sim 1 \, \unit{\micro\second}$ under specific ion-trajectory modulation~\cite{schafer2018} and coherence times of several minutes~\cite{wang2017} have been reported in trapped ion architectures.
In spite of the significant progress made over the past few decades, however, engineering scalable and controllable interactions free from the influence of the trap-induced vibrational modes remains a challenge~\cite{bruzewicz2019}.

In typical trapped ion systems, qubits are encoded within the energetically low-lying electronic states of the individual atomic ions~\cite{cirac1995, bruzewicz2019}.
The interactions between these states are then mediated by the phonon modes of the ion crystal via laser coupling of the electronic and vibrational degrees of freedom~\cite{schmidtkaler2003b, haljan2005, kim2009, muller2011, schneider2012, behrle2023}.
Established gate protocols utilizing this so-called ``phonon bus'' are, however, limited to smaller ion crystals because the vibrational spectrum becomes increasingly dense and complex as the number of ions increases.
Performing low-error gate operations in larger ion crystals can, therefore, become exceedingly challenging, since most proposed gate protocols require that the phase space trajectory of all of the vibrational modes closes simultaneously at the end of the laser pulse sequence~\cite{schafer2018}.
As a result, the speed of trapped ion gates is generally limited by the characteristic motional frequencies of the ions in the trap and the gate speeds further decrease as one increases the number of ions in the system.

Owing to their strong long-range electric dipolar forces, systems of optically trapped neutral Rydberg atoms offer a propitious alternative platform for quantum technologies~\cite{weimer2010, saffman2010, adams2019, browaeys2020}.
The electric dipolar interactions emerge when atoms are excited to energetically high-lying Rydberg states~\cite{gallagher1994}.
Under extreme conditions, these high excitations can result in the purported ``blockade'' or ``antiblockade'' effects, whereby the excitation of further atoms is impeded or facilitated, respectively~\cite{ates2007, urban2009, gaetan2009, amthor2010, dudin2012, marcuzzi2017}.
These phenomena have now been widely studied and have furthermore been demonstrated to be crucial for the implementation of scalable and fast quantum computation~\cite{jaksch2000, moller2008, isenhower2010, wilk2010, maller2015, theis2016, levine2019, graham2019, graham2022, evered2023, bluvstein2024, reichardt2024b, radnaev2024}.

The fast and strong dipole-induced interactions between Rydberg states, extensively used in neutral atom systems, can provide also a viable route to fast and high-fidelity gates for trapped ions:
Employing ions in highly excited Rydberg states promotes trapped ions to a new platform in which entangling gates are not naturally limited by the comparatively slower oscillation time scales of the ionic motion, but rather inherit the advantageous rapid dynamics of Rydberg interactions~\cite{muller2008, schmidtkaler2011, mokhberi2020, bao2024}.
By incorporating the precise control of trapped ions with the tunable interactions of Rydberg atoms, fast and robust quantum gate operations can be realized with strong dipole-dipole interactions via microwave dressing.
This finding~\cite{muller2008} led to the breakthrough experimental implementation of a submicrosecond entangling phase gate on a pair of trapped Rydberg ions~\cite{zhang2020}.
Furthermore, trapped ions excited to Rydberg states overcome some limitations met by their trapped atom counterparts~\cite{saffman2016}, which require trap switching before Rydberg excitation, resulting in coupling between electronic and vibrational modes.
In contrast, since the Rydberg ions are confined electromagnetically and interact through electric dipolar forces, the vibrational degrees of freedom can be used for the quantum simulation of molecular dynamics and the implementation of spin-laser models~\cite{wilkinson2024, zhang2022, bharti2023, gambetta2020, gambetta2021, magoni2023, martins2024}.

Following the breakthrough experiment in Ref.~\cite{zhang2020}, we theoretically investigate the interaction mechanism behind the implementation of an entangling controlled-Z (CZ) gate between two Rydberg ions and introduce a new optimization routine capable of fine-tuning key interaction parameters to maximize gate fidelities.
Our manuscript is structured as follows.
In section~\ref{sec:derivation}, we outline in detail the derivation of the Hamiltonian describing the dynamics of a chain of interacting trapped Rydberg ions.
We subsequently introduce the electronic states of the ions and the collective vibrational modes of the Coulomb crystal, restricting ourselves to an energetically well isolated subspace of electronic states to derive an effective interaction Hamiltonian for the purely electronic dynamics of the ion chain.
In section~\ref{sec:gates}, we transfer and generalize approaches for entangling-gate implementation in neutral atoms to the Rydberg-ion context.
We devise an optimization protocol that allows for numerical simulation of three different Rydberg-ion entangling-gate implementations with laser detuning and Rabi frequency fine-tuned for maximal fidelity.
This optimization scheme is general and not restricted to two-qubit gates.
We apply the optimization procedure within two regimes, each defining the parameters and bounds for the optimization, where in the first regime, we impose conservative parameter restrictions, while in the second regime, more optimistic bounds are chosen.
For the latter we show that fidelities of $>99 \, \%$ for a gate time of $0.2\,\unit{\micro\second}$ under consideration of finite radiative Rydberg lifetimes can be achieved.
The results highlight the promising potential of Rydberg ions for fast, high-accuracy quantum information processing.


\section{Derivation of the model Hamiltonian}\label{sec:derivation}

In this section, we derive  the Hamiltonian describing a chain of laser-excited singly-charged alkaline-earth Rydberg ions confined within the electric quadrupole potential of a linear Paul trap~\cite{muller2008, schmidtkaler2011, mokhberi2020, wilkinson2024}.
To start, we derive the Hamiltonian for a single Rydberg ion in a linear Paul trap (for experimental details, see Refs.~\cite{schmidtkaler2011, feldker2017, higgins2018, pokorny2020, vogel2021}).
We subsequently generalize the discussion to a chain of interacting trapped Rydberg ions and introduce the electrostatic interactions between them.
To close, we restrict the remainder of the discussion to an energetically well isolated subspace of electronic states relevant for the implementation of the desired quantum gate protocols and obtain the required model Hamiltonian used throughout the following sections.

\subsection{Single laser-excited Rydberg ion in a linear Paul trap}

We consider ions with a single valence electron orbiting in the modified Coulomb potential of a nucleus screened by the remaining core electrons~\cite{gallagher1988, gallagher1994, gallagher2023}.
In order to practically model the dynamics of such a system, it is useful to employ an approximation which reduces the many-body problem to an effective two-body problem.
This is feasible since the inner core electrons form closed shells around the nucleus that contribute negligibly to the dynamics and interactions.
Hence, the ions can be modeled as consisting of an ionic core (i.e., the nucleus and core electrons) with charge~$2e$ and mass~$m_{\mathrm{c}}$ and a valence electron of charge~$-e$ and mass~$m_{\mathrm{e}}$.
The Hamiltonian describing the effective dynamics of the ion is then approximated by
\begin{equation}
    H \approx \frac{\bm{p}_{\mathrm{c}}^{2}}{2 m_{\mathrm{c}}} + \frac{\bm{p}_{\mathrm{e}}^{2}}{2 m_{\mathrm{e}}} + V(\ab{\bm{r}_{\mathrm{e}} - \bm{r}_{\mathrm{c}}}),
\end{equation}
where $\bm{r}_{\mathrm{c}}$, $\bm{p}_{\mathrm{c}}$ and $\bm{r}_{\mathrm{e}}$, $\bm{p}_{\mathrm{e}}$ are the positions and momenta of the ionic core and valence electron, respectively.
The effective interaction between these charges is then approximated by a parametric model potential $V(\ab{\bm{r}_{\mathrm{e}} - \bm{r}_{\mathrm{c}}})$ that depends on the relative position $\bm{r}_{\mathrm{e}} - \bm{r}_{\mathrm{c}}$ of the valence electron with respect to the ionic core and on the orbital angular momentum quantum number $l$ of the electronic state (see App.~\ref{app:electronic-states} or Ref.~\cite{aymar1996}).
Here, the dependence on the orbital angular momentum accounts for the \textit{quantum defect}~\cite{seaton1983, gallagher1988, gallagher1994, gallagher2023} which quantifies the lowering of the energy of electronic states with low angular momentum quantum number (i.e., $l \leq 4$) due to the probing of the core electrons in the closed inner shells by the valence electron in the open outer shell.
While the parametric model potential is expected to accurately reproduce the measured energy level spectrum of the ions~\cite{nist2024}, it is not guaranteed to capture all relativistic and quantum field theoretic effects (e.g., electron-electron correlations~\cite{grant2023}), since it approximates a many-electron problem as an effective one-electron problem.
Therefore, we expect some minor discrepancies between experimentally measured and theoretically calculated values for the electronic wavefunctions and, thus, the electronic matrix elements~\cite{schmidtkaler2011, pawlak2020}.

The ions are confined within a \textit{linear Paul trap}~\cite{paul1990, brown1991, raizen1992}.
This provides three-dimensional confinement for charged particles through a combination of static and oscillating electric-field components~\cite{wineland1998, leibfried2003, major2005, hucul2008}.
These electric-field components generate an electric quadrupole potential at the trap center of the form
\begin{equation}\label{eq:quad-pot}
    \Phi(\bm{r}, t) = \alpha \cos(\nu t) [r_{x}^{2} - r_{y}^{2}] + \beta [3 r_{z}^{2} - \bm{r}^{2}].
\end{equation}
Here, $\alpha$ and $\beta$ are the oscillating and static electric-field gradients, which are determined by the geometry of the trap electrodes and applied voltages, and $\nu$ is the (radio) frequency.
For typical experimental parameters~\cite{mokhberi2020}, we find $\alpha \sim 10^{9} \, \unit{\volt\meter}^{-2}$ and $\beta \sim 10^{7} \, \unit{\volt\meter}^{-2}$, while the associated frequency $\nu \sim 2 \pi \times 10 \, \unit{\mega\hertz}$.
We note that in more general Paul traps, a dimensionless parameter can be introduced to break the axial symmetry and lift the degeneracy of the radial modes~\cite{mokhberi2020}.
However, for simplicity, we do not account for this breaking of symmetry and lifting of degeneracy here.
To facilitate transitions between the bound electronic states of the trapped ions, we additionally consider a time-dependent homogeneous laser or microwave electric-field contribution $\bm{E}(t)$ that we treat within the dipole approximation~\cite{foot2004}.

Taking into account the coupling between the charges of the ionic core and valence electron with the electric potential of the linear Paul trap and excitation field modes $\bm{E}(t)$, which is taken care of via the minimal coupling replacement~\cite{steck2024}, the Hamiltonian of the trapped Rydberg ion is well approximated by
\begin{equation}
    H \approx \underbrace{\frac{\bm{p}_{\mathrm{c}}^{2}}{2 m_{\mathrm{c}}} + \frac{\bm{p}_{\mathrm{e}}^{2}}{2 m_{\mathrm{e}}} + V(\ab{\bm{r}_{\mathrm{e}} - \bm{r}_{\mathrm{c}}})}_{\text{ion}} + \underbrace{\phantom{\frac{}{_{}}}\hspace{-2.5pt} 2 e \Phi(\bm{r}_{\mathrm{c}}, t) - e \Phi(\bm{r}_{\mathrm{e}}, t)}_{\text{ion-trap}} + \underbrace{\phantom{\frac{}{_{}}}\hspace{-2.5pt} 2 e \bm{r}_{\mathrm{c}} \cdot \bm{E}(t) - e \bm{r}_{\mathrm{e}} \cdot \bm{E}(t)}_{\text{ion-ext.\,field}} \!.
\end{equation}

It is convenient to treat the system within the center of mass frame.
As such, we introduce the center of mass and relative positions and momenta $\bm{R}$, $\bm{P}$ and $\bm{r}$, $\bm{p}$, which are defined in terms of the ionic core and valence electron positions and momenta by
\begin{equation}
    \bm{R} = \frac{m_{\mathrm{c}} \bm{r}_{\mathrm{c}} + m_{\mathrm{e}} \bm{r}_{\mathrm{e}}}{m_{\mathrm{c}} + m_{\mathrm{e}}}, \qquad
    \bm{P} = \bm{p}_{\mathrm{c}} + \bm{p}_{\mathrm{e}}, \qquad
    \bm{r} = \bm{r}_{\mathrm{e}} - \bm{r}_{\mathrm{c}}, \qquad
    \text{and} \qquad
    \bm{p} = \frac{m_{\mathrm{c}} \bm{p}_{\mathrm{e}} - m_{\mathrm{e}} \bm{p}_{\mathrm{c}}}{m_{\mathrm{c}} + m_{\mathrm{e}}}.
\end{equation}
The associated center of mass and relative masses (i.e., total mass $M$ and reduced mass $m$) are then
\begin{equation}
    M = m_{\mathrm{c}} + m_{\mathrm{e}} \qquad
    \text{and} \qquad
    m = \frac{m_{\mathrm{c}} m_{\mathrm{e}}}{m_{\mathrm{c}} + m_{\mathrm{e}}}.
\end{equation}
Represented in terms of the center of mass and relative coordinates, and exploiting that the mass of the ionic core is about five orders of magnitude larger than that of the valence electron (i.e., $m_{\mathrm{c}} \gg m_{\mathrm{e}}$), the trapped Rydberg ion Hamiltonian can be rewritten approximately as~\cite{muller2008}
\begin{equation}\label{eq:stationary-hamiltonian}
    H \approx \underbrace{\frac{\bm{P}^{2}}{2 M} + e \Phi(\bm{R}, t) + e \bm{R} \cdot \bm{E}(t)}_{\text{ion (i.e., center of mass)}} + \underbrace{\frac{\bm{p}^{2}}{2 m} + V(\ab{\bm{r}}) - e \Phi(\bm{r}, t) - e \bm{r} \cdot \bm{E}(t)}_{\text{electron (i.e., relative)}} - \underbrace{\frac{}{}\hspace{-2.5pt} e \bm{r} \cdot \boldsymbol{\nabla} \Phi(\bm{R}, t)}_{\text{ion-electron}} \!.
\end{equation}

From direct inspection of Eq.~\eqref{eq:quad-pot}, it follows that the electric quadrupole potential of the linear Paul trap provides static confinement along the trap axis (i.e., the $z$-axis), however, at no instant of time is static confinement present in the radial plane (i.e., the $xy$-plane).
Instead, the rapidly oscillating direction of the confining force creates a periodically oscillating potential saddle point located at the trap center~\cite{berkeland1998}.
As such, the ions experience an effectively static ponderomotive harmonic potential that provides radial confinement~\cite{cook1985}.
To manifest this separation of the static and oscillating motion, we unitarily transform the system into an oscillating frame at the electric-field frequency $\nu$ via the unitary operator
\begin{equation}
    U = \exp\!\bigg( \mathrm{i} \frac{e \alpha}{\hbar \omega} \sin(\nu t) [R_{x}^{2} - R_{y}^{2}] \bigg).
\end{equation}
Applying this to the trapped Rydberg ion Hamiltonian, we arrive at (see App.~\ref{app:unitary-transformation} for details)
\begin{equation}\label{eq:oscillating-hamiltonian}
    H \mapsto U H U^{\dagger} + \mathrm{i} \hbar \frac{\partial U}{\partial t} U^{\dagger} = H_{\mathrm{ex}} + H_{\mathrm{in}} + H_{\mathrm{co}},
\end{equation}
where $H_{\mathrm{ex}}$, $H_{\mathrm{in}}$, and $H_{\mathrm{co}}$ are the Hamiltonians describing the external dynamics of the ion, the internal dynamics of the electron, and the coupled dynamics between them that arises due to the electric potential of the linear Paul trap.
Note that in the following, we will often refer to the motion of the center of mass and relative coordinates of the ion as the external or vibrational dynamics and the internal or electronic dynamics, respectively.
Their nonseparable motion is then termed the coupled or vibronic dynamics.

The Hamiltonian governing the external dynamics of the trapped ion can be separated into the aforementioned static and oscillating terms as 
\begin{equation}
    H_{\mathrm{ex}} = \underbrace{\frac{\bm{P}^{2}}{2 M} + \frac{M}{2} \sum_{\mathclap{u}} \omega_{u}^{2} R_{u}^{2}}_{\text{secular motion}} - \underbrace{\phantom{\sum_{u}}\hspace{-14.5pt} \frac{2 e \alpha}{M \nu} \sin(\nu t) [R_{x} P_{x} - R_{y} P_{y}] - \frac{e^{2} \alpha^{2}}{M \nu^{2}} \cos(2 \nu t) [R_{x}^{2} + R_{y}^{2}] + e \bm{R} \cdot \bm{E}(t)}_{\text{micromotion}}
\end{equation}
with $u = x, y, z$ and where we have introduced the frequencies of the effectively static harmonic trap,
\begin{equation}
    \omega_{x} = \omega_{y} = \sqrt{\frac{2 e^{2} \alpha^{2}}{M^{2} \nu^{2}} - \frac{2 e \beta}{M}} \qquad
    \text{and} \qquad
    \omega_{z} = \sqrt{\frac{4 e \beta}{M}}.
\end{equation}
The former time-independent terms of the external Hamiltonian describe the slow \textit{secular motion} of the trapped ion and the latter time-dependent terms the fast driven motion, referred to as \textit{micromotion}, due to the oscillating electric potential~\cite{berkeland1998}.
For the experimental parameter ranges under consideration~\cite{mokhberi2020}, the fast driven motion occurs on much shorter timescales than the slow secular motion.
Therefore, the micromotion due to the oscillating electric-field mode of the linear Paul trap has a negligible effect on the motion of the trapped ions~\cite{cook1985}.
This follows from the fact that the coefficients of the periodic time-dependent functions vary slowly with time (i.e., with the slow secular motion of the ion) and as such the associated functions can be well approximated by their time-integrated values over the period of the secular motion.
For both $\sin(\nu t)$ and $\cos(2 \nu t)$, these evaluate to zero and can be safely neglected.
Further note that to avoid resonances, which is necessary to ensure stable trapping, the condition $M \nu^{2} \gg 2 \sqrt{2} e \alpha$, which implies $\nu \gg \omega_{u}$, must necessarily be satisfied~\cite{cook1985}, as it is for the experimental parameters under consideration.
The additional micromotion due to the oscillating external driving-field mode $\bm{E}(t)$ can also be omitted since the frequency of the mode is tuned to the transitions between the electronic states of the ions.
These have energy splittings of the order of $\unit{\giga\hertz}$ to $\unit{\tera\hertz}$, while those of the vibrational motion are of the order of $\unit{\mega\hertz}$.
This yields the approximate external Hamiltonian
\begin{equation}
    H_{\mathrm{ex}} \approx \underbrace{\frac{\bm{P}^{2}}{2 M} + \frac{M}{2} \sum_{\mathclap{u}} \omega_{u}^{2} R_{u}^{2}}_{\mathrm{ion}}\!.
\end{equation}

The internal-dynamics Hamiltonian describing  the motion of the valence electron in the parametric model potential of the screened ionic core together with the electric potential of the linear Paul trap and the external driving-field mode reads
\begin{equation}
    H_{\mathrm{in}} = \underbrace{\frac{\bm{p}^{2}}{2 m} + V(\ab{\bm{r}})}_{\text{electron}} - \underbrace{\frac{}{}\hspace{-2.5pt} e \alpha \cos(\nu t) [r_{x}^{2} - r_{y}^{2}] - e \beta [3 r_{z}^{2} - \bm{r}^{2}]}_{\text{electron-trap}} - \underbrace{\frac{}{}\hspace{-2.5pt} e \bm{r} \cdot \bm{E}(t)}_{\mathclap{\mathrm{electron\text{-}ext.\, field}}} \!.
\end{equation}
Here, the former terms are the kinetic and potential energy terms of the bound electron in the absence of external electromagnetic-field contributions which together constitute the field-free electronic Hamiltonian.
The latter terms account for the interaction of the charge of the valence electron with the electric potential of the linear Paul trap as well as with externally applied driving electric-field modes, such as an excitation laser or microwave.
Compared to the center of mass motion of the trapped ion the relative dynamics of the electron occurs on much shorter timescales.
However, due to the electric potential of the linear Paul trap, the vibrational and electronic degrees of freedom associated to the center of mass and relative motion are nonseparable.
Their coupled motion is accounted for by
\begin{equation}
    H_{\mathrm{co}} = \underbrace{- 2 e \alpha \cos(\nu t) [R_{x} r_{x} - R_{y} r_{y}] - 2 e \beta [3 R_{z} r_{z} - \bm{R} \cdot \bm{r}]}_{\text{electron-ion}} \!.
\end{equation}

\subsection{Electrostatic interactions between trapped Rydberg ions}

\begin{figure}
    \centering
    \includegraphics[scale=1]{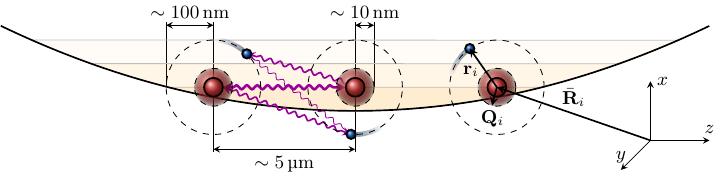}
    \caption{
        Relevant interactions and lengths for a one-dimensional chain of Rydberg ions within the effectively static harmonic potential of a linear Paul trap.
        At low temperatures and with tight radial trapping, the Rydberg ions align along the $z$-axis of the linear Paul trap to form a one-dimensional Coulomb crystal.
        The equilibrium positions and, therefore, the average distance between adjacent ions are determined by the balance between the repulsive Coulomb forces among the ions and attractive trapping forces confining them~\cite{james1998}.
        Typically (see main text for details), these are $\sim 5 \, \unit{\micro\meter}$.
        The amplitudes of the oscillations of the ground state ions about their equilibrium positions (i.e., the effective size of the vibrational wavepacket) are determined by the external trapping frequencies.
        Usually in experiments, these are of the order of $\unit{\mega\hertz}$, resulting in oscillation amplitudes of $\sim 10 \, \unit{\nano\meter}$.
        In contrast, the orbital radius of the valence electron, i.e., the effective size of the electronic wavepacket, scales with the principal quantum number $n$ of the electronic energy eigenstate.
        For the ground state, this is significantly smaller than the expected oscillations.
        However, for Rydberg states this can change drastically and assume values up to $\sim 100 \, \unit{\nano\meter}$.
        Shown on the left are the interactions (purple lines) between the charges of the ionic cores (red balls) and valence electrons (blue balls) of a pair of adjacent ions due to the Coulomb potential.
    }
    \label{fig:trapped-rydberg-ions}
\end{figure}

We now turn to the discussion of the electromagnetic interactions between the charges of the ions, as depicted in Fig.~\ref{fig:trapped-rydberg-ions}.
Denoting this potential between ions $i$ and $j$ by $V_{ij}$ and that of the Hamiltonian of ion $i$ [cf. Eq.~\eqref{eq:oscillating-hamiltonian}] by $H_{i}$, it follows that we can write the Hamiltonian for $N$ interacting trapped Rydberg ions as
\begin{equation}\label{eq:many-body-hamiltonian}
    H = \sum_{i = 1}^{N} H_{i} + \frac{1}{2} \sum_{\mathclap{\substack{i, j = 1 \\ j \neq i}}}^{N} V_{ij} = H_{\mathrm{ex}} + H_{\mathrm{in}} + H_{\mathrm{co}}.
\end{equation}
Similar to Eq.~\eqref{eq:oscillating-hamiltonian}, we have introduced the Hamiltonian terms $H_{\mathrm{ex}}$, $H_{\mathrm{in}}$, and $H_{\mathrm{co}}$ describing the external, internal, and coupled dynamics of the interacting trapped Rydberg ions. 
 
Following Refs.~\cite{muller2008, weber2017}, we neglect retardation effects~\cite{casimir1948}, since the average distance between adjacent ions is significantly smaller than the wavelengths associated to the Rydberg-Rydberg interactions.
Indeed, in current experiments where the harmonic frequencies are typically of the order of $2 \pi \times 1 \, \unit{\mega\hertz}$~\cite{mokhberi2020}, the separation between the ions is approximately $5 \, \unit{\micro\meter}$.
In contrast, the characteristic wavelengths corresponding to the energy scale of the electric dipole-dipole interaction between Rydberg states are at least of the order of $1 \, \unit{\meter}$.
Moreover, for the experimental parameter regimes under consideration, the distances between adjacent ions are sufficiently larger than the spatial extent of the electronic wavefunctions.
For example, for bound electronic states of a strontium $^{88}\mathrm{Sr}^{+}$ ion with principal quantum number $n \sim 50$, the LeRoy radius (which approximately defines the distance at which the electrostatic interactions between the ions can be treated classically) is about $400 \, \unit{\nano\meter}$~\cite{leroy1974}.
Hence, we can ignore the exchange and charge overlap interactions~\cite{weber2017}.
These assumptions drastically simplify the calculations of the interactions between the ions as they facilitate the treatment of the valence electrons and ionic cores as distinguishable particles.
This allows us to employ a multipole expansion to describe the interactions between the charges of the ions~\cite{friedrich2017}.

The potential describing the electrostatic interactions between the charges of the ionic cores and the valence electrons of ions $i$ and $j$ can be written as (see Fig.~\ref{fig:trapped-rydberg-ions})
\begin{equation}\label{eq:interaction-potential}
    \frac{V_{ij}}{C e^{2}} = \frac{4}{\ab{\bm{R}_{i} - \bm{R}_{j}}} - \frac{2}{\ab{\bm{R}_{i} - \bm{R}_{j} + \bm{r}_{i}}} - \frac{2}{\ab{\bm{R}_{i} - \bm{R}_{j} - \bm{r}_{j}}} + \frac{1}{\ab{\bm{R}_{i} - \bm{R}_{j} + \bm{r}_{i} - \bm{r}_{j}}},
\end{equation}
where $\bm{R}_{i}$, $\bm{r}_{i}$ and $\bm{R}_{j}$, $\bm{r}_{j}$ are the center of mass and relative positions of the ions $i$ and $j$ and $C = 1 / 4 \pi \epsilon_{0}$ is the Coulomb constant with $\epsilon_{0}$ the vacuum permittivity.
For the parameter regimes typically considered in current trapped Rydberg ion experiments, the center of mass distances between the ions are much greater than the relative distances between the ionic cores and their valence electrons~\cite{schmidtkaler2011, mokhberi2020}, that is, $\ab{\bm{R}_{i} - \bm{R}_{j}} \gg \ab{\bm{r}_{i}}$.
This is well fulfilled, since---as discussed prior---we are interested in the parameter regime wherein the distances between adjacent trapped ions of $\sim 5 \, \unit{\micro\meter}$, are much greater than the spatial extension of their electronic wavefunctions (i.e., the orbital size of the Rydberg ion, which due to the scaling with the principal quantum number $n$ can assume values up to $\sim 100 \, \unit{\nano\meter}$~\cite{higgins2019}).
Therefore, we can sufficiently well approximate the potential between the ions by neglecting the higher order corrections of the multipole expansion.
For simplicity, we only consider terms up to second order.
Accordingly, the interaction potential can be approximated by (see App.~\ref{app:multipole-expansion} for a derivation)
\begin{equation}\label{eq:multipole-expansion}
    \frac{V_{ij}}{C e^{2}} = \frac{1}{\ab{\bm{R}_{ij}}} + \frac{\bm{n}_{ij} \cdot \bm{r}_{i}}{\ab{\bm{R}_{ij}}^{2}} - \frac{\bm{n}_{ij} \cdot \bm{r}_{j}}{\ab{\bm{R}_{ij}}^{2}} - \frac{3 [\bm{n}_{ij} \cdot \bm{r}_{i}]^{2} - \bm{r}_{i}^{2}}{2 \ab{\bm{R}_{ij}}^{3}} - \frac{3 [\bm{n}_{ij} \cdot \bm{r}_{j}]^{2} - \bm{r}_{j}^{2}}{2 \ab{\bm{R}_{ij}}^{3}} - \frac{3 [\bm{n}_{ij} \cdot \bm{r}_{i}] [\bm{n}_{ij} \cdot \bm{r}_{j}] - \bm{r}_{i} \cdot \bm{r}_{j}}{\ab{\bm{R}_{ij}}^{3}},
\end{equation}
where we have introduced the following shorthand notations for the center of mass positions,
\begin{equation}
    \bm{n}_{ij} = \frac{\bm{R}_{ij}}{\ab{\bm{R}_{ij}}} \qquad
    \text{and} \qquad
    \bm{R}_{ij} = \bm{R}_{i} - \bm{R}_{j}.
\end{equation}
The first term describes the Coulomb repulsion between two charges, i.e., the interaction between the electric charge of ion $i$ with that of ion $j$.
The second and third terms are the dipole-charge and charge-dipole interactions, namely, the interactions between the electric dipole moment of ion $i$ with the charge of ion $j$ and vice versa.
These arise due to the displacement of the orbiting valence electron from its ionic core which leads to the induction of an electric dipole moment that interacts with the charge of the other ion.
Similarly, the fourth and fifth terms, which are the quadrupole-charge and charge-quadrupole interactions, respectively, emerge from the induced electric quadrupole moment of each ion interacting with the charge of the other ion, with the former arising analogously to the electric dipole moment.
It can be shown (e.g., by retaining the effective charge number of the ionic core, see App.~\ref{app:multipole-expansion}) that each of these terms is absent for the case of interacting neutral Rydberg atoms.
The sixth and final term is the well-known dipole-dipole interaction.

At sufficiently low temperatures $k_{\mathrm{B}} T \ll \hbar \omega_{u}$ and under tight radial trapping $\alpha \gg \beta$ (i.e., $\omega_{x}, \omega_{y} > \omega_{z}$), the ions align along the trap axis (i.e., $z$-axis) to form a one-dimensional Coulomb crystal~\cite{bollinger1994}.
In such structures, the ions vibrate about their equilibrium positions, which are determined by the balance of the repulsive electrostatic forces between the ions and the attractive trapping forces confining the ions~\cite{james1998}.
Under typical trapping conditions, the amplitudes of the oscillations of the ions about their equilibrium positions, usually $\sim 10 \, \unit{\nano\meter}$, are significantly smaller than the distances between the ions, yet much larger than the orbital radii of the ground state ions~\cite{mokhberi2020}.
However, when excited to Rydberg states, the orbital radius dramatically increases, which necessitates that the ions be modeled as \textit{composite} objects~\cite{muller2008}.

The equilibrium positions follow from the stationary point of the potential of the external dynamics, which governs the center of mass motion of the ions, and are calculated by solving the coupled equations
\begin{equation}\label{eq:equilibrium-position-ocondition}
    \boldsymbol{\nabla}_{i} V_{\mathrm{ex}}|_{\bm{R}_{i} = \bar{\bm{R}}_{i}} = \bm{0},
\end{equation}
where $\bar{\bm{R}}_{i}$ is the equilibrium position of ion $i$ and the external potential $V_{\mathrm{ex}}$ is given by
\begin{equation}
    V_{\mathrm{ex}} = \frac{M}{2} \sum_{\mathclap{i = 1}}^{N} \sum_{\mathclap{u}} \omega_{u}^{2} R_{i; u}^{2} + \frac{C e^{2}}{2} \sum_{\mathclap{\substack{i, j = 1 \\ j \neq i}}}^{N} \frac{1}{\ab{\bm{R}_{ij}}}.
\end{equation}
In what follows, it will prove convenient to express the trap frequencies as $\omega_{u} = \gamma_{u} \omega$ with $\gamma_{x} = \gamma_{y} = \gamma$ and $\gamma_{z} = 1$, where we have introduced the characteristic frequency of the trap $\omega$ and associated anisotropy $\gamma$ characterizing the relative strength of the radial to axial trapping defined by
\begin{equation}\label{eq:trap-frequency}
    \omega = \sqrt{\frac{4 e \beta}{M}} \qquad
    \text{and} \qquad
    \gamma = \sqrt{\frac{2 e^{2} \alpha^{2}}{M^{2} \omega^{2} \nu^{2}} - \frac{1}{2}}.
\end{equation}
For $\gamma < \gamma_{*}$, where $\gamma_{*}$ is the critical anisotropy of the trap, which scales with the number of trapped ions $N$ as $\gamma_{*} \sim 0.583 \smash{N^{0.9}}$~\cite{enzer2000}, the trapped ions undergo a structural phase transition from a one- to two-dimensional Coulomb crystal~\cite{fishman2008}.
Here, however, we only consider the regime where $\gamma > \gamma_{*}$ within which the ions align along the trap axis with equilibrium positions $\bar{\bm{R}}_{i} = (0, 0, \bar{R}_{i; z})$.
Introducing the characteristic length $L$ associated to the equilibrium distance between the ions~\cite{james1998},
\begin{equation}\label{eq:ion-distance}
    L = \sqrt[3]{\frac{C e^{2}}{M \omega^{2}}},
\end{equation}
and the corresponding dimensionless equilibrium positions $Z_{i}$, defined by $\bar{R}_{i; z} = L Z_{i}$, the coupled system of equations in Eq.~\eqref{eq:equilibrium-position-ocondition} can be succinctly recast as (see App.~\ref{app:equilibrium-positions})
\begin{equation}\label{eq:equilibrium-position}
    Z_{i} = \sum_{\mathclap{\substack{j = 1 \\ j \neq i}}}^{N} \frac{Z_{ij}}{\ab{Z_{ij}}^{3}},
\end{equation}
where $Z_{ij} = Z_{i} - Z_{j}$ with the implicit assumption that $Z_{1} < Z_{2} < \cdots < Z_{N}$.

Following Ref.~\cite{james1998}, we now perform a harmonic expansion of the center of mass positions about their equilibrium positions.
This will ultimately allow us to express the external vibrational dynamics in terms of normal phonon modes.
The Hamiltonian governing said center of mass motion is given by
\begin{equation}\label{eq:hamiltonian-external}
    H_{\mathrm{ex}} = \frac{1}{2 M} \sum_{\mathclap{i = 1}}^{N} \bm{P}_{i}^{2} + \frac{M \omega^{2}}{2} \sum_{\mathclap{i = 1}}^{N} \sum_{\mathclap{u}} \gamma_{u}^{2} R_{i; u}^{2} + \frac{C e^{2}}{2} \sum_{\mathclap{\substack{i, j = 1 \\ j \neq i}}}^{N} \frac{1}{\ab{\bm{R}_{ij}}}.
\end{equation}
Neglecting terms higher than second order, which is well justified by the typical lengths of the equilibrium distance between adjacent ions $\sim 5 \, \unit{\micro\meter}$ compared to their oscillation amplitudes $\sim 10 \, \unit{\nano\meter}$~\cite{muller2008}, the Hamiltonian of the external vibrational dynamics reads (see App.~\ref{app:harmonic-expansion} for details)
\begin{equation}\label{eq:hamiltonian-external-expanded}
    H_{\mathrm{ex}} = \underbrace{\frac{1}{2 M} \sum_{\mathclap{i = 1}}^{N} \bm{P}_{i}^{2} + \frac{M \omega^{2}}{2} \sum_{\mathclap{i, j = 1}}^{N} \sum_{u} K_{ij; u} Q_{i; u} Q_{j; u}}_{\mathrm{ion}} \!.
\end{equation}
Here, $\bm{Q}_{i} = \bm{R}_{i} - \bar{\bm{R}}_{i}$ is the displacement of the center of mass position $\bm{R}_{i}$ of ion $i$ from its equilibrium position $\bar{\bm{R}}_{i}$.
The coefficients $K_{ij; u}$ are components of the Hessian matrix, defined by
\begin{equation}
    K_{ij; x} = K_{ij; y} = \delta_{ij} \gamma^{2} - K_{ij} \qquad
    \text{and} \qquad
    K_{ij; z} = 2 K_{ij} + \delta_{ij},
\end{equation}
where we have introduced the generalized Hessian matrix components $K_{ij}$ which are in turn defined by
\begin{equation}\label{eq:generalized-hessian}
    K_{ij} = \delta_{ij} \sum_{\mathclap{k = 1}}^{N} \frac{1}{\ab{Z_{ik}}^{3}} - \frac{1}{\ab{Z_{ij}}^{3}}.
\end{equation}

Let us now consider the Hamiltonian of the internal electronic dynamics, which describes the relative motion of the valence electrons in the modified Coulomb potential of the screened nuclei superposed by the electric potentials of the linear Paul trap, external driving electric-field modes, and other valence electrons.
It reads
\begin{equation}\label{eq:hamiltonian-internal}
\begin{aligned}
    H_{\mathrm{in}} & = \sum_{\mathclap{i = 1}}^{N} \bigg[ \frac{\bm{p}_{i}^{2}}{2 m} + V(\ab{\bm{r}_{i}}) \bigg] - e \alpha \cos(\nu t) \sum_{\mathclap{i = 1}}^{N} [r_{i; x}^{2} - r_{i; y}^{2}] - e \beta \sum_{\mathclap{i = 1}}^{N} [3 r_{i; z}^{2} - \bm{r}_{i}^{2}] - e \sum_{\mathclap{i = 1}}^{N} \bm{r}_{i} \cdot \bm{E}(t) \\
    & \qquad - \frac{C e^{2}}{2} \sum_{\mathclap{\substack{i, j = 1 \\ j \neq i}}}^{N} \bigg[ \frac{3 [\bm{n}_{ij} \cdot \bm{r}_{i}]^{2} - \bm{r}_{i}^{2}}{2 \ab{\bm{R}_{ij}}^{3}} + \frac{3 [\bm{n}_{ij} \cdot \bm{r}_{j}]^{2} - \bm{r}_{j}^{2}}{2 \ab{\bm{R}_{ij}}^{3}} + \frac{3 [\bm{n}_{ij} \cdot \bm{r}_{i}][\bm{n}_{ij} \cdot \bm{r}_{j}] - \bm{r}_{i} \cdot \bm{r}_{j}}{\ab{\bm{R}_{ij}}^{3}} \bigg].
\end{aligned}
\end{equation}
After similarly performing the harmonic expansion about the equilibrium positions of the ions, we find that the Hamiltonian of the internal electronic dynamics can be written as (see App.~\ref{app:harmonic-expansion})
\begin{equation}\label{eq:hamiltonian-internal-expanded}
\begin{aligned}
    H_{\mathrm{in}} & = \underbrace{\sum_{\mathclap{i = 1}}^{N} \bigg[ \frac{\bm{p}_{i}^{2}}{2 m} + V(\ab{\bm{r}_{i}}) \bigg]}_{\mathrm{electron}} - \underbrace{e \alpha \cos(\nu t) \sum_{\mathclap{i = 1}}^{N} [r_{i; x}^{2} - r_{i; y}^{2}] - e \beta \sum_{\mathclap{i = 1}}^{N} K_{ii; z} [3 r_{i; z}^{2} - \bm{r}_{i}^{2}]}_{\mathrm{electron\text{-}trap}} - \underbrace{e \sum_{\mathclap{i = 1}}^{N} \bm{r}_{i} \cdot \bm{E}(t)}_{\mathrm{electron\text{-}ext.\,field}} \\
    & \qquad + \underbrace{\frac{M \omega^{2}}{4} \sum_{\mathclap{\substack{i, j = 1 \\ j \neq i}}}^{N} K_{ij; z} [3 r_{i; z} r_{j; z} - \bm{r}_{i} \cdot \bm{r}_{j}]}_{\mathrm{electron\text{-}electron}}.
\end{aligned}
\end{equation}
Here, we can identify the first term as the field-free electronic Hamiltonian and the last term as the electric dipole-dipole interaction~\cite{muller2008}.
Note that apart from the introduction of the electric dipole-dipole interaction, the second order terms from the multipole expansion of the interaction potential only result in a position-dependent modification to the static electric field gradient of the linear Paul trap.

Finally, we turn to the coupling between the external vibrational motion and internal electronic states.
This term constitutes the coupled dynamics of the single trapped ions, which arises due to the coupling of the motion of the charges of the ions with the electric quadrupole potential of the linear Paul trap, and the dipolar terms of the multipole expansion of the interaction potential.
It follows as
\begin{equation}\label{eq:hamiltonian-coupled}
\begin{aligned}
    H_{\mathrm{co}} & = - 2 e \alpha \cos(\nu t) \sum_{\mathclap{i = 1}}^{N} [R_{i; x} r_{i; x} - R_{i; y} r_{i; y}] - 2 e \beta \sum_{\mathclap{i = 1}}^{N} [3 R_{i; z} r_{i; z} - \bm{R}_{i} \cdot \bm{r}_{i}] + \frac{C e^{2}}{2} \sum_{\mathclap{\substack{i, j = 1 \\ j \neq i}}}^{N} \bigg[ \frac{\bm{n}_{ij} \cdot \bm{r}_{i}}{\ab{\bm{R}_{ij}}^{2}} - \frac{\bm{n}_{ij} \cdot \bm{r}_{j}}{\ab{\bm{R}_{ij}}^{2}} \bigg].
\end{aligned}
\end{equation}
Similarly performing the harmonic expansion, we obtain (see App.~\ref{app:harmonic-expansion})
\begin{equation}\label{eq:hamiltonian-coupled-expanded}
    \qo{H}_{\mathrm{co}} = \underbrace{- 2 e \alpha \cos(\nu t) \sum_{\mathclap{i = 1}}^{N} [Q_{i; x} r_{i; x} - Q_{i; y} r_{i; y}] - 2 e  \beta \sum_{\mathclap{i, j = 1}}^{N} K_{ij; z} [3 Q_{i; z} r_{j; z} - \bm{Q}_{i} \cdot \bm{r}_{j}]}_{\mathrm{electron\text{-}ion}} \!.
\end{equation}

From here, we transform the Hamiltonian describing the external vibrational dynamics into a diagonal form by introducing the phonon modes via the ladder operator method.
To do so, we express the canonical coordinates, that is, the center of mass displacement and momentum $Q_{i; u}$ and $P_{i; u}$ in terms of bosonic creation and annihilation operators $a_{p; u}^{\dagger}$ and $a_{p; u}$,
\begin{equation}
    Q_{i; u} = l \sum_{\mathclap{p = 1}}^{N} \frac{1}{\sqrt{\gamma_{p; u}}} \Gamma_{ip; u} [a_{p; u}^{\dagger} + a_{p; u}], \qquad
    P_{i; u} = \mathrm{i} l M \omega \sum_{\mathclap{p = 1}}^{N} \sqrt{\gamma_{p; u}}\, \Gamma_{ip; u} [a_{p; u}^{\dagger} - a_{p; u}],
\end{equation}
where we have introduced the characteristic length $l$ associated to the equilibrium oscillations of the ions given by
\begin{equation}
    l = \sqrt{\frac{\hbar}{2 M \omega}}.
\end{equation}
The dimensionless phonon mode frequencies $\gamma_{p; u}$ are computed by numerically diagonalizing the generalized Hessian matrix, the eigenequation of which can be written in component form as (see App.~\ref{app:phonon-modes})
\begin{equation}
    \sum_{\mathclap{j = 1}}^{N} K_{ij} \Gamma_{jp} = \gamma_{p}^{2} \Gamma_{ip}.
\end{equation}
Similar to the Hessian matrix components $K_{ij; u}$, the generalized dimensionless phonon mode frequencies $\gamma_{p}$ are defined via the relations
\begin{equation}
    \gamma_{p; x}^{2} = \gamma_{p; y}^{2} = \gamma^{2} - \gamma_{p}^{2}, \qquad
    \gamma_{p; z}^{2} = 2 \gamma_{p}^{2} + 1
\end{equation}
with the generalized orthogonal matrix components $\Gamma_{ip}$ similarly related to the orthogonal matrix components $\Gamma_{ip; u}$ by $\Gamma_{ip; x} = \Gamma_{ip; y} = \Gamma_{ip; z} = \Gamma_{ip}$.
In diagonal form, the Hamiltonian governing the external vibrational dynamics is then given by (see App.~\ref{app:phonon-modes})
\begin{equation}\label{eq:hamiltonian-external-phonon}
    H_{\mathrm{ex}} = \hbar \omega \sum_{\mathclap{p = 1}}^{N} \sum_{u} \gamma_{p; u} a_{p; u}^{\dagger} a_{p; u}.
\end{equation}

In the absence of the electric potential of the trap, the external driving-field mode, and other ions, the electronic quantum state of each Rydberg ion can be characterized by the principal $n$, orbital angular momentum $l$, spin angular momentum $s$, total angular momentum $j$, and total magnetic $m_{j}$ quantum numbers.
This formally defines a basis within which we can represent the electronic degrees of freedom, namely, the fine structure basis (i.e., the eigenbasis of the field-free electronic Hamiltonian, see App.~\ref{app:electronic-states}),
\begin{equation}
    \bigg[ \frac{\bm{p}_{i}^{2}}{2 m} + V(\ab{\bm{r}_{i}}) \bigg] \! \ket{n, l, j, m_{j}}_{i} = E_{n l j} \! \ket{n, l, j, m_{j}}_{i} \!,
\end{equation}
with $E_{n l j}$ the energy associated to the state $\ket{n, l, j, m_{j}}$.
Here, for simplicity we have omitted the spin quantum number $s = 1/2$.
In order to express the internal relative coordinates in terms of the fine structure basis, it proves convenient to transform the relative position operators from Cartesian to spherical polar coordinates, $\bm{r}_{i} = (r_{i; x}, r_{i; y}, r_{i; z}) \mapsto (r_{i}, \theta_{i}, \varphi_{i})$.
Explicitly expanding the relative coordinates in the fine structure basis, we can write
\begin{equation}
    r_{i; u} = \sum_{\mathclap{\bm{n}, \bm{n}\smash{'}}} \! \me{\bm{n}'}{r_{i; u}}{\bm{n}} \! \op{\bm{n}'}{\bm{n}}_{i} \!,
\end{equation}
where we have introduced the ``super'' quantum number $\bm{n} = (n, l, j, m_{j})$.
It follows that in spherical polar coordinates the dipolar terms are given by
\begin{equation}
\begin{gathered}
    \me{\bm{n}'}{r_{i; x}}{\bm{n}} \equiv \sqrt{\frac{2 \pi}{3}} \! \me{\bm{n}'}{r}{\bm{n}} \! \big[ \! \me{\bm{n}'}{Y_{1}^{-1}}{\bm{n}} - \me{\bm{n}'}{Y_{1}^{1}}{\bm{n}} \! \big], \\
    \me{\bm{n}'}{r_{i; y}}{\bm{n}} \equiv \i \sqrt{\frac{2 \pi}{3}} \! \me{\bm{n}'}{r}{\bm{n}} \! \big[ \! \me{\bm{n}'}{Y_{1}^{-1}}{\bm{n}} + \me{\bm{n}'}{Y_{1}^{1}}{\bm{n}} \! \big], \\
    \me{\bm{n}'}{r_{i; z}}{\bm{n}} \equiv \sqrt{\frac{4 \pi}{3}} \! \me{\bm{n}'}{r}{\bm{n}} \! \me{\bm{n}'}{Y_{1}^{0}}{\bm{n}} \!,
\end{gathered}
\end{equation}
while the relevant quadrupolar terms are
\begin{equation}
\begin{gathered}
    \me{\bm{n}'}{[r_{i; x}^{2} - r_{i; y}^{2}]}{\bm{n}} \equiv \sqrt{\frac{8 \pi}{15}} \! \me{\bm{n}'}{r^{2}}{\bm{n}} \! \big[ \! \me{\bm{n}'}{Y_{2}^{-2}}{\bm{n}} + \me{\bm{n}'}{Y_{2}^{2}}{\bm{n}} \! \big], \\
    \me{\bm{n}'}{[3 r_{i; z}^{2} - \bm{r}_{i}^{2}]}{\bm{n}} \equiv \sqrt{\frac{16 \pi}{5}} \! \me{\bm{n}'}{r^{2}}{\bm{n}} \! \me{\bm{n}'}{Y_{2}^{0}}{\bm{n}} \!.
\end{gathered}
\end{equation}
Note that to obtain these expressions, we have employed the separation of variables, which permits the factorization of the states into a radial and angular part (i.e., $\ket{n, l, j, m_{j}}_{i} = \ket{n, l, j}_{i} \! \ket{l, j, m_{j}}_{i}$) with
\begin{equation}
    \me{\bm{n}'}{r}{\bm{n}} \equiv \me{n'\!, l'\!, j'}{r_{i}}{n, l, j} \qquad
    \text{and} \qquad
    \me{\bm{n}'}{Y_{k}^{m_{k}}}{\bm{n}} \equiv \me{l'\!, j'\!, m_{j'}}{Y_{k}^{m_{k}}(\theta_{i}, \varphi_{i})}{l, j, m_{j}} \!.
\end{equation}
We have also made use of the notation for the spherical harmonic operators $\qo{Y}_{l}^{m_{l}} \equiv Y_{l}^{m_{l}}(\qo{\theta}_{i}, \qo{\varphi}_{i})$ with $\qo{\theta}_{i}$ and $\qo{\varphi}_{i}$ the polar and azimuthal angle operators of ion $i$ and $\ro_{i}$ the associated radial position operator.
Further notice that within matrix elements we have dropped the ion index $i$ since the eigenvalues and eigenstates of the field-free electronic Hamiltonian are independent of the ion.
However, to avoid ambiguity we  retain the index on the associated operators (i.e., $\op{\bm{n}'}{\bm{n}}_{i} \equiv \op{n'\!, l'\!, j'\!, m_{j'}}{n, l, j, m_{j}}_{i}$) to distinguish which Hilbert space they act upon.
The angular matrix elements are calculated analytically using standard angular momentum algebra, as detailed in App.~\ref{app:matrix-elements} (see also Ref.~\cite{louck2023}).
In contrast, the radial matrix elements are computed numerically using the radial wavefunctions obtained by solving the radial Schr{\"o}dinger equation for the field-free electronic Hamiltonian with the parametric model potential $V(r_{i})$ used in Ref.~\cite{schmidtkaler2011} (see App.~\ref{app:matrix-elements} and the detailed discussion in Ref.~\cite{aymar1996}).

In order to calculate the effects of the interactions of the electrons with the electric potential due to the linear Paul trap, external driving-field mode, and other electrons, one must, in principle, express the full system Hamiltonian in Eq.~\eqref{eq:many-body-hamiltonian} in the fine structure basis and subsequently diagonalize it~\cite{schmidtkaler2011}.
This poses an analytically intractable problem, particularly when treating highly excited Rydberg states due to the high density of states.
To overcome this challenge, we exploit the fact that the states relevant for the implementation of the envisioned quantum gates are energetically well isolated from the remaining states.
We therefore neglect these couplings since they contribute negligibly to the dynamics of the energetically well isolated subspace of states~\cite{schmidtkaler2011}.

\subsection{Energetically isolated electronic state subspace}\label{subsec:electronic_subspace}

\begin{figure}[t!]
    \centering
    \includegraphics[scale=1]{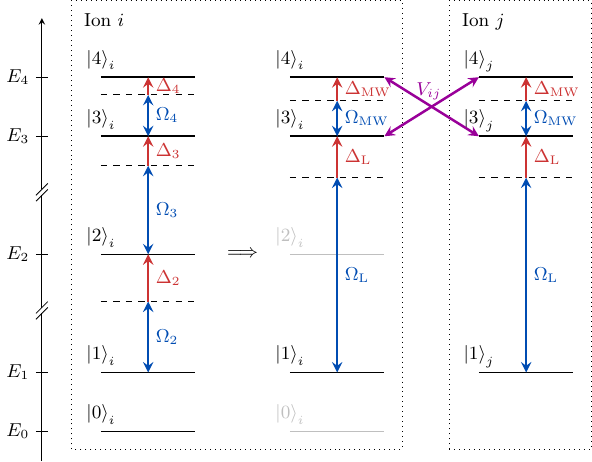}
    \caption{
        Energy level diagrams of the relevant electronic subspaces before (left) and after (middle) adiabatically eliminating the excited state $\ket{2}_{i}$ and neglecting the (decoupled) ground state $\ket{0}_{i}$.
        The metastable state $\ket{1}_{i}$ is coupled to a Rydberg state $\ket{3}_{i}$ by a two-photon excitation scheme via the excited state $\ket{2}_{i}$ (see main text for details).
        The states $\ket{1}_{i}$ and $\ket{3}_{i}$ are resonantly coupled while the state $\ket{2}_{i}$ is far detuned and, hence, can be adiabatically eliminated~\cite{zhang2020}.
        The coupling strength~is then given by the effective Rabi frequency $\Omega_{\mathrm{L}} = \Omega_{3} \Omega_{2} / 2 \Delta_{2}$.
        To induce stronger interactions between the trapped ions, we couple the Rydberg states $\ket{3}_{i}$ and $\ket{4}_{i}$ using a microwave mode of Rabi frequency $\Omega_{\mathrm{MW}} = \Omega_{4}$ (see main text for the detunings $\Delta_{\mathrm{L}}$ and $\Delta_{\mathrm{MW}}$).
        When excited to the microwave-dressed Rydberg states, the ions interact through electric dipole-dipole interactions with strength $V_{ij}$.
    }
    \label{fig:energy-levels}
\end{figure}

Following Refs.~\cite{zhang2020, higgins2017a, higgins2017b} and taking into account the above considerations, we restrict ourselves to a subspace of five electronic states that represent energy levels of a strontium $^{88}\mathrm{Sr}^{+}$ ion, as illustrated in Fig.~\ref{fig:energy-levels}.
Note that this choice is made for the sake of concreteness, but our analysis is not generally limited to strontium ions.
For simplicity, we represent these states using the following notation,
\begin{equation}
\begin{aligned}
    \ket{0} & \equiv \ket{5, 0, 1/2, -1/2} \equiv \ket{5 S_{1/2}(-1/2)} \!, \\
    \ket{1} & \equiv \ket{4, 2, 5/2, -5/2} \equiv \ket{4 D_{5/2}(-5/2)} \!, \\
    \ket{2} & \equiv \ket{6, 1, 3/2, -3/2} \equiv \ket{6 P_{3/2}(-3/2)} \!, \\
    \ket{3} & \equiv \ket{n, 0, 1/2, -1/2} \equiv \ket{n S_{1/2}(-1/2)} \!, \\
    \ket{4} & \equiv \ket{n, 1, 1/2, +1/2} \equiv \ket{n P_{1/2}(+1/2)} \!,
\end{aligned}
\end{equation}
where we set the Rydberg state principal quantum number $n \gg 1$.
The computational subspace of the qubits we implement the quantum gates on is encoded in the ground state $\ket{0}$ and a long-lived metastable state $\ket{1}$ of the ion~\cite{zhang2020}.
The state $\ket{1}$ is coupled to the Rydberg state $\ket{3}$ by a two-photon excitation scheme via a lossy intermediate state $\ket{2}$.
A microwave (MW) field couples the state $\ket{3}$ to another Rydberg state $\ket{4}$.
Further details on the experimental setup can be found in the review in Ref.~\cite{mokhberi2020}. 

Decomposing the Hamiltonian of the internal electronic dynamics in Eq.~\eqref{eq:hamiltonian-internal-expanded} into its respective terms yields
\begin{equation}
    H_{\mathrm{in}} = H_{\mathrm{in}}^{\mathrm{el}} + H_{\mathrm{in}}^{\mathrm{el\text{-}tr}} + H_{\mathrm{in}}^{\mathrm{el\text{-}df}} + H_{\mathrm{in}}^{\mathrm{el\text{-}el}},
\end{equation}
where $H_{\mathrm{in}}^{\mathrm{el}}$ contains the field-free electronic Hamiltonians of the valence electrons, which written in terms of the fine structure basis in the isolated subspace (i.e., $\bm{n} = 0, 1, 2, 3, 4$) reads
\begin{equation}
    H_{\mathrm{in}}^{\mathrm{el}} = \sum_{\mathclap{i = 1}}^{N} \sum_{\bm{n}} E_{\bm{n}} \! \op{\bm{n}}{\bm{n}}_{i} \!.
\end{equation}
The remaining terms $H_{\mathrm{in}}^{\mathrm{el\text{-}tr}}$, $H_{\mathrm{in}}^{\mathrm{el\text{-}df}}$, and $H_{\mathrm{in}}^{\mathrm{el\text{-}el}}$ then detail the interactions of the valence electrons with the electric potential of the linear Paul trap, external driving-field mode, and other valence electrons, respectively.

Considering first the electric quadrupolar coupling terms describing the electron-trap interactions,
\begin{equation}
\begin{aligned}
    H_{\mathrm{in}}^{\mathrm{el\text{-}tr}} & = - \sqrt{\frac{8 \pi}{15}} e \alpha \cos(\nu t) \sum_{\mathclap{i = 1}}^{N} \sum_{\mathclap{\bm{n}, \bm{n}\smash{'}}} \! \me{\bm{n}'}{r^{2}}{\bm{n}} \! \big[ \! \me{\bm{n}'}{Y_{2}^{-2}}{\bm{n}} + \me{\bm{n}'}{Y_{2}^{2}}{\bm{n}} \! \big] \! \op{\bm{n}'}{\bm{n}}_{i} \\ & \qquad
    - \sqrt{\frac{16 \pi}{5}} e \beta \sum_{\mathclap{i = 1}}^{N} [2 K_{ii} + 1] \sum_{\mathclap{\bm{n}, \bm{n}\smash{'}}} \! \me{\bm{n}'}{r^{2}}{\bm{n}} \! \me{\bm{n}'}{Y_{2}^{0}}{\bm{n}} \! \op{\bm{n}'}{\bm{n}}_{i} \!,
\end{aligned}
\end{equation}
it follows from selection rules that the angular matrix elements are only nonzero when the total angular momentum quantum number of either the initial or final state $j > 1/2$~\cite{higgins2018}.
Accordingly, only the states $\ket{1}$ and $\ket{2}$ couple to other states via electric quadrupolar transitions.
However, as shown in App.~\ref{app:hamiltonian-reduction}, this coupling is sufficiently small in comparison to the energy splittings between states and, therefore, can henceforth be neglected~\cite{schmidtkaler2011}.

\subsection{Microwave dressed Rydberg states}

In addition to the quadrupolar couplings, the states are coupled via electric dipole transitions by the time-dependent external driving-field mode and the electric dipole-dipole interactions between the ions.
For typical trap parameters currently considered in experiments and throughout this work~\cite{mokhberi2020}, the interactions between the Rydberg ions are relatively weak compared to the vibrational energy of the external center of mass motion of the ions~\cite{li2014}.
To circumvent this, we implement the method of Rydberg state dressing used in Ref.~\cite{muller2008}, whereby a microwave frequency electric-field mode couples the Rydberg states.
This induces oscillating dipole moments in the dressed Rydberg states that result in strong and controllable long-range interactions between the ions.
Furthermore, for sufficiently strong electric dipolar interactions, it can be shown that the external vibrational motion of the ionic core and the internal electronic dynamics of the valence electron approximately decouple~\cite{muller2008, wilkinson2024}.

The electric field representing the microwave and laser is of the form
\begin{equation}
    \bm{E}(t) = \sum_{\mathclap{j = 2}}^{4} E_{j; x} \boldsymbol{\epsilon}_{j; x} \cos(\nu_{j} t),
\end{equation}
where $E_{j; u}$ and $\boldsymbol{\epsilon}_{j; u}$ are respectively the amplitudes and polarization vectors of the electric field driving transitions between states $\ket{j - 1}$ and $\ket{j}$ along the $u$-axis, with $\nu_{j}$ the associated frequency.
In particular, the laser modes with frequencies $\nu_{2}$ and $\nu_{3}$ enact the Rydberg excitation via the intermediate state $\ket{2}$~\cite{zhang2020}, while the microwave mode with frequency $\nu_{4}$ dresses the Rydberg states.
To couple the states, we choose the polarization to be along the $x$-axis (i.e., $\boldsymbol{\epsilon}_{j; y} = \boldsymbol{\epsilon}_{j; z} = \bm{0}$) for all of the modes of the electric field, such that
\begin{equation}
    H_{\mathrm{in}}^{\mathrm{el\text{-}df}} = - \sqrt{\frac{2 \pi}{3}} e \sum_{\mathclap{i = 1}}^{N} \sum_{\mathclap{j = 2}}^{4} E_{j; x} \cos(\nu_{j} t) \sum_{\mathclap{\bm{n}, \bm{n}\smash{'}}} \! \me{\bm{n}'}{r}{\bm{n}} \! \big[ \! \me{\bm{n}'}{Y_{1}^{-1}}{\bm{n}} - \me{\bm{n}'}{Y_{1}^{1}}{\bm{n}} \! \big] \! \op{\bm{n}'}{\bm{n}}_{i} \!.
\end{equation}
In order to bring this expression into a form appropriate for the implementation of a quantum gate, we move into the rotating frame of the electric field $\bm{E}(t)$, which makes manifest the desired coupling between states. 
However, before proceeding with this, let us consider the electron-electron interactions due to the electric dipole-dipole potential which reads
\begin{equation}
\begin{aligned}
    H_{\mathrm{in}}^{\mathrm{el\text{-}el}} & = \frac{2 \pi}{3} M \omega^{2} \sum_{\mathclap{\substack{i, j = 1 \\ j \neq i}}}^{N} K_{ij} \sum_{\mathclap{\bm{n}, \bm{n}\smash{'}\!\!, \bm{m}, \bm{m}\smash{'}}} \! \me{\bm{n}'}{r}{\bm{n}} \! \me{\bm{m}'}{r}{\bm{m}} \! \big[ 2 \! \me{\bm{n}'}{Y_{1}^{0}}{\bm{n}} \! \me{\bm{m}'}{Y_{1}^{0}}{\bm{m}} \\
    & \qquad + \me{\bm{n}'}{Y_{1}^{-1}}{\bm{n}} \! \me{\bm{m}'}{Y_{1}^{1}}{\bm{m}} + \me{\bm{n}'}{Y_{1}^{1}}{\bm{n}} \! \me{\bm{m}'}{Y_{1}^{-1}}{\bm{m}} \! \big] \! \op{\bm{n}'}{\bm{n}}_{i} \! \op{\bm{m}'}{\bm{m}}_{j} \!.
\end{aligned}
\end{equation}
From direct inspection of the radial matrix elements (see App.~\ref{app:hamiltonian-reduction}), it is evident that the electron-electron interaction is dominated by the Rydberg states (i.e., the interaction between the states $\ket{3}$ and $\ket{4}$).
Therefore, it can be well approximated solely by this contribution such that~\cite{wilkinson2024}
\begin{equation}
    H_{\mathrm{in}}^{\mathrm{el\text{-}el}} \approx -\frac{1}{9} M \omega^{2} \ab{\!\me{4}{\ro}{3}\!}^{2} \sum_{\mathclap{\substack{i, j = 1 \\ j \neq i}}}^{N} K_{ij} [\op{4}{3}_{i} \! \op{3}{4}_{j} + \op{3}{4}_{i} \! \op{4}{3}_{j}].
    \label{neededlabel1}
\end{equation}

Let us now employ the aforementioned transformation into the rotating frame via the unitary
\begin{equation}
    U = \e^{\i E_{1} t / \hbar} \sum_{\mathclap{i = 1}}^{N} [\e^{\i [\nu_{4} + \nu_{3} + \nu_{2}] t} \! \op{4}{4}_{i} + \e^{\i [\nu_{3} + \nu_{2}] t} \! \op{3}{3}_{i} + \e^{\i \nu_{2} t} \! \op{2}{2}_{i} + \op{1}{1}_{i} + \op{0}{0}_{i}],
    \label{neededlabel2}
\end{equation}
where the global phase $E_{1} t / \hbar$ simply shifts the zero-point energy such that $E_{k} \mapsto E_{k} - E_{1}$.
Notice that Eq.~\eqref{neededlabel1} is invariant under the unitary transformationin Eq.~\eqref{neededlabel2}.
Therefore, within the rotating frame and upon applying the rotating wave approximation to eliminate the rapidly oscillating time-dependent terms, the Hamiltonian governing the internal electronic dynamics in the isolated subspace reads
\begin{equation}
\begin{aligned}
    H_{\mathrm{in}} & = \hbar \sum_{\mathclap{i = 1}}^{N} \bigg[[\Delta_{4} + \Delta_{3} + \Delta_{2}] \! \op{4}{4}_{i} + [\Delta_{3} + \Delta_{2}] \! \op{3}{3}_{i} + \Delta_{2} \! \op{2}{2}_{i} - \frac{\Omega_{4}}{2} [\op{4}{3}_{i} + \op{3}{4}_{i}] \\
    & \qquad - \frac{\Omega_{3}}{2} [\op{3}{2}_{i} + \op{2}{3}_{i}] + \frac{\Omega_{2}}{2} [\op{2}{1}_{i} + \op{1}{2}_{i}]\bigg] + \frac{1}{2} \sum_{\mathclap{\substack{i, j = 1 \\ j \neq i}}}^{N} V_{ij} [\op{4}{3}_{i} \! \op{3}{4}_{j} + \op{3}{4}_{i} \! \op{4}{3}_{j}].
\end{aligned}
\end{equation}
where for simplicity we neglect the decoupled qubit states $\ket{0}_{i}$.
Here, we have introduced the detunings $\Delta_{k}$, Rabi frequencies $\Omega_{k}$, which are chosen to be strictly positive real, and electric dipole-dipole interaction strengths $V_{ij}$, which are given by
\begin{equation}\label{eq:laser-parameters}
\begin{gathered}
    \Delta_{4} = \frac{E_{4} - E_{3}}{\hbar} - \nu_{4}, \qquad
    \Delta_{3} = \frac{E_{3} - E_{2}}{\hbar} - \nu_{3}, \qquad
    \Delta_{2} = \frac{E_{2} - E_{1}}{\hbar} - \nu_{2}, \\
    \Omega_{4} = \frac{e E_{4; x}}{\hbar} \frac{\ab{\!\me{4}{\qo{r}}{3}\!}}{3}, \qquad
    \Omega_{3} = \frac{e E_{3; x}}{\hbar} \frac{\ab{\!\me{3}{\qo{r}}{2}\!}}{\sqrt{6}}, \qquad
    \Omega_{2} = \frac{e E_{2; x}}{\hbar} \frac{\ab{\!\me{2}{\qo{r}}{1}\!}}{\sqrt{5}}, \\
    V_{ij} = -\frac{2}{9} M \omega^{2} \ab{\!\me{4}{\qo{r}}{3}\!}^{2} K_{ij}.
\end{gathered}
\end{equation}
Note that the interaction strengths $V_{ij}$ are strictly positive, since the $K_{ij}$ are strictly negative for $i \neq j$ [cf. Eq.~\eqref{eq:generalized-hessian}].
We remark that the rotating wave approximations hold since all frequencies are of the order of $\unit{\giga\hertz}$ to $\unit{\tera\hertz}$ so that for typical experimental timescales (of the order of $\unit{\micro\second}$) the time-dependent phases rapidly average to zero.
As mentioned, upon moving into the rotating frame and applying the rotating wave approximation, we realize that to zeroth order the vibrational and electronic degrees of freedom decouple (i.e., the time-integrated contributions from the coupling term are negligible; see Ref.~\cite{wilkinson2024}).
Given that we are only interested in the electronic degrees of freedom for the envisioned quantum gates, we henceforth neglect the decoupled vibrational degrees of freedom, such that within the rotating frame the full Hamiltonian can be approximated solely by the internal Hamiltonian.

The energetically low-lying qubit state $\ket{1}$ is coupled to the high-lying Rydberg state $\ket{3}$ through the metastable excited state $\ket{2}$, as shown in Fig.~\ref{fig:energy-levels}, via a near-resonant two-photon process.
Coherent population transfer between the states $\ket{1}$ and $\ket{3}$ can be achieved if the intermediate coupling to the state $\ket{2}$ remains negligible~\cite{higgins2017b, higgins2019, mokhberi2020}.
Within this regime, the state $\ket{2}$ can be adiabatically eliminated, such that under the adiabatic approximation, the Hamiltonian can then be rewritten as
\begin{equation}
\begin{aligned}
    H & = \hbar \sum_{\mathclap{i = 1}}^{N} \bigg[ [\Delta_{\mathrm{MW}} + \Delta_{\mathrm{L}}] \! \op{4}{4}_{i} + \Delta_{\mathrm{L}} \! \op{3}{3}_{i} - \frac{\Omega_{\mathrm{MW}}}{2} [\op{4}{3}_{i} + \op{3}{4}_{i}] + \frac{\Omega_{\mathrm{L}}}{2} [\op{3}{1}_{i} + \op{1}{3}_{i}] \bigg] \\
    & \qquad + \frac{1}{2} \sum_{\mathclap{\substack{i, j = 1 \\ j \neq i}}}^{N} V_{ij} [\op{4}{3}_{i} \! \op{3}{4}_{j} + \op{3}{4}_{i} \! \op{4}{3}_{j}],
\end{aligned}
\end{equation}
where we have introduced the effective detunings $\Delta_{\mathrm{MW}}$, $\Delta_{\mathrm{L}}$ and Rabi frequencies $\Omega_{\mathrm{MW}}$, $\Omega_{\mathrm{L}}$, defined by
\begin{equation}
    \Delta_{\mathrm{MW}} = \Delta_{4} + \frac{\Omega_{3}^{2}}{4 \Delta_{2}}, \qquad
    \Delta_{\mathrm{L}} = \Delta_{2} + \Delta_{3} - \frac{\Omega_{3}^{2} - \Omega_{2}^{2}}{4 \Delta_{2}}, \qquad
    \Omega_{\mathrm{MW}} = \Omega_{4}, \qquad
    \text{and} \qquad
    \Omega_{\mathrm{L}} = \frac{\Omega_{3} \Omega_{2}}{2 \Delta_{2}},
\end{equation}
with the labels $\mathrm{MW}$ and $\mathrm{L}$ indicating that the detuning and Rabi frequency are associated to microwave and laser modes, respectively. 

To make manifest the microwave dressing, we diagonalize the manifold of Rydberg states,
\begin{equation}
    H_{\mathrm{Ry}} = \hbar \bigg[[\Delta_{\mathrm{MW}} + \Delta_{\mathrm{L}}] \! \op{4}{4}_{i} + \Delta_{\mathrm{L}} \! \op{3}{3}_{i} - \frac{\Omega_{\mathrm{MW}}}{2} [\op{4}{3}_{i} + \op{3}{4}_{i}]\bigg] = \hbar [\Delta_{+} \! \op{+}{+}_{i} + \Delta_{-} \! \op{-}{-}_{i}],
\end{equation}
where the microwave dressed frequencies (i.e., eigenvalues) and normalized eigenstates are
\begin{equation}
    \Delta_{\pm} = \Delta_{\mathrm{L}} + \frac{\Delta_{\mathrm{MW}} \pm \sqrt{\Delta_{\mathrm{MW}}^{2} + \Omega_{\mathrm{MW}}^{2}}}{2} \qquad
    \text{and} \qquad
    \ket{\pm}_{i} = \frac{N_{\mp} \! \ket{3}_{i} \mp N_{\pm} \! \ket{4}_{i}}{\sqrt{2}}
\end{equation}
with the normalization constants $N_{\smash{\pm}}$ given by
\begin{equation}
    N_{\pm} = \sqrt{1 \pm \frac{\Delta_{\mathrm{MW}}}{\sqrt{\Delta_{\mathrm{MW}}^{2} + \Omega_{\mathrm{MW}}^{2}}}}.
\end{equation}
Inverting the transformation between the undressed and dressed states yields
\begin{equation}
    \ket{4}_{i} = \frac{N_{-} \! \ket{-}_{i} - N_{+} \! \ket{+}_{i}}{\sqrt{2}} \qquad
    \text{and} \qquad
    \ket{3}_{i} = \frac{N_{+} \! \ket{-}_{i} + N_{-} \! \ket{+}_{i}}{\sqrt{2}},
\end{equation}
such that in the microwave dressed frame, the full Hamiltonian reads
\begin{equation}
\begin{aligned}
    \qo{H} & = \hbar \sum_{\mathclap{i = 1}}^{N} \bigg[[\Delta_{+} \! \op{+}{+}_{i} + \Delta_{-} \! \op{-}{-}_{i}] + \frac{\Omega_{+}}{2} [\op{+}{1}_{i} + \op{1}{+}_{i}] + \frac{\Omega_{-}}{2} [\op{-}{1}_{i} + \op{1}{-}_{i}]\bigg] \\
    & \qquad + \frac{1}{2} \sum_{\mathclap{\substack{i, j = 1 \\ j \neq i}}}^{N} \frac{V_{ij}}{2} \bigg[\frac{\Omega_{\mathrm{MW}}^{2}}{\Delta_{\mathrm{MW}}^{2} + \Omega_{\mathrm{MW}}^{2}} \bigg[[\op{+}{+}_{i} - \op{-}{-}_{i}] + \frac{\Delta_{\mathrm{MW}}}{\Omega_{\mathrm{MW}}} [\op{+}{-}_{i} + \op{-}{+}_{i}] \bigg] \\
    & \qquad \qquad \times \bigg[[\op{+}{+}_{j} - \op{-}{-}_{j}] + \frac{\Delta_{\mathrm{MW}}}{\Omega_{\mathrm{MW}}} [\op{+}{-}_{j} + \op{-}{+}_{j}]\bigg] - [\op{+}{-}_{i} - \op{-}{+}_{i}] [\op{+}{-}_{j} - \op{-}{+}_{j}]\bigg],
\end{aligned}
\end{equation}
where we have introduced the microwave dressed Rabi frequencies,
\begin{equation}
    \Omega_{\pm} = \frac{N_{\mp} \Omega_{\mathrm{L}}}{\sqrt{2}}.
\end{equation}
Owing to their large electric polarizabilities, which are made manifest by treating the coupling between the electronic and vibrational degrees of freedom perturbatively~\cite{muller2008, schmidtkaler2011}, trapped ions excited to Rydberg states experience state-dependent mechanical forces that modify the trap frequencies $\omega_{u}$~\cite{wilkinson2024}.
It was theoretically proposed in Ref.~\cite{li2014} and later experimentally demonstrated in Ref.~\cite{pokorny2020} that these unwanted effects could be overcome by coupling Rydberg states with a microwave mode.
For a particular choice of the microwave detuning $\Delta_{\mathrm{MW}}$ and associated Rabi frequency $\Omega_{\mathrm{MW}}$, the normalization constants $N_{\pm}$ can be set such that the microwave dressed states $\ket{\pm}$ exhibit vanishing polarizabilities.
In practice, this would fix the ratio of the microwave detuning and Rabi frequency.
However, since we have eliminated the coupling that induces these effects by transforming into the oscillating frame and applying the rotating wave approximation, we relax this restriction and instead choose a microwave detuning that maximizes the interaction strength~\cite{zhang2020}, specifically, $\Delta_{\mathrm{MW}} = 0$.
In this limit, the Hamiltonian reads
\begin{equation}\label{eq:model-hamiltonian}
\begin{aligned}
    H & = \hbar \sum_{\mathclap{i = 1}}^{N} \bigg[ \Delta_{+} \! \op{+}{+}_{i} + \Delta_{-} \! \op{-}{-}_{i} + \frac{\Omega_{+}}{2} [\op{+}{1}_{i} + \op{1}{+}_{i}] + \frac{\Omega_{-}}{2} [\op{-}{1}_{i} + \op{1}{-}_{i}]\bigg] \\
    & \qquad + \frac{1}{2} \sum_{\mathclap{\substack{i, j = 1 \\ j \neq i}}}^{N} \frac{V_{ij}}{2} \bigg[ [\op{+}{+}_{i} - \op{-}{-}_{i}] [\op{+}{+}_{j} - \op{-}{-}_{j}] - [\op{+}{-}_{i} - \op{-}{+}_{i}] [\op{+}{-}_{j} - \op{-}{+}_{j}] \bigg],
\end{aligned}
\end{equation}
where now the microwave dressed detunings $\Delta_{\pm}$, Rabi frequencies $\Omega_{\pm}$, and states $\ket{\pm}$ are
\begin{equation}\label{eq:dressed-frequencies}
    \Delta_{\pm} = \Delta_{\mathrm{L}} \pm \frac{\Omega_{\mathrm{MW}}}{2}, \qquad
    \Omega_{\pm} = \frac{\Omega_{\mathrm{L}}}{\sqrt{2}}, \qquad
    \text{and} \qquad
    \ket{\pm}_{i} = \frac{\ket{3}_{i} \mp \ket{4}_{i}}{\sqrt{2}}.
\end{equation}
Hamiltonian~\eqref{eq:model-hamiltonian} is the main result of this section and in the following it will be employed to implement the envisioned quantum gate protocols.


\section{Implementation of fast entangling phase gate protocols}\label{sec:gates}

With the derivation of the model Hamiltonian complete, we now address the implementation of the trapped Rydberg ion quantum gate protocols. 
We first discuss the constraints which need to be fulfilled in order to implement a CZ gate and analyse three different gate protocols with regard to their capability of realizing fast entangling gates according to these constraints. 
Then, we show that in a particular regime the dimension of the Hamiltonian can be reduced significantly via adiabatic elimination, facilitating the comprehension of the phenomenology underlying the gate implementation. 
Lastly, we discuss the impact of imperfections, e.g., the finite lifetime of the Rydberg states, on the gate fidelity and potential improvements to enable better performance.

In the following, we demonstrate how a CZ gate can be implemented by employing the strong electric dipole-dipole interaction between the microwave-dressed trapped Rydberg ions.
For this purpose, we particularize the Hamiltonian in Eq.~\eqref{eq:model-hamiltonian} to a system with $N = 2$ ions such that it can be written as
\begin{equation}\label{eq:two-ion-hamiltonian}
\begin{aligned}
    H & = \sum_{\mathclap{i = 1}}^{2} \bigg[\Big[\Delta_{\mathrm{L}} + \frac{\Omega_{\mathrm{MW}}}{2}\Big] \! \op{+}{+}_{i} + \Big[\Delta_{\mathrm{L}} - \frac{\Omega_{\mathrm{MW}}}{2}\Big] \! \op{-}{-}_{i} + \frac{\Omega_{\mathrm{L}}}{2 \sqrt{2}} [\op{+}{1}_{i} + \op{1}{+}_{i} + \op{-}{1}_{i} + \op{1}{-}_{i}]\bigg] \\ & \qquad
    + \frac{V}{2} \bigg[[\op{+}{+}_{1} - \op{-}{-}_{1}] [\op{+}{+}_{2} - \op{-}{-}_{2}] - [\op{+}{-}_{1} - \op{-}{+}_{1}] [\op{+}{-}_{2} - \op{-}{+}_{2}]\bigg] \!
\end{aligned}
\end{equation}
with the interaction strength $V \equiv V_{12} = V_{21}$ and where we have set $\hbar = 1$. 
For the gate implementation, we assume the Rabi frequency and detuning of the excitation laser to be given by time-dependent functions,
\begin{equation}\label{eqn:pulses}
    \Delta_{\mathrm{L}} = \Delta_{\mathrm{L}}(\mathbf{\Delta}_{0}, t), \qquad
    \Omega_{\mathrm{L}} = \Omega_{\mathrm{L}}(\mathbf{\Omega}_{0}, t),
\end{equation}
each of which depends on a set of parameters, $\mathbf{\Delta}_{0}$ and $\mathbf{\Omega}_{0}$, respectively.
For any initial state $\ket{\Psi(t = 0)}$, after a pulse-driven interaction of duration $t = \tau$, the final state can always be written as
\begin{equation}\label{eqn:time_evol++}
    \ket{\Psi(\tau)} = \sum_{\mathclap{a, \smash{b}}} c_{ab}(\tau) \e^{\i \varphi_{ab}(\tau)} \! \ket{ab} \!
\end{equation}
with $a, b \in \{0, 1, -, +\}$, $c_{ab}(t) \in [0, 1]$, and $\varphi_{ab}(t) \in [0, 2 \pi)$. 
Here, $p_{ab}(t) = \ab{c_{ab}(t)}^{2}$ and $\varphi_{ab}(t)$ are the time-dependent populations and accumulated phases, respectively. 
To find an optimized parameter set $\mathbf{\Omega}_{0}^{\mathrm{opt}}$ and $\mathbf{\Delta}_{0}^{\mathrm{opt}}$, it is useful to assume the initial state $\ket{\Psi(t=0)} = [\ket{00} + \ket{01} + \ket{10} + \ket{11}]/2$ and, irrespective of the specific pulse shapes chosen, maximize the fidelity~\cite{nielsen2010}
\begin{equation}\label{eq:fidelity}
    \mathcal{F}_{\mathrm{B}} = \ab{\!\ip{\Psi_{\mathrm{B}}}{\Psi(\tau)}\!}^{2},
\end{equation}
where $\ket{\Psi_{\mathrm{B}}} = [\ket{00} + \ket{01} + \ket{10} - \ket{11}]/2$ is the target state associated to the CZ gate. 
Note that $\ket{\Psi_{\mathrm{B}}}$ is equivalent to a Bell state up to a single-qubit rotation, which is why we call $\mathcal{F}_{\mathrm{B}}$ \textit{Bell state fidelity}.
Accordingly, this imposes the following constraints on the desired probability amplitudes and accumulated phases,
\begin{equation}\label{eqn:constraints_CZ}
\begin{aligned}
    &c_{00}(\tau) = c_{01}(\tau) = c_{10}(\tau) = c_{11}(\tau) = \frac{1}{2}, \\
    &\varphi_{00}(\tau) = \varphi_{01}(\tau) = \varphi_{10}(\tau) = \varphi_{11}(\tau) - \pi = 0.
\end{aligned}
\end{equation}
Due to the invariance of the system Hamiltonian \eqref{eq:two-ion-hamiltonian} under the exchange of the two ions, the states $\ket{10}$ and $\ket{01}$ undergo the same evolution meaning that $c_{10}(t) = c_{01}(t)$ and $\varphi_{10}(t) = \varphi_{01}(t)$ hold for~${t \in [0, \tau]}$. 
This reduces the number of constraints we must fulfill in order to realize a CZ gate from eight to six. 
Note that we later will also consider one pulse scheme requiring single-ion addressing, where these assumption generally do not apply.
Further noting that the ground state $\ket{00}$ is completely decoupled from the rest of the considered Hilbert space, we also have that $c_{00}(\tau) = c_{00}(0)$ and $\varphi_{00}(\tau) = \varphi_{00}(0)$, reducing the total number of constraints to four. 
Due to experimental limitations, particularly concerning the achievability of large laser Rabi frequencies~\cite{zhang2020}, it is more feasible to construct a CZ gate up to trivial single-qubit rotations.
Consequently, the resulting entangling phase needs to fulfill $\varphi_{*}(\tau)=[{\varphi_{11}(\tau) - 2 \varphi_{10}(\tau)]\bmod{2\pi} = \pi}$, thus reducing the total number of constraints to three. 
In this case, we generalize the Bell state fidelity in Eq.~\eqref{eq:fidelity} to
\begin{align}\label{eq:fidelity-sqr}
    \mathcal{F}_\text{B} = \ab{\!\me{\Psi_{\mathrm{B}}}{\qo{R}_{2}(\varphi_{01}) \qo{R}_{1}(\varphi_{10})}{\Psi(\tau)}\!}^{2} \!,
\end{align}
where we have introduced an additional single-qubit phase shift on each qubit generated by the operator $\qo{R}_{i}(\varphi) = \e^{-\i \varphi} \! \op{1}{1}_{i} + \op{0}{0}_{i}$. 
To specify sources of infidelity more precisely, we define the population error~$\bar{p}_{*}$ and entangling phase error~$\bar{\varphi}_{*}$ as
\begin{equation}\label{eqn:rel-errors}
    \bar{p}_{*} = 1 - \frac{1}{4} \Big[\sum_{a, b} c_{ab}(\tau)\Big ]^2, \qquad
    \bar{\varphi}_{*} = 1-\frac{1}{16}\big|3 - \e^{\i \varphi_{*}(\tau)} \big|^2 .
\end{equation}

Note that the problem is simplified significantly if we assume that the pulses excite the electronic states of the ions adiabatically~\cite{moller2008}.
In this regime, the pulse length is effectively infinitely long compared to the internal time scale of the system, causing the occupation of the instantaneous eigenstates to remain constant throughout the pulse-driven interaction, which guarantees that the populations always return to their initial states.
Nevertheless, for finite-time processes, the population of the instantaneous eigenstates may vary during the pulse-driven evolution, meaning that the exact eigenstates of the system deviate from the instantaneous eigenstates.
It has been shown, however, that in near-adiabatic processes, i.e., for sufficiently long smooth pulses, the populations will approximately completely return to their initial configurations even though the exact eigenstates deviate from the instantaneous eigenstates during the process \cite{benseny2021}.
Hence, the population constraints are approximately fulfilled as long as the duration of the external laser pulse is long compared to the system's internal time scale.

In order to find an optimized parameter set $\mathbf{\Omega}_{0}^{\mathrm{opt}}$ and $\mathbf{\Delta}_{0}^{\mathrm{opt}}$ realizing a CZ gate for a given time-dependent pulse shape \eqref{eqn:pulses}, we maximize the fidelity $ \mathcal{F}_{\mathrm{B}}$ definied in Eq.~\eqref{eq:fidelity-sqr}.
For optimization we use the \texttt{differential\_evolution} method from the \texttt{scipy.optimize} library~\cite{scipy2020}:
a stochastic, population-based optimization algorithm which searches for the global maximum within a given parameter range~\cite{storn1997}.

\subsection{Analysis of two-qubit quantum gate schemes}\label{sec:gate-schemes}
\begin{figure}[t!]
    \centering
    \includegraphics[scale=1]{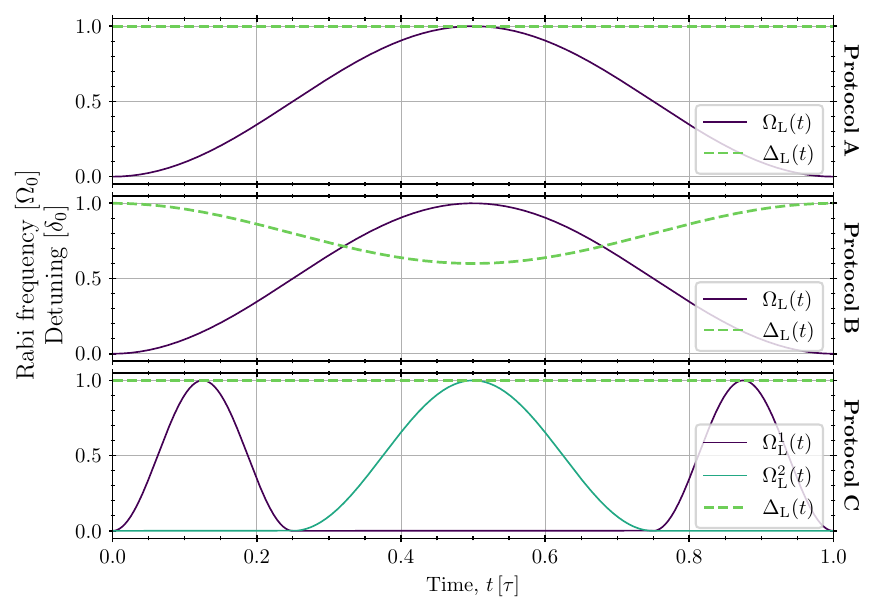}
    \caption{
        Pulse schemes for the Rabi frequency $\Omega_{\mathrm{L}}(t)$ and detuning $\Delta_{\mathrm{L}}(t)$ of the excitation laser for the proposed gate protocols. 
        The Rabi frequency and detuning are normalized with respect to their maxima and the time scale is normalized by the pulse duration $\tau$. 
        The simplest pulse scheme is Protocol A, given by Eq.~\eqref{eqn:protocolA}, which only requires modulation of the laser Rabi frequency, while the detuning is kept constant (upper plot). 
        Protocol B, defined in Eq.~\eqref{eqn:protocolB}, is an extension of this through an additional modulation of the laser detuning (center plot).
        Protocol C represents the $\pi$-$2 \pi$-$\pi$ gate protocol, given in Eq.~\eqref{eqn:protocolC}, where the ions are addressed individually by three laser pulses of the same type as in Protocol~A (lower plot).
    }
    \label{fig:pulse-schemes}
\end{figure}

\begin{table}[t!]
\centering
\begin{tabular}{ll||ccc|ccc}
\multicolumn{1}{c}{}          &                              & \multicolumn{3}{c|}{\textbf{Conservative}}                                  & \multicolumn{3}{c}{\textbf{Optimistic}}   \\ \hline \hline 
                              &                              & \multicolumn{3}{c|}{}                                                       & \multicolumn{3}{c}{}              \\[-0.2cm]
\textbf{Fixed}                & $V$                          & \multicolumn{3}{c|}{10}                                                     & \multicolumn{3}{c}{25}            \\
\textbf{parameters}           & $\Omega_\mathrm{MW}$         & \multicolumn{3}{c|}{100}                                                    & \multicolumn{3}{c}{250}           \\
                              & $\tau$                       & \multicolumn{3}{c|}{1}                                                      & \multicolumn{3}{c}{0.3}           \\
                              &                              & \multicolumn{3}{c|}{}                                                       & \multicolumn{3}{c}{}              \\[-0.2cm]
\textbf{Bounds}               & $\Omega_0$                   & \multicolumn{3}{c|}{$[0, 10]$}                                              & \multicolumn{3}{c}{$[0, 100]$}    \\
                              & $\delta_0$                   & \multicolumn{3}{c|}{$[0, 100]$}                                             & \multicolumn{3}{c}{$[0, 250]$}    \\
                              & $\Delta_0$                   & \multicolumn{3}{c|}{$[-100, 100]$}                                          & \multicolumn{3}{c}{$[-250, 250]$} \\[-0.2cm]
                              &                              & \multicolumn{3}{c|}{}                                                       & \multicolumn{3}{c}{}              \\ \hline
\textbf{Protocol}             &                              & \textbf{A}        & \textbf{B}        & \textbf{C}       & \textbf{A}       & \textbf{B}       & \textbf{C}      \\ \hline 
                              &                              &                   &                   &                  &                  &                  &                 \\[-0.2cm]
\textbf{Gate}                 & $\Omega_0^{\mathrm{opt}}$    & $7.78$            & $9.80$            & $5.66$           & $92.04$          & $84.37$          & $18.86$         \\
\textbf{parameters}           & $\delta_0^{\mathrm{opt}}$    & $47.61$           & $37.44$           & $50$             & $114.07$         & $39.94$          & $125$           \\
                              & $\Delta_0^{\mathrm{opt}}$    & $-$               & $-12.10$          & $-$              & $-$              & $197.13$         & $-$             \\[-0.2cm]
                              &                              &                   &                   &                  &                  &                  &                 \\
\textbf{Performance}          & $\bar{p}_{*}$                & $3.2 \magn{-2}$   & $2.2 \magn{-4}$   & $1.4\magn{-3}$   & $2.1\magn{-2}$   & $2.3\magn{-6}$   &  $2.1\magn{-2}$  \\
                              & $\bar{\varphi}_{*}$          & $5.7 \magn{-6}$   & $5.1 \magn{-8}$   & $1.6\magn{-1}$   & $1.8\magn{-3}$   & $8.4\magn{-12}$   &  $2.4\magn{-1}$  \\
                              & $\mathcal{F}_\text{B}$       & $96.81\,\%$       & $99.98\,\%$       & $84.36\,\%$      & $97.72\,\%$      & $> 99.99\,\%$    & $74.95\,\%$    \\
\end{tabular}
\caption{Parameter choices and best optimization outcomes for the conservative and optimistic regimes in protocols A, B and C. 
In both regimes, the microwave Rabi frequency $\Omega_\mathrm{MW}$, interaction strength $V$, and gate duration $\tau$ are fixed to a constant value for all three protocols.
Protocol A and B, see Eqs.~\eqref{eqn:protocolA} and \eqref{eqn:protocolB}, contain free gate parameters which are optimized according to the bounds in the corresponding regime.
For Protocol C, see Eq.~\eqref{eqn:protocolC}, the gate parameters are not optimized, but are chosen following the $\pi$-$2 \pi$-$\pi$ gate protocol specifications, therefore leading to substantially lower fidelity values (cf. main text).
The fidelities are calculated according to Eq.~\eqref{eq:fidelity-sqr} and the relative population and phase errors are defined by Eqs.~\eqref{eqn:rel-errors}.
Note that we did not consider any kind of incoherent error sources here and, thus, the given fidelities should be interpreted as a benchmark for performance of the different protocols, but not as realistic fidelities reachable in experiments.
The gate duration $\tau$ is given in $\unit{\micro\second}$ while all other parameters, with the exception of performance outcomes, are given in units of $2 \pi \times \unit{\mega\hertz}$.}
\label{tab:parameter_table}
\end{table}

Many entangling gate protocols already exist in trapped ion systems which make use of the all-to-all connectivity provided by the phonon modes of the Coulomb crystal \cite{cirac1995, ballance2016, benhelm2008, schmidtkaler2003, leibfried2003b}.
While certain fast gate protocols have been demonstrated, they face challenges such as the requirement for the phase-space trajectory of all vibrational modes to close simultaneously at the end of the gate, which limits their scalability with the number of ions~\cite{schafer2018}.
On the other hand, the gate speed in more commonly used protocols is fundamentally limited by the frequency of the phonon modes \cite{leibfried2003b, benhelm2008}, what prevents reaching fast entangling-gate times when compared to, e.g., optically trapped neutral atom systems.
For the latter, highly excited Rydberg atoms interact strongly via dipole-dipole interactions, allowing the implementation of entangling gates on nanosecond timescales \cite{evered2023, jaksch2000, levine2019, saffman2020}.
In the following, we demonstrate how the speed limitation in trapped ion systems can be overcome by utilizing the strong dipole-dipole interaction between microwave-dressed ions excited to Rydberg states.
We consider three experimentally realizable pulse schemes, which are depicted in Fig.~\ref{fig:pulse-schemes}, and investigate which of them is most suitable for realizing a CZ gate with trapped Rydberg ions.
To this end, we optimize the free pulse parameters within two regimes, one with conservative and one with rather optimistic parameter bounds, and compare the fidelities reached with the different pulse schemes.
All parameters defining the conservative and optimistic regimes, where the former can be understood as experimental parameter ranges readily achievable in current experiments~\cite{mokhberi2020} and the latter represents more ambitious, yet still technically feasible ranges, can be found in Tab.~\ref{tab:parameter_table}.

\subsubsection*{Protocol A}

\begin{figure}[t!]
    \centering
    \includegraphics[scale=1]{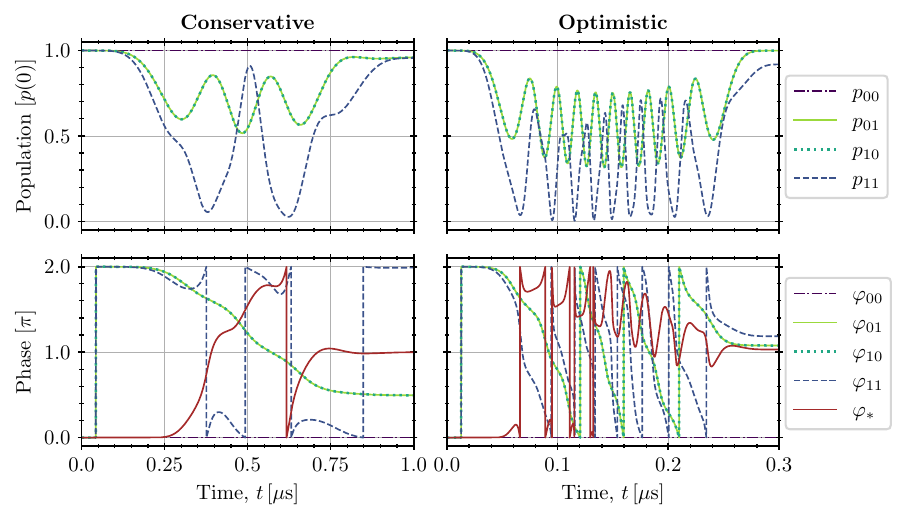}
    \caption{
        Numerical simulations of the gate Protocol A in the conservative (left) and optimistic (right) regime.
        The pulse is defined in Eq.~\eqref{eqn:protocolA} and the optimized gate parameters in both regimes are listed in Tab.~\ref{tab:parameter_table}.
        The upper plots demonstrate how the populations of the computational basis states evolve during the gate implementation and in the lower plots the time-evolution of the accumulated phases and the resulting entangling phase, $\varphi_{*}(t) = [\varphi_{11}(t)-2\varphi_{10}(t)]\bmod{2\pi}$, are depicted.
        We see that in both cases the populations do not completely return to their initial configurations, which explains the relatively high population errors $\bar{p}_{*} = 3.2 \magn{-2}$ in the conservative and $\bar{p}_{*} = 2.1 \magn{-2}$ in the optimistic case.
        Also apparent is the larger phase error occurring in the optimistic regime compared to the conservative one.
        These gates achieve fidelities of $96.81\,\%$ in the conservative and $97.72\,\%$ in the optimistic case.
    }
    \label{fig:gate-comparison-A}
\end{figure}

The first protocol, depicted in the upper plot in Fig.~\ref{fig:pulse-schemes}, utilizes a simple sinusoidal pulse which only requires modulation of the Rabi frequency $\Omega_{\mathrm{L}}(t)$.
Specifically, we choose
\begin{equation}\label{eqn:protocolA}
    \Omega_{\mathrm{L}}(t) = \Omega_{0} \sin^{2}(\pi t / \tau), \qquad
    \Delta_{\mathrm{L}}(t) = \delta_{0},
\end{equation}
where, as before, $\tau$ is the gate duration and $\Omega_{0}$ and $\delta_{0}$ are constants constituting the parameter set to be optimized.
In Fig.~\ref{fig:gate-comparison-A}, we plot the populations $p_{ab}(t)$ of the computational-basis states, accumulated phases $\varphi_{ab}(t)$, and entangling phase $\varphi_{*}(t)$ for the optimal parameters found within the conservative and optimistic regime.
The population error $\bar{p}_{*}$, entangling phase error $\bar{\varphi}_{*}$, and Bell state fidelity $\mathcal{F}_{\mathrm{B}}$ resulting from the gate protocol, along with the optimal parameters $\Omega_{0}^{\mathrm{opt}}$ and $\delta_{0}^{\mathrm{opt}}$, are presented in Tab.~\ref{tab:parameter_table}.

Despite the simplicity of the pulse shape, the optimization scheme successfully found parameter sets which achieve fidelities of $96.81 \, \%$ in the conservative and $97.72 \, \%$ in the optimistic regime, with most of the infidelity arising from the loss of state population during the gate protocol.
We also find that in spite of the extended parameter ranges and increased interaction and coupling strengths, the decreased pulse time essentially prevents a significant improvement in the fidelity in the optimistic case.

\begin{figure}[t!]
    \centering
    \includegraphics[scale=1]{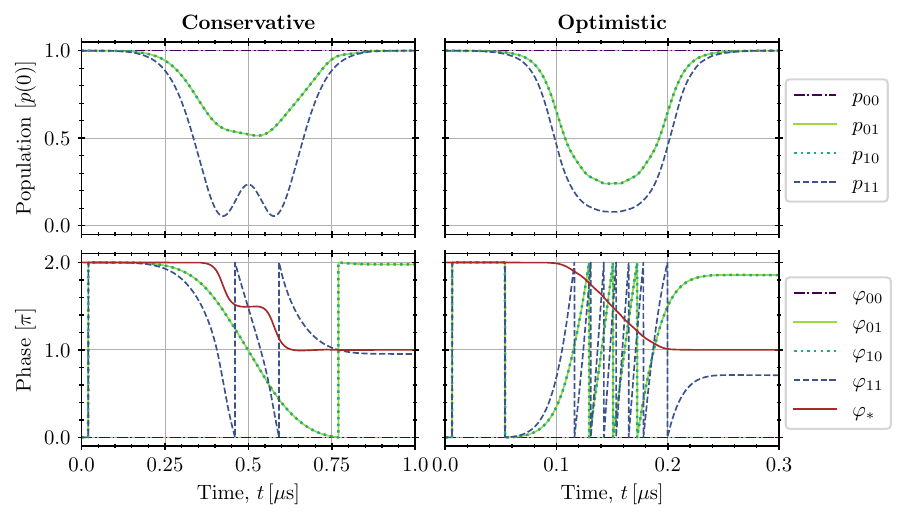}
    \caption{
        Numerical simulations of gate Protocol B, given in Eq.~\eqref{eqn:protocolB}, optimized within the conservative (left) and optimistic (right) parameter ranges.
        The resulting optimal gate parameters are listed in Tab.~\ref{tab:parameter_table}.
        The upper plots show the time-evolution of the populations of the computational basis states, while the lower plots depict how the accumulated phases and the resulting entangling phase, $\varphi_{*}(t)=[\varphi_{11}(t)- 2\varphi_{10}(t)]\bmod{2\pi}$, evolve during the gate implementation.
        Based on these plots, an error source is not clearly apparent in either the conservative or the optimistic regime, which is also reflected in the high fidelities of $99.98\,\%$ and $>99.99\,\%$ and the low error rates given in Tab.~\ref{tab:parameter_table}.
        However, the gate in the optimistic regime is not only a lot faster, with an infidelity of the order of~$\sim 10^{-6}$ it outperforms the gate in the conservative regime by two orders of magnitude.
    }
    \label{fig:gate-comparison-B}
\end{figure}

\subsubsection*{Protocol B}

The second protocol, shown in the center plot in Fig~\ref{fig:pulse-schemes}, generalizes the scheme implemented in Protocol A by additionally modulating the laser detuning $\Delta_{\mathrm{L}}(t)$.
While experimentally more challenging to implement due to the smooth variation of the detuning, we expect a significant improvement in performance compared to the previous pulse scheme, as already observed in neutral atom systems~\cite{sun2020}.
In particular, we consider the following time-dependent laser Rabi frequency and detuning
\begin{equation}\label{eqn:protocolB}
    \Omega_{\mathrm{L}}(t) = \Omega_{0} \sin^{2}(\pi t / \tau), \qquad
    \Delta_{\mathrm{L}}(t) = \delta_{0} - \Delta_{0} \sin^{2}(\pi t / \tau),
\end{equation}
where, compared to Protocol A, we have an additional optimization parameter $\Delta_0$ defining the modulation amplitude of the laser detuning.
As before, we consider the parameter regimes with conservative and optimistic value ranges, given in Tab.~\ref{tab:parameter_table}.
The time evolution of the relative populations, accumulated phases, and entangling phase for the optimized parameter set, tabulated in Tab.~\ref{tab:parameter_table}, are plotted in Fig.~\ref{fig:gate-comparison-B}.
Compared to Protocol A, both the populations and the phases evolve more smoothly during the pulse and exhibit less fluctuations.
In both regimes, Protocol B outperforms the previous protocol in terms of fidelity and error rates as demonstrated in Tab.~\ref{tab:parameter_table}.
While already performing very well in the conservative regime with a fidelity of $99.98\,\%$, the protocol reaches a remarkable performance with a fidelity of over $99.99\,\%$ and an infidelity $\sim\!10^{-6}$ in the optimistic regime.
This demonstrates the feasibility of implementing fast quantum gates on trapped Rydberg ion platforms using this protocol. 
Moreover, this illustrates how a larger set of optimization parameters results in greater tunability allowing for notably better performance and perhaps even shorter gate times if one implements more sophisticated pulse schemes~\cite{theis2016, mohan2023}.
However, since here we only intend to analyse the capability of the different protocols to realize fast high-fidelity entangling gates, we do not yet consider any kind of incoherent error sources.
Therefore, the given fidelities should be interpreted as a benchmark for performance of the various protocols, but not as realistic fidelities reachable in experiments.

\subsubsection*{Protocol C}

The third protocol is the seminal $\pi$-$2 \pi$-$\pi$ gate protocol proposed in Ref.~\cite{jaksch2000}, which for consistency we call Protocol C. 
As shown in Fig.~\ref{fig:pulse-schemes}, this scheme generalizes Protocol A \eqref{eqn:protocolA} by addressing the ions independently and applying three sinusoidal pulses with constant detuning. 
More precisely, the laser Rabi frequencies $\Omega_{\mathrm{L}}^{i}(t)$, where $i$ denotes the addressed ion, are modulated such that the pulses correspond to a $\pi$-, $2 \pi$-, and $\pi$-pulse, i.e.,
\begin{equation}\label{eqn:protocolC}
    \Omega_{\mathrm{L}}^{1}(t) =
    \begin{cases}
        \Omega_{0} \sin^{2}(4 \pi t / \tau), & \text{if $0 \leq t \leq \tau / 4$}, \\
        0, & \text{otherwise}, \\
        \Omega_{0} \sin^{2}(4 \pi t / \tau), & \text{if $3 \tau / 4 \leq t \leq \tau$}, \\
    \end{cases} \qquad
    \Omega_{\mathrm{L}}^{2}(t) =
    \begin{cases}
        0, & \text{if $0 \leq t \leq \tau / 4$}, \\
        \Omega_{0} \cos^{2}(2 \pi t / \tau), & \text{otherwise}, \\
        0, & \text{if $3 \tau / 4 \leq t \leq \tau$} \\
    \end{cases}
\end{equation}
with $\Omega_{0} = 8 \sqrt{2} \pi / \tau$. 
For the protocol to work properly, each $\pi$-pulse needs to fully transfer the population from the qubit state to the Rydberg manifold or vice versa~\cite{jaksch2000}.
Therefore, we fix the detuning to 
\begin{equation}
    \Delta_{\mathrm{L}}^{1}(t) = \Delta_{\mathrm{L}}^{2}(t) = \frac{\Omega_\mathrm{MW}}{2}
\end{equation}
leading to a resonant coupling between $\ket{1}$ and $\ket{-}$ [cf. Eq.~\eqref{eq:two-ion-hamiltonian}].
Consequently, the other Rydberg state $\ket{+}$ is highly detuned and, for large enough microwave Rabi frequencies $\Omega_\mathrm{MW}\gg\Omega_0$, it does not get populated during the entire pulse sequence.
Applying pulse scheme \eqref{eqn:protocolC}, the resulting entangling phase is ${\varphi_{*} = \pi[1 -{\Omega_0}/{(\sqrt{2}V)}]}$~\cite{jaksch2000}, which is effectively equivalent to the implementation of a CZ-gate for $V\gg\Omega_0$.

Since there are no free parameters left in this protocol, its performance does not rely on any optimization and we can directly plug in the fixed parameters $V$, $\Omega_\mathrm{MW}$, and $\tau$ as they are defined for the two regimes in Tab.~\ref{tab:parameter_table}.
In contrast to Protocol A and B, the gate performs relatively poorly with fidelities of $84.26\,\%$ in the conservative and $74.80\,\%$ in the optimistic regime.
The reason for this is that the amplitude of the laser Rabi frequency, $\Omega_{0} \approx 2 \pi \times 5.66 \, \unit{\mega\hertz}$ in the conservative and $\Omega_{0} \approx 2 \pi \times 18.86 \, \unit{\mega\hertz}$ in the optimistic case, is not small enough compared to the interaction strength $V$ in the corresponding regimes, $V = 2 \pi \times 10 \, \unit{\mega\hertz}$ and $V = 2 \pi \times 25 \, \unit{\mega\hertz}$, respectively.
Therefore, in order to achieve higher fidelities, we need to increase either the gate duration $\tau$ or the interaction strength $V$.
For example, by increasing the gate time to $\tau=5\,\unit{\micro\second}$ in the optimistic case, we find a fidelity of $99.88\,\%$, which is indeed a significant improvement but still does not reach the performance of Protocol B.
A further increase in gate time to achieve higher fidelities contradicts our objective of realizing fast entangling gates that outperform common phonon-based trapped ion gates in terms of speed~\cite{ballance2016}.
Alternatively, improving the gate performance by increasing the interaction strength faces experimental limitations as can be seen by recalling the scaling of the interaction strength,
\begin{equation}\label{eqn:interaction_strength}
   V_{ij} \propto \frac{\ab{\!\me{n P_{\smash{1/2}}}{\qo{r}}{n S_{\smash{1/2}}}\!}^{2}}{\ab{\bar{\mathbf{R}}_{ij}}^{3}},
\end{equation}
where $\bar{\mathbf{R}}_{ij}$ denotes the distance between the equilibrium positions of ion $i$ and $j$.
Since the matrix element in Eq.~\eqref{eqn:interaction_strength} scales quadratically with the principal quantum number~$n$~\cite{mokhberi2020}, one would expect that choosing more energetic Rydberg states would enable ever higher interaction strengths and, thus, arbitrarily high fidelities.
This, however, introduces additional problems since for too large principal quantum numbers, population losses of the electronic states occur due to double ionization of the ions via the trapping field modes~\cite{muller2008}.
Similarly, the likelihood of double ionization due to blackbody radiation increases with higher principal quantum numbers~\cite{higgins2018}.
On the other hand, higher interaction strengths can be realized by reducing the inter-ion distance, e.g., by increasing the total number of ions in the trap or using higher axial trapping frequencies.
For instance, for trapped strontium $^{88}\mathrm{Sr}^{+}$ ions separated by $2.3\,\unit{\micro\meter}$ and excited to $n=60$, the interaction strength is $V_{ij}\approx 2\pi\times 21.9\,\unit{\mega\hertz}$~\cite{zhang2020}.
However, due to the finite spot size of the laser beams, such small inter-ion distances are difficult to combine with the single-ion addressing and laser Rabi frequencies needed for this protocol~\cite{bachor2016}.
Accordingly, we conclude that while successful in the realm of quantum information processing with neutral Rydberg atoms, Protocol C proves inadequate for the implementation of fast quantum gates with trapped Rydberg ions.

\subsection{Simplification of the Hamiltonian via adiabatic elimination}

\begin{figure}[t!]
    \centering
    \includegraphics[scale=1]{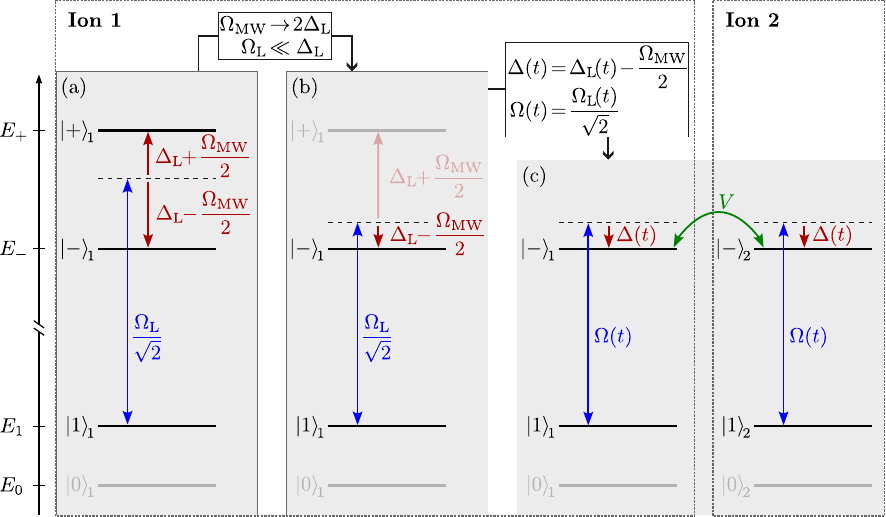}
    \caption{
        Energy level scheme of the states relevant for the implementation of the CZ gate protocols. 
        (a) The microwave-dressed Rydberg states $\ket{\pm}$ are coupled to the ground state $\ket{1}$ via a two-photon excitation process with effective detunings $\Delta_{\pm} = \Delta_{\mathrm{L}} \pm \tfrac{\Omega_{\mathrm{MW}}}{2}$ and Rabi frequencies $\Omega_{\pm} = \tfrac{\Omega_{\mathrm{L}}}{\sqrt{2}}$. 
        (b)~As~$\Omega_{\mathrm{MW}}\!\approx 2 \Delta_{\mathrm{L}}$, the detuning $\Delta_{-}$ decreases while $\Delta_{+}$ increases. 
        Within this regime we can therefore adiabatically eliminate the state $\ket{+}$ whenever its coupling strength to state $\ket{1}$ is sufficiently small compared to its energy splitting, i.e., $\Omega_{\mathrm{L}} \ll \Delta_{\mathrm{L}}$. 
        This leads to an energy level scheme wherein the ground state $\ket{1}$ is coupled to only one Rydberg state $\ket{-}$.
        (c)~Assuming the corresponding reduced two-ion system, the application of a time-dependent laser pulse with detuning $\Delta(t)$ and Rabi frequency $\Omega(t)$, transfers part of the population of each ion from the state $\ket{1}$ to the Rydberg state $\ket{-}$.
        Only when both ions are in the Rydberg state, they interact via an electric dipole-dipole interaction and accumulate an entanglement phase $\varphi_{*}$, which depends on the interaction strength $V$.
    }
    \label{fig:adiabatic-elimination}
\end{figure}

Given that the energetically low-lying state $\ket{0}$ is not coupled to any other state (see Fig.~\ref{fig:energy-levels}), it follows that the $16$-dimensional Hilbert space of the two-ion Hamiltonian~\eqref{eq:two-ion-hamiltonian} can be decomposed into four distinct subspaces.
These subspaces correspond to the one-, three-, three-, and nine-dimensional subspaces associated with the computational basis states $\ket{00}$, $\ket{01}$, $\ket{10}$, and $\ket{11}$, respectively.  
In the following, we demonstrate that within certain parameter regimes, the dimensions of these subspaces can be reduced significantly enabling deeper insight into the gate dynamics.
As shown in Fig.~\ref{fig:adiabatic-elimination}, this can be accomplished by considering a regime within which the Rydberg state $\ket{+}$ is far off-resonant, allowing it to be adiabatically eliminated~\cite{paulisch2014}.

Assuming the system to be invariant under the interchange of the ions, i.e., when a global laser pulse is applied as in protocols A and B, it is reasonable to perform a basis transformation by introducing the superpositions
\begin{equation}
\begin{aligned}
    \ket{\mathcal{S}_{R}} & = \frac{1}{\sqrt{2}} \big[\ket{+-} + \ket{-+}\!\big], & \qquad
    \ket{\mathcal{A}_{R}} & = \frac{1}{\sqrt{2}} \big[\ket{+-} - \ket{-+}\!\big], \\
    \ket{\mathcal{S}_{\pm}} & = \frac{1}{\sqrt{2}} \big[\ket{\pm 1} + \ket{1 \pm}\!\big], & \qquad
    \ket{\mathcal{A}_{\pm}} & = \frac{1}{\sqrt{2}} \big[\ket{\pm 1} - \ket{1 \pm}\!\big], \\
\end{aligned}
\end{equation}
where the antisymmetric superpositions $\ket{\mathcal{A}_{R}}$ and $\ket{\mathcal{A}_{\pm}}$ are completely decoupled from the system.
Considering the subspace associated with $\ket{11}$, in the new basis $\mathcal{B}_{11}=\{\ket{++} \!, \ket{\mathcal{S}_{R}} \!, \ket{\mathcal{S}_{+}} \!, \ket{--} \!, \ket{\mathcal{S}_{-}} \!, \ket{11}\}$ the Hamiltonian expressed in terms of the laser detuning $\Delta_{\mathrm{L}}$ reads
\begin{equation}\label{eqn:HamiltonianDelta}
    \frac{H_{11}}{\Delta_{\mathrm{L}}} = \frac{1}{2}\!
    \begin{bmatrix}
        4 \varepsilon_{+} + \eta & 0 & \delta & - \eta & 0 & 0 \\
        0 & 2(\varepsilon_{+} + \varepsilon_{-}) & \frac{\delta}{\sqrt{2}} & 0 & \frac{\delta}{\sqrt{2}} & 0 \\
        \delta & \frac{\delta}{\sqrt{2}} & 2\varepsilon_{+} & 0 & 0 & \delta \\
        - \eta & 0 & 0 & 4 \varepsilon_{-} + \eta & \delta & 0 \\
        0 & \frac{\delta}{\sqrt{2}} & 0 & \delta & 2\varepsilon_{-} & \delta \\
        0 & 0 & \delta & 0 & \delta & 0 \\
    \end{bmatrix} \!,
\end{equation}
where we have defined
\begin{equation}
    \varepsilon_{\pm} = \frac{2 \Delta_{\mathrm{L}} \pm \Omega_{\mathrm{MW}}}{2 \Delta_{\mathrm{L}}} = \frac{\Delta_{\pm}}{\Delta_{\mathrm{L}}}, \qquad
    \delta = \frac{\Omega_{\mathrm{L}}}{\Delta_{\mathrm{L}}}, \qquad
    \eta = \frac{V}{\Delta_{\mathrm{L}}}.
\end{equation}
The remaining six-dimensional Hilbert space can be further reduced by considering a regime where we can adiabatically eliminate one of the dressed Rydberg states.
To visualize in which regime this holds, Fig.~\ref{fig:adiabatic-elimination} depicts the coupling of the energy levels for a single ion.
As illustrated, we find that for $\Omega_{\mathrm{MW}} \approx  2 \Delta_{\mathrm{L}}$, i.e., $\varepsilon_{-}\!\ll 1$, the coupling between the Rydberg state $\ket{-}$ and the qubit state $\ket{1}$ becomes near on-resonant, while the coupling to $\ket{+}$ becomes far off-resonant.
In a regime where the latter coupling is sufficiently small compared to the detuning, i.e., $\Omega_{\mathrm{L}}\!\ll\!\Delta_{\mathrm{L}}$ or $\delta \ll \varepsilon_{+}$, respectively, we can adiabatically eliminate the Rydberg state $\ket{+}$ \cite{brion2007}.
Note that we could analogously perform an elimination of $\ket{-}$ by assuming $\Omega_{\mathrm{MW}} \approx  -2 \Delta_{\mathrm{L}}$.

\begin{figure}[t!]
    \centering
    \includegraphics[scale=1]{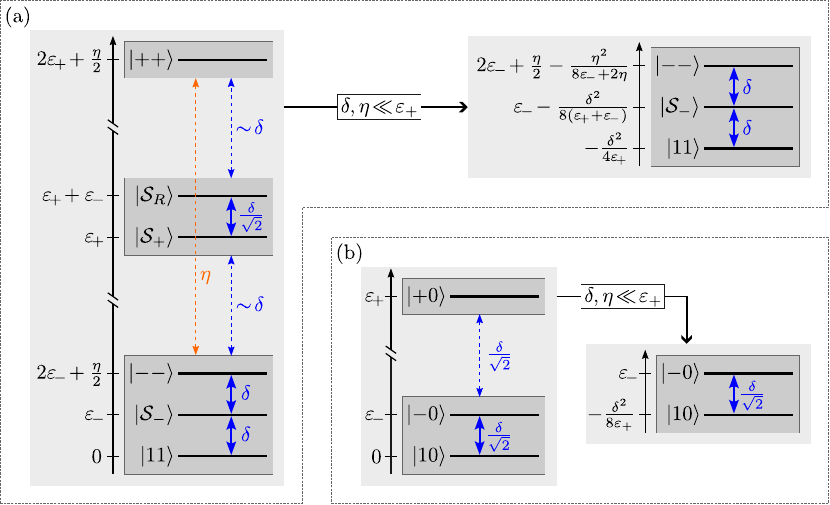}
    \caption{
        Energy level diagram of the two-ion Hamiltonian for $\varepsilon_{-}\!\ll 1$. 
        (a) The subspace associated to the state $\ket{11}$ described by Hamiltonian \eqref{eqn:HamiltonianDelta} decomposes into three energetically separated subspaces.
        Assuming that the coupling between these subspaces is small compared to the energy separation between them, i.e., $\delta, \eta \ll \varepsilon_{+}$, we can adiabatically eliminate the states $\ket{++} \!, \ket{\mathcal{S}_{R}}$, and $\ket{\mathcal{S}_{+}}$.
        The three remaining states experience an energy shift $\sim\mathcal{O}(\frac{\delta^{2}}{\varepsilon_{+}}), \mathcal{O}(\frac{\eta^{2}}{\varepsilon_{+}})$.
        (b) This procedure can also be applied for the Hilbert space associated to the state $\ket{10}$.
        In the same regime, the state $\ket{+0}$ is highly detuned and just weakly coupled to the other two states, and thus, can be adiabatically eliminated.
        In the resulting two-dimensional system, the state $\ket{10}$ experiences an energy shift $\sim\mathcal{O}(\frac{\delta^{2}}{\varepsilon_{+}})$ as a consequence of the adiabatic elimination.
    }
    \label{fig:adiabatic-regime}
\end{figure}

Now we aim to extend this concept to the interacting two-ion case.
To this end, Fig.~\ref{fig:adiabatic-regime} shows the energy levels and couplings of the subspace associated to $\ket{11}$ given by the Hamiltonian \eqref{eqn:HamiltonianDelta} for $\varepsilon_{-}\!\ll 1$.
As illustrated, within this regime, the system comprises three energetically separated subspaces.
For $\delta,\eta\ll \varepsilon_{+}$, these subspaces are just weakly coupled considering the energy separations between them, allowing to adiabatically eliminate the states $\ket{++} \!, \ket{\mathcal{S}_{R}}$, and $\ket{\mathcal{S}_{+}}$.
For this, we assume an arbitrary state $\ket{\Psi(t)}=\sum_{\ket{a}\in\mathcal{B}_{11}}\xi_a(t)\ket{a}$, apply the Schrödinger equation defined by Hamiltonian~\eqref{eqn:HamiltonianDelta}, and enforce
\begin{align}
    \partial_{t} \xi_{++}(t) &= \partial_{t} \xi_{\mathcal{S}_{R}}(t) = \partial_{t} \xi_{\mathcal{S}_{+}}(t) = 0,
\end{align}
where only first-order couplings are taken into account.
The resulting three-dimensional Hamiltonian approximating the relevant dynamics within the reduced subspace $\{\ket{--} \!, \ket{\mathcal{S}_{-}} \!, \ket{11}\}$ reads
\begin{equation}\label{eq:reduced-hamiltonian-11}
    \frac{H_{11}^{\mathrm{eff}}}{\Delta_{\mathrm{L}}} = \frac{1}{2}\!
    \begin{bmatrix}
        4 \varepsilon_{-} + \eta - \frac{\eta^{2}}{4 \varepsilon_{+} + \eta} & \delta & 0 \\
        \delta & 2\varepsilon_{-} - \frac{\delta^{2}}{4(\varepsilon_{+} + \varepsilon_{-})} & \delta \\
        0 & \delta & - \frac{\delta^{2}}{2\varepsilon_{+}} \\
    \end{bmatrix} \!
\end{equation}
with the remaining states experiencing an energy shift $\sim\mathcal{O}(\frac{\delta^{2}}{\varepsilon_{+}}), \mathcal{O}(\frac{\eta^{2}}{\varepsilon_{+}})$ as a consequence of the adiabatic elimination.

As illustrated in Fig~\ref{fig:adiabatic-regime}, in the same regime, the three-dimensional subspace associated to the state $\ket{10}$ can be similarly reduced to the two-dimensional subspace $\{\ket{-0} \!, \ket{10}\}$ by eliminating the state $\ket{+0}$, giving
\begin{equation}\label{eq:reduced-hamiltonian-10}
    \frac{H_{10}^{\mathrm{eff}}}{\Delta_{\mathrm{L}}} =
    \begin{bmatrix}
        \varepsilon_{-} & \frac{\delta}{2\sqrt{2}} \\[5pt]
        \frac{\delta}{2\sqrt{2}} & - \frac{\delta^{2}}{8\varepsilon_{+}} \\
    \end{bmatrix} \!.
\end{equation}
Note that due to the system's symmetry, the two states $\ket{10}$ and $\ket{01}$ behave equivalently, therefore, we treat the subspace associated to $\ket{10}$ as representative for both cases.

\begin{figure}[t!]
    \centering
    \includegraphics[scale=1]{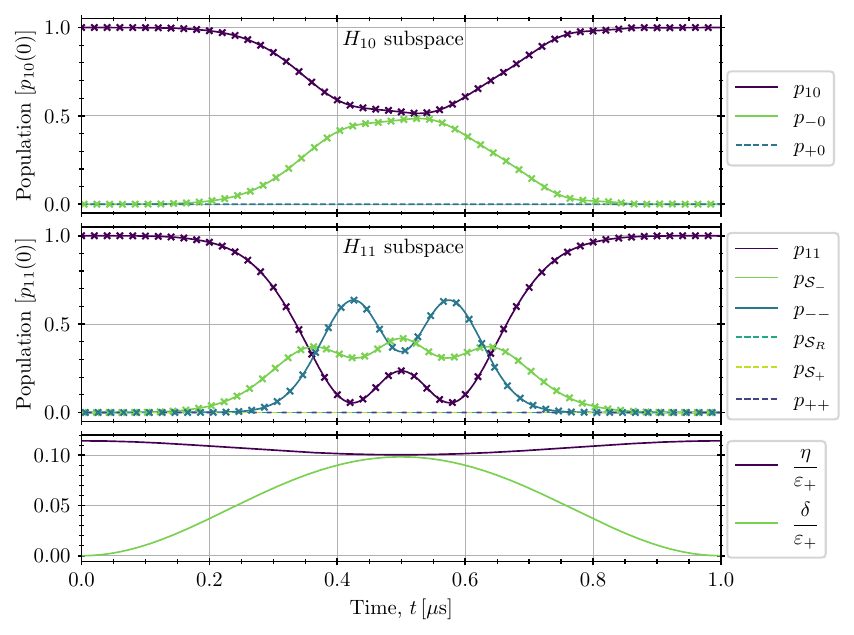}
    \caption{
        Comparison of the time evolution of the state populations during the CZ gate implemented using Protocol B with the conservative parameters tabulated in Fig.~\ref{tab:parameter_table}.
        The lines denote the populations calculated using the full Hamiltonian in Eq.~\eqref{eq:two-ion-hamiltonian}, while the markers ($\times$) indicate those computed using the reduced Hamiltonians $H^{\text{eff}}_{10}$~\eqref{eq:reduced-hamiltonian-10}(top) and $H^{\text{eff}}_{11}$ \eqref{eq:reduced-hamiltonian-11}(middle).
        As expected, the occupancy of the eliminated states (dashed lines) remains negligible throughout the duration of the gate, since the parameters satisfy the adiabatic condition, $\delta, \eta \ll \varepsilon_{+}$, during the entire pulse (bottom). 
        This is evidenced by the fact that the dynamics of the reduced Hamiltonian closely match those of the full Hamiltonian, as demonstrated by the very good overlap of the markers and lines.
    }
    \label{fig:adiabatic-comparison}
\end{figure}

In the context of the effective Hamiltonians in Eqs.~\eqref{eq:reduced-hamiltonian-11} and \eqref{eq:reduced-hamiltonian-10}, the application of a time-dependent pulse, as employed for the protocols in Sec.~\ref{sec:gate-schemes}, implies that $\varepsilon_{\pm}$, $\delta$, and $\eta$ become time-dependent. 
The conservative parameters used in Protocol B reside within the regime for which the adiabatic elimination can be applied (cf. Tab.~\ref{tab:parameter_table}). 
To demonstrate this, we plot the dynamics of the populations of the subspaces associated to the states $\ket{10}$ and $\ket{11}$ during the gate implementation in Fig.~\ref{fig:adiabatic-comparison}. 
Evidently, the occupation of the $\ket{+}$ state by either ion is negligible during the protocol, as expected within this regime. 
Hence, the approximate dynamics described by the effective Hamiltonians in Eqs.~\eqref{eq:reduced-hamiltonian-11} and \eqref{eq:reduced-hamiltonian-10} closely matches the exact dynamics resulting from solving the Schr{\"o}dinger equation of the full $16$-dimensional Hamiltonian~\eqref{eq:two-ion-hamiltonian}. 
This demonstrates that the approximation is also applicable for time-dependent pulses, provided that the perturbation parameters remain sufficiently small throughout the pulse-driven interaction, as shown in the lower plot of Fig.~\ref{fig:adiabatic-comparison}. 
Note that since, as demonstrated in Fig.~\ref{fig:adiabatic-elimination}, the decoupling of the $\ket{+}$ state is independent of the number of ions, this approximation is not limited to two ions and can be generalized to the $N$-ion case.

The advantage of considering this parameter regime is, on the one hand, faster numerical simulations due to the reduced size of the Hilbert space, and on the other hand, deeper insight into the processes during the gate application.
As illustrated in Fig.~\ref{fig:adiabatic-comparison}, during the laser pulse, part of the electronic population in~$\ket{1}$ is excited to the Rydberg state $\ket{-}$.
Only when both ions are in the Rydberg state, i.e., when population is transferred to the state $\ket{--}$, they interact via dipole-dipole interaction and thereby pick up a non-trivial phase.
In a near-adiabatic regime (see discussion at the beginning of Sec.~\ref{sec:gates}), the accumulated entangling phase can be approximated via
\begin{equation}\label{eq:ent_phase_inst_eigenenergies}
    \varphi_{*} \approx \int_{0}^{\tau} \d t \, [\epsilon_{11}(t) - \epsilon_{10}(t) - \epsilon_{01}(t)]
\end{equation}
with $\epsilon_{ab}(t)$ the eigenenergy of the instantaneous eigenstate adiabatically connected to the state $\ket{ab}$ \cite{jaksch2000}.
Although not discussed in this work, an analytical expression for the entangling phase $\varphi_{*}$ can be obtained within the adiabatic elimination regime by formally integrating Eq.~\eqref{eq:ent_phase_inst_eigenenergies}.
Such an expression could then enable deeper understanding of the correlations between the experimental parameters and accumulated phases, facilitating better gate designs and further optimized pulse shapes.

\subsection{Imperfections and possible improvements}

\begin{figure}[t!]
    \centering
    \includegraphics[scale=1]{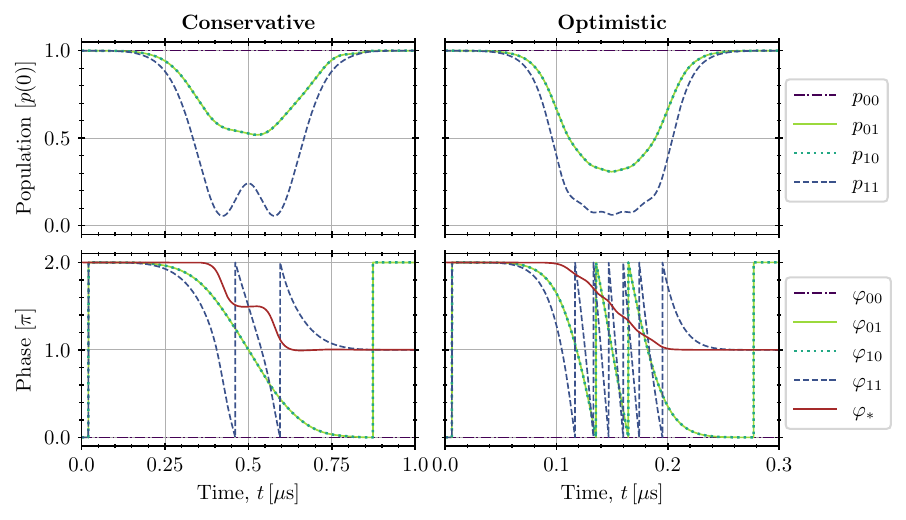}
    \caption{
        Gates utilizing Protocol B~\eqref{eqn:protocolB} optimized within the conservative (left) and optimistic~(right) parameter ranges without allowing for additional single-qubit rotations.
        The resulting optimized gate parameters are ${\Delta_0 = -2\pi \times 11.50\, \unit{\mega\hertz}}$, ${\delta_0 = 2\pi \times 37.97\, \unit{\mega\hertz}}$, and ${\Omega_0 = 2\pi \times 9.71\, \unit{\mega\hertz}}$ in the conservative and ${\Delta_0 = -2\pi \times 134.37\, \unit{\mega\hertz}}$, ${\delta_0 = 2\pi \times 8.39\, \unit{\mega\hertz}}$, and ${\Omega_0 = 2\pi \times 72.72\, \unit{\mega\hertz}}$ in the optimistic case.
        The time-evolution of the populations of the computational basis states is shown in the upper plot, while the lower plot depicts how the accumulated phases evolve during the gate implementation.
        There are no obvious deviations from the conditions for a gate without requiring additional single-qubit rotations as defined in Eq.~\eqref{eqn:constraints_CZ}, which is evidenced by high fidelities of $99.98\%$ in the conservative and $>99.99\%$ in the optimistic regime.
    }
    \label{fig:Gate-noSQR}
\end{figure}

In Sec.~\ref{sec:gate-schemes}, we have simulated CZ gates up to additional single qubit rotations by maximizing the fidelity~in Eq.~\eqref{eq:fidelity-sqr} within given parameter bounds. 
Since in an algorithm it is preferable to use as few gates as possible to reduce the error rate~\cite{preskill1998}, we generally favor gates that do not require additional single-qubit rotations. 
However, for this we need to fulfill stricter constraints defined by Eq.~\eqref{eqn:constraints_CZ}.
For Protocol~B~\eqref{eqn:protocolB}, within the conservative parameter limits given in Tab.~\ref{tab:parameter_table}, the fidelity achieved by applying our optimization scheme for a CZ gate without single-qubit rotations is $99.98\,\%$. 
In this case, we have used the optimized parameters from Sec.~\ref{sec:gate-schemes} as an initial guess in the optimization protocol since the conservative gate in Fig.~\ref{fig:gate-comparison-B} is already close to working without applying additional single-qubit rotations.
In the optimistic regime, a fidelity of $> 99.99\,\%$ can be achieved for a $300\, \unit{\nano\second}$ CZ gate without single-qubit rotations.
The corresponding dynamics of the populations of the computational basis states and the accumulated phases during both gates are depicted in Fig.~\ref{fig:Gate-noSQR}.

\begin{figure}[t!]
    \centering
    \includegraphics[scale=1]{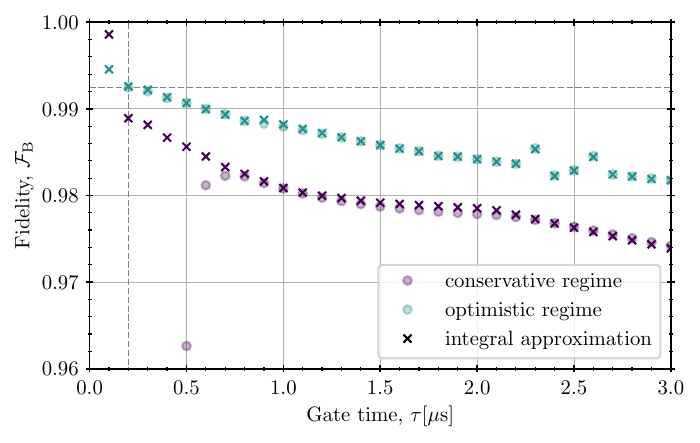}
    \caption{
        Fidelities of the CZ gate implemented with Protocol B \eqref{eqn:protocolB} optimized for different gate times~$\tau$ within conservative (purple) and optimistic (green) parameter bounds (cf. Tab.~\ref{tab:parameter_table}). 
        As long as the decay is the dominant error source, the approximate fidelities calculated via the integral~\eqref{eq:fidelity-decay}~($\times$) agree well with the values obtained by integrating the Hamiltonian~\eqref{eq:two-ion-hamiltonian} with the addition of the non-Hermitian term~\eqref{eq:non-hermitian-hamiltonian} ($\circ$). 
        For the Rydberg state lifetime $\tau_{\mathrm{R}} \approx 7.8 \, \unit{\micro\second}$~\cite{zhang2020}, the Bell state fidelities of the gates in Fig.\ref{fig:gate-comparison-B} drop from~$99.98\,\%$ to~$98.10\%$ in the conservative and from~$>99.99\,\%$ to~$99.20\,\%$ in the optimistic regime.
        With a fidelity of $99.25\,\%$, the best performance is achieved in the optimistic regime for $\tau = 200\,\unit{\nano\second}$ as indicated by the dashed lines.
    }
    \label{fig:fidelity-decay}
\end{figure}

The focus so far has been on demonstrating how effective our general approach of fidelity optimization is for different protocols. 
For this we have neglected error sources like the finite lifetime of the Rydberg states, and therefore we have been able to obtain gate fidelities close to one.
However, for physically realistic pulse schemes, the finite lifetime of the Rydberg states is a primary factor contributing to the loss of fidelity in trapped Rydberg ion platforms~\cite{mokhberi2020}. 
Hence, in the following we briefly examine the influence of the Rydberg state decay on the fidelity of Protocol B~\eqref{eqn:protocolB}. 
To this end, we augment the Hamiltonian in Eq.~\eqref{eq:two-ion-hamiltonian} by appending a non-Hermitian term,
\begin{equation}\label{eqn:non-herm-Hamiltonian}
    H_{\mathrm{R}} = - \frac{\i}{2} \sum_{\mathclap{i = 1}}^{2} [\gamma_{3} \! \op{3}{3}_{i} + \gamma_{4} \! \op{4}{4}_{i}],
\end{equation}
where $\gamma_{3}$ and $\gamma_{4}$ represent the decay rates from the electronic Rydberg states ${\ket{3} \equiv \ket{n S_{\smash{1/2}}^{\smash{-1/2}}}}$ and \linebreak ${\ket{4} \equiv \ket{n P_{\smash{1/2}}^{\smash{1/2}}}}$, respectively.
For the sake of simplicity, we assume that the Rydberg state decay rates are equal such that $\gamma_{3} \approx \gamma_{4} \equiv \gamma_{\mathrm{R}}$. 
For an approximate lifetime of the Rydberg manifold $\tau_{\mathrm{R}} \approx 7.8 \, \unit{\micro\second}$~\cite{zhang2020}, we then have $\gamma_{\mathrm{R}} = 1 / \tau_{\mathrm{R}}\approx 0.13 \, \unit{\per\micro\second}$. 
As such, we can succinctly rewrite Eq.~\eqref{eqn:non-herm-Hamiltonian} in terms of the microwave dressed Rydberg states as
\begin{equation}\label{eq:non-hermitian-hamiltonian}
    H_{\mathrm{R}} \approx -\frac{\i \gamma_{\mathrm{R}}}{2} \sum_{\mathclap{i = 1}}^{2} [\op{+}{+}_{i} + \op{-}{-}_{i}].
\end{equation}
The benefit of this simple approach is that specific information regarding the decay (i.e., the decay channels) is not required. 
Nevertheless, we emphasize that a more precise description of the decay effects requires employing the Lindbladian formalism~\cite{zhang2020}. 
However, due to the relatively long lifetime compared to the gate duration, the approximate description by means of the non-Hermitian Hamiltonian in Eq.~\eqref{eq:non-hermitian-hamiltonian} is sufficient to get a quantitatively accurate approximation of the effect of the finite lifetime on the gate fidelity. 
Since the decay probability is directly related to the time the ion remains in the Rydberg states, we can approximate the reduced Bell-state fidelity by
\begin{equation}\label{eq:fidelity-decay}
    \mathcal{F}_{\mathrm{B}} \approx \bigg| 1 - \frac{\gamma_{\mathrm{R}}}{2} \int_{0}^{\tau} \d t \sum_{\mathclap{i = 1}}^{2} [p_{+}^{i}(t) + p_{-}^{i}(t)] \bigg|^{2},
\end{equation}
where $p_{\pm}^{i}(t)$ denotes the population of ion $i$ in the state $\ket{\pm}$.
Figure~\ref{fig:fidelity-decay} illustrates how the gate fidelity~\eqref{eq:fidelity-sqr} changes as we vary the gate time $\tau$. 
We find that for longer gate times the fidelity determined by integral~\eqref{eq:fidelity-decay} coincides remarkably well with that calculated by the non-Hermitian Hamiltonian. 
This holds as long as the Rydberg state decay is the dominant source of fidelity loss, since the approximation in Eq.~\eqref{eq:fidelity-decay} assumes unit fidelity whenever the decay is absent.
Since the optimization protocol is not able to achieve fidelities close to one for very short gate times, i.e., $\tau < 0.8\,\unit{\micro\second}$ in the conservative and $\tau < 0.2\,\unit{\micro\second}$ in the optimistic regime, the approximation is not valid in these regions.
We conjecture that the reduced fidelities are due to the short gate durations combined with insufficient interaction strengths.
Therefore, the time spent in the Rydberg states is too short to accumulate the desired phases.
For the two gates in Fig.~\ref{fig:gate-comparison-B} utilizing Protocol B, we observe that the gate fidelities achieved by applying the optimization procedure drop from $99.98\,\%$ to $98.10\%$ in the conservative and from $>99.99\,\%$ to $99.20\,\%$ in the optimistic regime. 
This demonstrates that the finite lifetime of the Rydberg states significantly impacts the gate performance, even for submicrosecond gates. 
Nonetheless, we find that, despite decay, an even faster gate with $\tau = 200\,\unit{\nano\second}$ and a fidelity of $99.25\,\%$ can be realized, clearly demonstrating that fast, high-fidelity quantum gates are feasible on trapped Rydberg ion platforms.

While our optimization procedure already considers the impact of the finite Rydberg lifetime, an alternative approach to reduce the fidelity loss due to decay is to consider pulse shapes in which the ions spend less time in the Rydberg states. 
This has already been investigated for neutral Rydberg atoms, wherein a gate fidelity of more than $99.9\,\%$ was achieved, despite decay, by using optimized pulse shapes~\cite{pagano22}. 
Since the natural lifetime of Rydberg ions increases with the principal quantum number as $n^3$ and the blackbody radiation-limited lifetime as $n^2$ \cite{higgins2018}, exciting the ions to higher Rydberg states might also mitigate the impact of the decay on the gate fidelity. 
However, besides the fact that due to limitations in laser power, this remains an experimentally challenging task~\cite{mokhberi2020}, it also increases the probability of the ion being ionized by the trap potential, which in turn would lead to fidelity loss.

Another source of decoherence is the two-photon excitation mechanism (cf. Sec.~\ref{subsec:electronic_subspace}), which is utilized to excite the trapped ions to Rydberg states via a detuned intermediate state~\cite{zhang2020}. 
This way it is possible to reach Rydberg states with high principal quantum numbers. 
However, this also leads to additional photon scattering, preventing the ion from occupying the desired Rydberg state, which in turn results in a loss of fidelity. 
An additional factor contributing to imperfect gate operation in experiments is the ionization of ions in the Rydberg state. 
Since the Rydberg states are energetically close to the ionization limit, black body radiation can cause ionization in insufficiently cooled systems~\cite{pokorny2020, higgins2018}. 
However, adequate cooling can significantly suppress this effect~\cite{zhang2020, beterov2009, magalhaes2000, brandl2016}, and in our simulations, it is already included in the finite Rydberg lifetime.

 
\section{Conclusions and outlook}

In this work, we have explored the feasibility of laser-excited Rydberg ions confined within a linear Paul trap for quantum information processing via an extensive study of which two-qubit controlled phase gates can be optimally achieved. 
We have shown via a detailed derivation that, assuming realistic experimental conditions, an effective Hamiltonian can be obtained capturing the dynamics of the system.  
For each ion we consider two energetically low-lying electronic states, constituting the qubit states, and two high-lying Rydberg states.
We have demonstrated that a microwave mode coupling the two Rydberg states enables strong and long-range dipole-dipole interactions between ions excited to the Rydberg manifold.
Making use of the derived Hamiltonian, we have simulated the evolution of a two-ion system driven by a time-dependent laser pulse (i.e., time-dependent Rabi frequency and detuning) that couples the qubit states to the Rydberg manifold, effectively implementing fast and robust CZ gates.
In order to find optimal pulse parameters, which maximize the fidelity of the CZ gate, we have developed an optimization protocol using a stochastic, population-based method, that searches for the global optimum within a certain parameter range.
With this, we have theoretically investigated three gate protocols by numerically optimizing the gate fidelities and tracking the qubit-state populations and accumulated phases during the gate implementation within two experimental parameter regimes. 
In the absence of decoherence, we have shown that submicrosecond, near unity fidelity gates can be realized for a particularly simple and experimentally feasible pulse scheme. 
We have extended our analysis to include the finite lifetime of the Rydberg states, introducing  a realistic source of fidelity loss.
Notwithstanding, our optimization protocol reveals that a $200\, \unit{\nano\second}$ CZ-gate with $99.25\%$ fidelity should be experimentally realizable, thereby reaching an accuracy required for scalable, error-corrected quantum computing, for which fidelities above $99\,\%$ are essential.

The optimization scheme employed here, as well as the analytical Hilbert-space reduction via adiabatic elimination, can also be used for models with larger numbers of ions. This should allow for design and optimization of multi-ion gates with Rydberg ions, further extending the set of available native gates in this platform and potentially allowing for the development of specific quantum-information-processing protocols that profit from the use of this native gate set.


\section*{Acknowledgments}

We thank M. Mallweger, M. Hennrich, and F. Schmidt-Kaler for fruitful discussions. 
We gratefully acknowledge the support by the European Union’s Horizon Europe Research and Innovation Program under Grant No. 101046968 (BRISQ) and funding from the Deutsche Forschungsgemeinschaft (DFG, German Research Foundation), Projects No. 428276754 and No. 435696605, and through the Research Unit FOR 5413/1, Grant No. 465199066. 
KB, TLMG, and MM gratefully acknowledge funding and support provided by the Deutsche Forschungsgemeinschaft (DFG, German Research Foundation) under Germany’s Excellence Strategy ‘Cluster of Excellence Matter and Light for Quantum Computing (ML4Q) EXC 2004/1’ 390534769, the ERC Starting Grant QNets through Grant No. 804247, the German Federal Ministry of Education and Research (BMBF) through the project IQuAn, and the Entangled Logical Qubits (ELQ) program, managed by the Intelligence Advanced Research Projects Activity (IARPA).
This research is also part of the Munich Quantum Valley (K-8), which is supported by the Bavarian state government with funds from the Hightech Agenda Bayern Plus. 
This work was also supported by funding as part of the Excellence Strategy of the German Federal and State Governments, in close collaboration with the University of T{\" u}bingen and the University of Nottingham. 
This work also received support from the Engineering and Physical Sciences Research Council, Grant No. EP/V031201/1 and EP/W015641/1. JW was supported by the University of T{\" u}bingen through a Research@T{\" u}bingen fellowship. The code used to produce the data supporting the findings of this article is available on Zenodo~\cite{zenodo2024}.


\appendix

\section{Numerical diagonalization of the field-free electronic Hamiltonian}\label{app:electronic-states}

In this appendix, we outline some of the details behind the numerical computation of the eigenenergies and eigenfunctions of the field-free electronic Hamiltonian,
\begin{equation}
    H_{\mathrm{in}}^{\mathrm{free}} = \frac{\bm{p}^{2}}{2 m} + V(\ab{\bm{r}}).
\end{equation}
In general, this constitutes a complicated many-electron system.
However, since all but one electron are core electrons and form closed shells around the nucleus, the ion can be treated within an effective two-body approximation~\cite{gallagher1988, gallagher1994, gallagher2023}.
Here, the ions are modeled by a valence electron and an ionic core, composed of the screened nucleus and core electrons.
The model potential describing their interaction is parametrized by~\cite{aymar1996}
\begin{equation}\label{eq:model-potential}
    V(r) = V_{\mathrm{C}}(r) + V_{\mathrm{P}}(r) + V_{\mathrm{R}}(r),
\end{equation}
where $r = \ab{\bm{r}}$ denotes the relative radial coordinate.
The first term, $V_{\mathrm{C}}(r)$, is a modified Coulomb potential that describes the effective central potential experienced by the valence electron due to the screening of the charge of the nucleus by the core electrons.
The second term, $V_{\mathrm{P}}(r)$, is an induced polarization potential that accounts for the effects of the dipole moment of the screened ionic core due to the presence of the valence electron.
The third term, $V_{\mathrm{R}}(r)$, is then a relativistic spin-orbit interaction potential which specifies the coupling of the magnetic moment of the valence electron with the magnetic field of the screened nucleus and core electrons.
All three terms depend on both the position of the valence electron relative to the ionic core $r$ and the orbital angular momentum through the orbital angular momentum quantum number $l$.
This latter dependence accounts for the \textit{quantum defect}~\cite{gallagher2023}, which quantifies the lowering of the energy of electronic states with low angular momentum due to the probing of the core electrons by the valence electron.
Following Ref.~\cite{aymar1996}, the modified Coulomb potential $V_{\mathrm{C}}(r)$ can be expressed as
\begin{equation}\label{eq:modified-coulomb-potential}
    V_{\mathrm{C}}(r) = - C e^{2} \frac{Z_{\mathrm{n}}(r)}{r},
\end{equation}
where $Z_{\mathrm{n}}(r)$ is the modified nuclear charge number given by
\begin{equation}
    Z_{\mathrm{n}}(r) = Z_{\mathrm{c}} + [Z - Z_{\mathrm{c}}] \e^{-k_{1} r} + k_{2} r \e^{-k_{3} r}
\end{equation}
with $Z$ the nuclear charge number and $Z_{\mathrm{c}} = 2$ the effective charge number of the screened ionic core.
The $k_{i}$ are empirically determined fitting parameters that depend on the ion and orbital angular momentum quantum number of the electronic state.
These are obtained by solving the radial Schr{\"o}dinger equation and requiring that the field-free electronic eigenenergies calculated theoretically agree with those measured experimentally.
The induced core polarizability potential $V_{\mathrm{P}}(r)$ reads
\begin{equation}\label{eq:induced-polarization-potential}
    V_{\mathrm{P}}(r) = - C^{2} e^{2} \frac{\alpha_{\mathrm{d}}}{2 r^{4}} \big[1 - \e^{-[r / r_{\mathrm{c}}]^{6}}\big],
\end{equation}
where $\alpha_{\mathrm{d}}$ is the static electric dipole polarizability of the doubly-positively-charged ionic core and $r_{\mathrm{c}}$ the cutoff radius (i.e., the effective size of the ionic core).
This cutoff is introduced to ensure that the polarization potential is physically well behaved near the origin~\cite{aymar1996}.
Explicit values for the parameters $k_{i}$ and cutoff radius $r_{\mathrm{c}}$ taken from Ref.~\cite{aymar1996} for the alkaline earth metal ions are presented in Tab.~\ref{tab:empirical-fitting-parameters}.\footnote{Note that those for the alkaline metals can be found in Ref.~\cite{marinescu1994}.}
The relativistic spin-orbit interaction potential $V_{\mathrm{R}}(r)$ then follow as
\begin{equation}\label{eq:relativistic-spin-orbit-potential}
    V_{\mathrm{R}}(r) = \frac{1}{2 m^{2} c^{2}} \frac{1}{r N(r)} \frac{\partial V_{\mathrm{NR}}(r)}{\partial r} \bm{L} \cdot \bm{S},
\end{equation}
where $V_{\mathrm{NR}}(r) = V_{\mathrm{C}}(r) + V_{\mathrm{P}}(r)$ is the nonrelativistic model potential with $\bm{L}$ and $\bm{S}$ the orbital and spin angular momentum operators.
Note that their dot product can be written more conveniently as
\begin{equation}
    \bm{L} \cdot \bm{S} = \frac{1}{2} [\bm{J}^{2} - \bm{L}^{2} - \bm{S}^{2}],
\end{equation}
where $\bm{J} = \bm{L} + \bm{S}$ is the total angular momentum operator.
The normalization factor $N(r)$, defined by
\begin{equation}
    N(r) = \bigg[ 1 - \frac{V_{\mathrm{NR}}(r)}{2 m c^{2}} \bigg]^{2},
\end{equation}
and suggested by relativistic first-order perturbation theory~\cite{condon1935}, is included to regularize the electronic wavefunction at the origin (i.e., to guarantee that the solutions of the radial Schr{\"o}dinger equation are well defined near the origin as $r \to 0$).

\begin{table}[t!]
    \begin{center}
    \begin{tabular}{cccccccc}
        \hline\hline
        $\quad \mathrm{Ion} \quad$ & $\quad Z \quad$ & $\quad \alpha_{\mathrm{d}} \quad$ & $\quad l \quad$ & $\quad k_{1} \quad$ & $\quad k_{2} \quad$ & $\quad k_{3} \quad$ & $\quad r_{\mathrm{c}} \quad$ \\
        \hline
        $^{40}\mathrm{Ca}^{+}$ & 20 & 3.5 & s & 4.0616 & 13.4912 & 2.1539 & 1.5736 \\
         & & & p & 5.3368 & 26.2477 & 2.8233 & 1.0290 \\
         & & & d & 5.5262 & 29.2059 & 2.9216 & 1.1717 \\
         & & & f & 5.0687 & 24.3421 & 6.2170 & 0.4072 \\
        \hline
        $^{88}\mathrm{Sr}^{+}$ & 38 & 7.5 & s & 3.4187 & 4.7332 & 1.5915 & 1.7965 \\
         & & & p & 3.3235 & 2.2539 & 1.5712 & 1.3960 \\
         & & & d & 3.2533 & 3.2330 & 1.5996 & 1.6820 \\
         & & & f & 5.3540 & 7.9517 & 5.6624 & 1.0057 \\
        \hline
        $^{138}\mathrm{Ba}^{+}$ & 56 & 11.4 & s & 3.0751 & 2.6107 & 1.2026 & 2.6004 \\
         & & & p & 3.2304 & 2.9561 & 1.1923 & 2.0497 \\
         & & & d & 3.2961 & 3.0248 & 1.2943 & 1.8946 \\
         & & & f & 3.6237 & 6.7416 & 2.0379 & 1.0473 \\
        \hline
        $^{226}\mathrm{Ra}^{+}$ & 88 & 18.0 & s & 3.7702 & 4.9928 & 1.5179 & 1.3691 \\
         & & & p & 3.9430 & 5.0552 & 3.6770 & 1.0924 \\
         & & & d & 3.7008 & 4.7748 & 1.4956 & 2.2784 \\
         & & & f & 3.8125 & 5.0332 & 2.1016 & 1.2707 \\
        \hline\hline
    \end{tabular}
    \end{center}
    \caption{Empirically determined dimensionless fitting parameters for the parametric model potential $V(r)$ in Eq.~\eqref{eq:model-potential} for the alkaline earth metal ions, taken from Ref.~\cite{aymar1996}.
    For readability, parameter values are expressed in atomic units.
    In particular, the static core electric dipole polarizability $\alpha_{\mathrm{d}}$ is in units of $m_{e} e^{2} a_{0}^{4} / \hbar^{2}$, the coefficients $k_{i}$ are in units of $1 / a_{0}$, and the cutoff radius $r_{\mathrm{c}}$ is in units of $a_{0}$, where $a_{0} = \hbar^{2} / m_{e} C e^{2}$ is the Bohr radius.}
    \label{tab:empirical-fitting-parameters}
\end{table}

Under the effective two-body approximation and in the absence of the electric potential of the linear Paul trap, excitation laser, and remaining valence electrons, the bound electronic quantum states of the ions can be uniquely characterized by the principal $n$, orbital angular momentum $l$, spin angular momentum $s$, total angular momentum $j$, and total magnetic $m_{j}$ quantum numbers.
Since the model potential $V(r)$ commutes with the total angular momentum operator $\bm{J}$ and, therefore, conserves the total angular momentum (i.e., it preserves the spherical symmetry), the electronic states can be factored into radial and angular parts,
\begin{equation}
    \ket{n, l, s, j, m_{j}} \equiv \underbrace{\ket{n, l, j}}_{\text{radial}} \! \underbrace{\ket{l, s, j, m_{j}}}_{\text{angular}},
\end{equation}
where for simplicity we omit the spin angular momentum quantum number $s = 1/2$ when unnecessary (e.g., in the radial part of the electronic state).
The angular part can be further decomposed into orbital and spin angular components using standard angular momentum theory~\cite{louck2023},
\begin{equation}
    \ket{l, s, j, m_{j}} = \sum_{\mathclap{m_{s} = -s}}^{s} \ip{l, m_{j} - m_{s}, s, m_{s}}{j, m_{j}} \! \underbrace{\ket{l, m_{j} - m_{s}}}_{\text{orbital}} \! \underbrace{\ket{s, m_{s}}}_{\text{spin}},
\end{equation}
where $\ip{l, m_{l}, s, m_{s}}{j, m_{j}}$ are Clebsch-Gordon coefficients with $m_{l}$ and $m_{s}$ the orbital and spin magnetic quantum numbers, respectively.
Note that due to the selection rules, the angular momentum quantum numbers must satisfy $m_{l} + m_{s} = m_{j}$, hence the missing summation over $m_{l}$.
For generality, however, we will include this summation, yet omit the standard bounds of the summation (i.e., $m_{s} = -s, ..., s$).

The Schr{\"o}dinger equation for the field-free electronic Hamiltonian is given by
\begin{equation}
    \bigg[\frac{\bm{p}^{2}}{2 m} + V(\ab{\bm{r}})\bigg] \! \ket{n, l, s, j, m_{j}} = E_{n l j} \! \ket{n, l, s, j, m_{j}} \!,
\end{equation}
where $E_{n l j}$ is the eigenenergy associated to the eigenstate $\ket{n, l, s, j, m_{j}}$.
In order to transform this into the wavefunction representation, we multiply by the position-spin vector $\ket{\bm{r}, \sigma} = \ket{r, \theta, \varphi, \sigma}$, with $\ket{\sigma} = \ket{\uparrow}\!, \ket{\downarrow}$ denoting the spin projections (i.e., $m_{s} = \pm 1/2$).
Then the Schr{\"o}dinger equation reads
\begin{equation}
    \bigg[ \!-\! \frac{\hbar^{2}}{2 m} \boldsymbol{\nabla}^{2} + V(r) \bigg] \Psi_{n l j}^{m_{j}}(\bm{r}, \sigma) = E_{n l j} \Psi_{n l j}^{m_{j}}(\bm{r}, \sigma)
\end{equation}
with $\Psi_{n l j}^{m_{j}}(\bm{r}, \sigma) = \ip{\bm{r}, \sigma}{n, l, s, j, m_{j}}$ the electronic position-spin wavefunction, the components of which are position wavefunctions corresponding to the valence electron in the spin-up and spin-down states.
Here, the position-spin wavefunction is expressed as
\begin{equation}
    \Psi_{n l j}^{m_{j}}(\bm{r}, \sigma) = R_{n l j}(r) \sum_{\mathclap{m_{l}, m_{s}}} \ip{l, m_{l}, s, m_{s}}{j, m_{j}}  Y_{l}^{m_{l}}(\theta, \varphi) \chi_{s}^{m_{s}}(\sigma)
\end{equation}
with the radial, orbital angular, and spin angular components given by
\begin{equation}
    R_{n l j}(r) = \ip{r}{n, l, j} \!, \qquad
    Y_{l}^{m_{l}}(\theta, \varphi) = \ip{\theta, \varphi}{l, m_{l}} \!, \qquad
    \text{and} \qquad
    \chi_{s}^{m_{s}}(\sigma) = \ip{\sigma}{s, m_{s}} \!,
\end{equation}
where $Y_{l}^{m_{l}}(\theta, \varphi)$ are spherical harmonics and $\chi_{s}^{m_{s}}(\sigma)$ are two-spinors~\cite{louck2023}.
Integrating over the angular components, we obtain the radial Schr{\" o}dinger equation,
\begin{equation}\label{eq:radial-equation}
    \bigg[ \!-\! \frac{\hbar^{2}}{2 m} \frac{\d^{2}}{\d r^{2}} + \frac{\hbar^{2} l (l + 1)}{2 m r^{2}} + V(r) \bigg] \Phi_{n l j}(r) = E_{n l j} \Phi_{n l j}(r),
\end{equation}
where we have made the substitution $\Phi_{n l j}(r) = r R_{n l j}(r)$ and impose the physically motivated boundary conditions that the radial wavefunction $\Phi_{n l j}(r) \to 0$ as $r \to 0$ and $r \to \infty$.
As expected, the theoretically calculated values for the eigenenergies are in good agreement with those measured experimentally~\cite{nist2024}.
We are, therefore, sufficiently confident that the associated electronic radial wavefunctions constitute a well grounded basis for the remaining analysis of the trapped Rydberg ions.

\section{Unitary transformation from the stationary to oscillating frame}\label{app:unitary-transformation}

In this appendix, we detail the derivation of the rotating frame Hamiltonian in Eq.~\eqref{eq:oscillating-hamiltonian} from the stationary frame Hamiltonian in Eq.~\eqref{eq:stationary-hamiltonian}.
Recall the unitary transformation under consideration,
\begin{equation}
    H \mapsto U H U^{\dagger} + \i \hbar \frac{\partial U}{\partial t} U^{\dagger} \qquad
    \text{where} \qquad
    U = \mathrm{exp}\bigg(\i \frac{e \alpha}{\hbar \nu} \sin(\nu t) [R_{x}^{2} - R_{y}^{2}]\bigg),
\end{equation}
and the canonical commutation relations for the center of mass and relative positions and momenta,
\begin{equation}
    [R_{u}, r_{v}] = [R_{u}, p_{v}] = [P_{u}, r_{v}] = [P_{u}, p_{v}] = 0, \qquad
    [R_{u}, P_{v}] = [r_{u}, p_{v}] = \i \hbar \delta_{u v},
\end{equation}
for $u, v = x, y, z$.
It follows that the only term within the Hamiltonian $H$ that does not commute with the unitary operator $U$ is the center of mass momentum $\bm{P}$.
To calculate expressions for this term, we employ the Baker–Campbell–Hausdorff (BCH) formula.
From this, one finds that third and higher order commutators vanish.
Hence, the center of mass momentum transforms exactly as
\begin{equation}
    \bm{P}^{2} \mapsto \bm{P}^{2} - \frac{4 e \alpha}{\nu} \sin(\nu t) [R_{x} P_{x} - R_{y} P_{y}] + \frac{4 e^{2} \alpha^{2}}{\nu^{2}} \sin^{2}(\nu t) [R_{x}^{2} + R_{y}^{2}].
\end{equation}
Recalling the trigonometric identity $2 \sin^{2}(\nu t) = 1 - \cos(2 \nu t)$, the Hamiltonian then transforms as
\begin{equation}
    H \mapsto H + \frac{e^{2} \alpha^{2}}{M \nu^{2}} [R_{x}^{2} + R_{y}^{2}] - \frac{2 e \alpha}{M \nu} \sin(\nu t) [R_{x} P_{x} - R_{y} P_{y}] - \frac{e^{2} \alpha^{2}}{M \nu^{2}} \cos(2 \nu t) [R_{x}^{2} + R_{y}^{2}].
\end{equation}
where we identify the latter two terms as the micromotion due to the linear Paul trap.
Note that the time derivative term from the unitary transformation is
\begin{equation}
    \i \hbar \frac{\partial U}{\partial t} U^{\dagger} = - e \alpha \cos(\nu t) [R_{x}^{2} - R_{y}^{2}],
\end{equation}
which cancels the amplitude of the oscillating electric-field mode of the linear Paul trap.
This then returns the expression for the rotating frame Hamiltonian in Eq.~\eqref{eq:oscillating-hamiltonian}.

\section{Multipole expansion of the electrostatic interaction potential}\label{app:multipole-expansion}

In this appendix, we derive an expression for the multipole expansion of the interaction potential $V_{ij}$ in Eq.~\eqref{eq:interaction-potential} between the charges of ions $i$ and $j$ about the center of mass distance between the ions.
In terms of the center of mass and relative positions $\Rv_{i}$ and $\rv_{i}$, the potential reads
\begin{equation}
    \frac{V_{ij}}{C e^{2}} = \frac{Z_{\mathrm{c}}^{2}}{\ab{\bm{R}_{ij}}} - \frac{Z_{\mathrm{c}}}{\ab{\bm{R}_{ij} + \bm{r}_{i}}} - \frac{Z_\mathrm{c}}{\ab{\bm{R}_{ij} - \bm{r}_{j}}} + \frac{1}{\ab{\bm{R}_{ij} + \bm{r}_{i} - \bm{r}_{j}}},
\end{equation}
where $Z_{\mathrm{c}} = 2$ denotes the effective charge number of the ionic core, $C = 1 / 4 \pi \epsilon_{0}$ the Coulomb constant, and $\epsilon_{0}$ the electric constant.
In general, the multipole expansion of an arbitrary interaction potential can be expressed as either a Taylor expansion in Cartesian coordinates (cf. Ref.~\cite{muller2008}) or a Laplace expansion in spherical polar coordinates (cf. Ref.~\cite{weber2017}).
Here, we utilize the former so that we can combine terms of the expansion of the interaction potential with those of the trapped ion Hamiltonian in Eq.~\eqref{eq:oscillating-hamiltonian}.

In order to simplify the calculations, we consider the terms of the interaction potential independently such that the multipole expansion of an arbitrary term can be written as
\begin{equation}
    \frac{1}{\ab{\bm{R}_{ij} + \rv_{k}}} = \sum_{\mathclap{n = 0}}^{\infty} \frac{[\bm{r}_{k} \cdot \boldsymbol{\nabla}_{k}]^{n}}{n!} \bigg(\frac{1}{\ab{\bm{R}_{ij}}}\bigg),
\end{equation}
where explicitly the relative position $\bm{r}_{k} = \bm{0}, \bm{r}_{i}, -\bm{r}_{j}, \bm{r}_{i} - \bm{r}_{j}$ and for clarity the derivatives $\boldsymbol{\nabla}_{i}$ are taken with respect to the center of mass positions $\bm{R}_{i}$.
From this, it follows that the multipole expansion of the arbitrary potential can be written, up to second order, as
\begin{equation}
    \frac{1}{\ab{\bm{R}_{ij} + \bm{r}_{k}}} = \frac{1}{\ab{\bm{R}_{ij}}} - \frac{\bm{n}_{ij} \cdot \bm{r}_{k}}{\ab{\bm{R}_{ij}}^{2}} + \frac{3 [\bm{n}_{ij} \cdot \bm{r}_{k}]^{2} - \bm{r}_{k}^{2}}{2 \ab{\bm{R}_{ij}}^{3}} + \cdots \!,
\end{equation}
where the normalized center of mass positions and center of mass distances,
\begin{equation}
    \bm{n}_{ij} = \frac{\bm{R}_{ij}}{\ab{\bm{R}_{ij}}} \qquad
    \text{and} \qquad
    \bm{R}_{ij} = \bm{R}_{i} - \bm{R}_{j}.
\end{equation}
Respectively setting $\bm{r}_{k}$ as above and truncating to second order then yields the following expressions,
\begin{equation}
\begin{aligned}
    \frac{1}{\ab{\bm{R}_{ij}}} & = \frac{1}{\ab{\bm{R}_{ij}}}, \\
    \frac{1}{\ab{\bm{R}_{ij} + \rv_{i}}} & = \frac{1}{\ab{\bm{R}_{ij}}} - \frac{\nv_{ij} \cdot \rv_{i}}{\ab{\bm{R}_{ij}}^{2}} + \frac{3 [\nv_{ij} \cdot \rv_{i}]^{2} - \rv_{i}^{2}}{2 \ab{\bm{R}_{ij}}^{3}}, \\
    \frac{1}{\ab{\bm{R}_{ij} - \rv_{j}}} & = \frac{1}{\ab{\bm{R}_{ij}}} + \frac{\nv_{ij} \cdot \rv_{j}}{\ab{\bm{R}_{ij}}^{2}} + \frac{3 [\nv_{ij} \cdot \rv_{j}]^{2} - \rv_{j}^{2}}{2 \ab{\bm{R}_{ij}}^{3}}, \\
    \frac{1}{\ab{\bm{R}_{ij} + \rv_{i} - \rv_{j}}} & = \frac{1}{\ab{\bm{R}_{ij}}} - \frac{\nv_{ij} \cdot \rv_{i}}{\ab{\bm{R}_{ij}}^{2}} + \frac{\nv_{ij} \cdot \rv_{j}}{\ab{\bm{R}_{ij}}^{2}} + \frac{3 [\nv_{ij} \cdot \rv_{i}]^{2} - \rv_{i}^{2}}{2 \ab{\bm{R}_{ij}}^{3}} \\
    & \qquad - \frac{3 [\nv_{ij} \cdot \rv_{i}] [\nv_{ij} \cdot \rv_{j}] - \rv_{i} \cdot \rv_{j}}{\ab{\bm{R}_{ij}}^{3}} + \frac{3 [\nv_{ij} \cdot \rv_{j}]^{2} - \rv_{j}^{2}}{2 \ab{\bm{R}_{ij}}^{3}}.
\end{aligned}
\end{equation}
Appropriately combining these terms returns the expression for the second order multipole expansion of the interaction potential in Eq.~\eqref{eq:multipole-expansion}.
Notice that if we do not implicitly set the effective charge number of the ionic core $Z_{\mathrm{c}} = 2$ (i.e., for Rydberg ions), instead we set $Z_{\mathrm{c}} = 1$ (i.e., for Rydberg atoms), then the only nonvanishing term is the electric dipole-dipole interaction.

\section{Computation of the center of mass equilibrium positions}\label{app:equilibrium-positions}

In this appendix, we calculate the center of mass equilibrium positions of the interacting trapped Rydberg ions.
These are determined by the competing potential of the electrostatic interactions between the ions and the electric-field modes trapping the ions.
The corresponding potential reads
\begin{equation}
    V_{\mathrm{ex}} = \frac{M \omega^{2}}{2} \sum_{\mathclap{i = 1}}^{N} \sum_{u} \gamma_{u}^{2} R_{i; u}^{2} + \frac{C e^{2}}{2} \sum_{\mathclap{\substack{i, j = 1 \\ j \neq i}}}^{N} \frac{1}{\ab{\bm{R}_{ij}}},
\end{equation}
where we have expressed the harmonic trap frequencies as $\omega_{u} = \gamma_{u} \omega$ for $\gamma_{x} = \gamma_{y} = \gamma$ and $\gamma_{z} = 1$ with $\omega$ the characteristic trap frequency and $\gamma$ the associated anisotropy given by
\begin{equation}
    \omega = \sqrt{\frac{4 e \beta}{M}} \qquad
    \text{and} \qquad
    \gamma = \sqrt{\frac{2 e^{2} \alpha^{2}}{M^{2} \omega^{2} \nu^{2}} - \frac{1}{2}}.
\end{equation}
The equilibrium positions of the ions are determined by the stationary point of this external potential and are calculated by solving the coupled system of equations,
\begin{equation}
    \boldsymbol{\nabla}_{i} V_{\mathrm{ex}} |_{\bm{0}} = \bm{0},
\end{equation}
where $|_{\bm{0}}$ denotes evaluation of the center of mass position $\bm{R}_{i} = \bar{\bm{R}}_{i} + \bm{Q}_{i}$ at its equilibrium position $\bar{\bm{R}}_{i}$, with $\bm{Q}_{i}$ the associated displacement and $\boldsymbol{\nabla}_{i}$ the derivative with respect to the center of mass position.
Taking the derivative and evaluating at the equilibrium position $\bar{\bm{R}}_{i} = (0, 0, \bar{R}_{i; z})$ gives
\begin{equation}
    \frac{\partial V_{\mathrm{ex}}}{\partial R_{i; x}} \bigg|_{0} \! = \frac{\partial V_{\mathrm{ex}}}{\partial R_{i; y}} \bigg|_{0} \! = 0 \qquad
    \text{and} \qquad
    \frac{\partial V_{\mathrm{ex}}}{\partial R_{i; z}} \bigg|_{0} \! = M \omega^{2} \bar{R}_{i; z} - C e^{2} \sum_{\mathclap{\substack{j = 1 \\ j \neq i}}}^{N} \frac{\bar{R}_{ij; z}}{\ab{\bar{R}_{ij; z}}^{3}}
\end{equation}
with $\bar{R}_{ij; z} = \bar{R}_{i; z} - \bar{R}_{j; z}$.
Introducing the dimensionless equilibrium position~$Z_{i}$ of ion $i$ and the length~$L$, defined by the formulae
\begin{equation}\label{eq:app:dimensionless-equilibrium-position}
    \bar{R}_{i; z} = L Z_{i} \qquad
    \text{and} \qquad
    L^{3} = \frac{C e^{2}}{M \omega^{2}},
\end{equation}
we can recast the condition for the dimensionless center of mass equilibrium positions compactly as 
\begin{equation}\label{eq:equilibrium-position-condition}
    Z_{i} = \sum_{\mathclap{\substack{j = 1 \\ j \neq i}}}^{N} \frac{Z_{ij}}{\ab{Z_{ij}}^{3}},
\end{equation}
where $Z_{ij} = Z_{i} - Z_{j}$ with the implicit assumption that $Z_{i} < Z_{i + 1}$.

\begin{figure}[t!]
    \centering
    \includegraphics[width=0.9\textwidth]{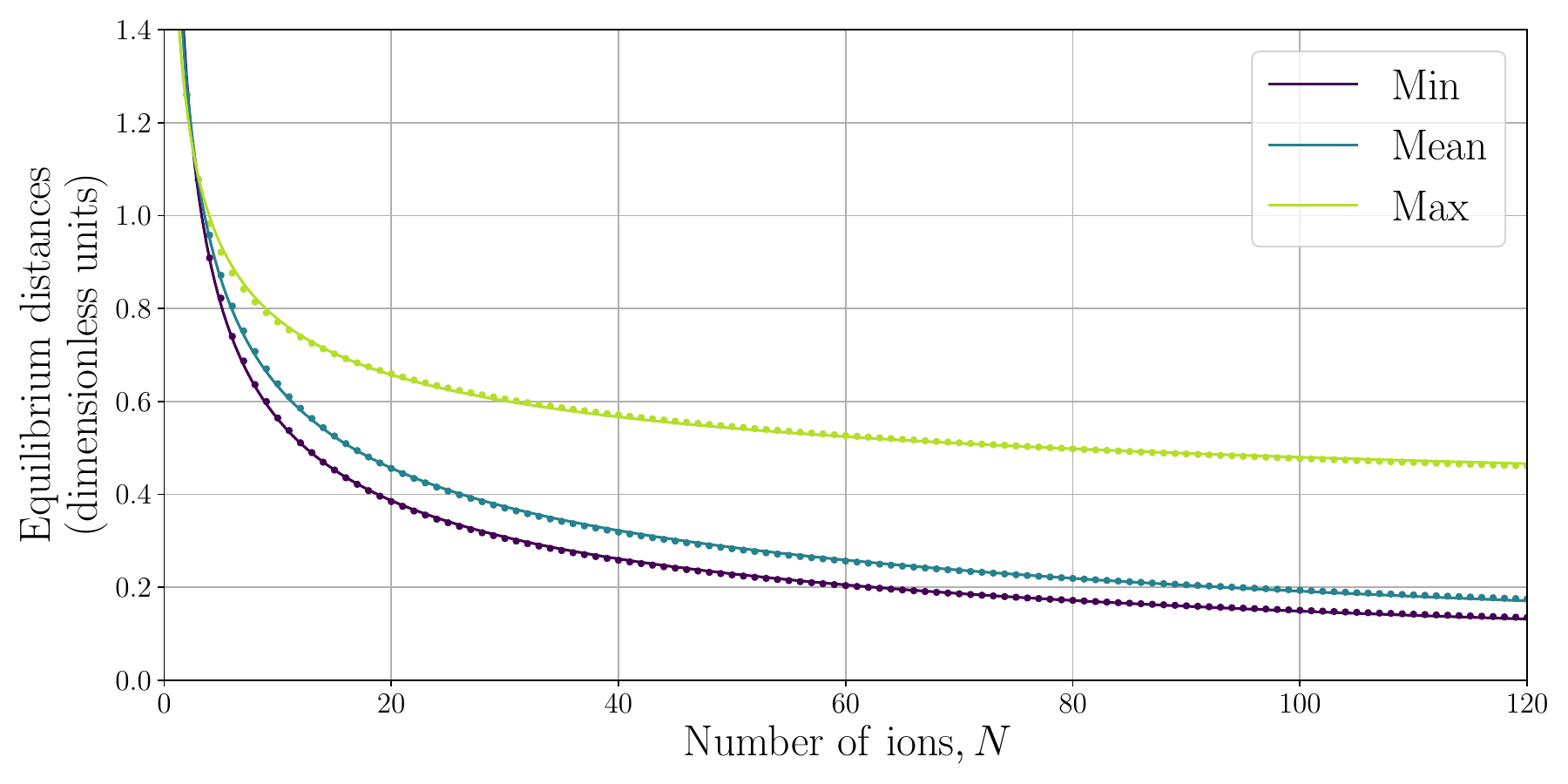}
    \caption{
        Scaling of the minimum, mean, and maximum dimensionless equilibrium distances between the trapped Rydberg ions with the number of ions $N$.
        The points are calculated using Eq.~\eqref{eq:equilibrium-position-condition} while the curves are computed with Eq.~\eqref{eq:equilibrium-distances-equations}.
    }
    \label{fig:equilibrium-distances}
\end{figure}

For $N \leq 3$, we can solve these equations analytically and find~\cite{james1998}:
\begin{equation}
\begin{aligned}
    N & = 1: \qquad
    & Z_{1} & = 0, \\
    N & = 2: \qquad
    & Z_{1} & = - \sqrt[3]{\frac{1}{4}}, \qquad
    & Z_{2} & = \sqrt[3]{\frac{1}{4}}, \\
    N & = 3: \qquad
    & Z_{1} & = - \sqrt[3]{\frac{5}{4}}, \qquad
    & Z_{2} & = 0, \qquad
    & Z_{3} & = \sqrt[3]{\frac{5}{4}}.
\end{aligned}
\end{equation}
Nevertheless, for $N > 3$ they must be solved numerically.
In Fig.~\ref{fig:equilibrium-distances} we plot the minimum, mean, and maximum dimensionless equilibrium distances for systems with up to $N = 120$ trapped Rydberg ions.
These are, for example, of relevance for scalable quantum computation and quantum many-body system simulation~\cite{bruzewicz2019}.
Compiling the numerical solutions to the dimensionless equilibrium position condition in Eq.~\eqref{eq:equilibrium-position-condition}, we obtain the following scaling relations for the minimum, mean, and maximum dimensionless equilibrium distances,
\begin{equation}\label{eq:equilibrium-distances-equations}
    Z_{\mathrm{min}} \sim \frac{1.880}{N^{0.486}} - 0.052, \qquad
    Z_{\mathrm{mean}} \sim \frac{1.823}{N^{0.388}} - 0.115, \qquad
    \text{and} \qquad
    Z_{\mathrm{max}} \sim \frac{1.246}{N^{0.404}} + 0.286.
\end{equation}

\section{Harmonic expansion of the center of mass positions}\label{app:harmonic-expansion}

In these appendices, we outline the second order harmonic expansion of the Hamiltonians describing the external, internal, and coupled dynamics in Eqs.~\eqref{eq:hamiltonian-external},~\eqref{eq:hamiltonian-internal}, and~\eqref{eq:hamiltonian-coupled}, respectively.
Following Ref.~\cite{muller2008} and the approximations made for the multipole expansion about the center of mass positions $\bm{R}_{i}$ in App.~\ref{app:multipole-expansion}, we truncate the harmonic expansion about the center of mass equilibrium positions $\bar{\bm{R}}_{i}$ at second order.
This is well justified by the typical length scales of the system, namely, of the distances between the ions $\ab{\bar{\bm{R}}_{ij}} \sim 5 \, \unit{\micro\meter}$, the extension of the Rydberg wavefunction $\ab{\bm{r}_{i}} \sim 100 \, \unit{\nano\meter}$, and the oscillations of the ions $\ab{\bm{Q}_{i}} \sim 10 \, \unit{\nano\meter}$~\cite{mokhberi2020}.

\subsection{External vibrational dynamics}

To start, we recall the Hamiltonian governing the external vibrational dynamics in Eq.~\eqref{eq:hamiltonian-external} which reads
\begin{equation}\label{app:eq:external-hamiltonian}
    H_{\mathrm{ex}} = \frac{1}{2 M} \sum_{\mathclap{i = 1}}^{N} \bm{P}_{i}^{2} + V_{\mathrm{ex}} \qquad
    \text{with} \qquad
    V_{\mathrm{ex}} = \frac{M \omega^{2}}{2} \sum_{\mathclap{i = 1}}^{N} \sum_{\mathclap{u}} \gamma_{u}^{2} R_{i; u}^{2} + \frac{C e^{2}}{2} \sum_{\mathclap{\substack{i, j = 1 \\ j \neq i}}}^{N} \frac{1}{\ab{\bm{R}_{ij}}}.
\end{equation}
The former kinetic term is invariant under the harmonic expansion and evaluation at the center of mass equilibrium positions, since it is independent of the center of mass positions.
Hence, we only need consider the latter potential term.
Noting that we can calculate the terms of this harmonic expansion as we did for the interaction potential in App.~\ref{app:multipole-expansion}, it follows that we can write
\begin{equation}\label{eq:app:eq:potential-recursion}
    V_{\mathrm{ex}} = \sum_{\mathclap{n = 0}}^{\infty} V_{\mathrm{ex}}^{(n)} |_{\bm{0}} \qquad
    \text{where} \qquad
    V_{\mathrm{ex}}^{(n)} = \frac{1}{n} \sum_{\mathclap{i = 1}}^{N} \bm{Q}_{i} \cdot \boldsymbol{\nabla}_{i} V_{\mathrm{ex}}^{(n - 1)},
\end{equation}
with $|_{\bm{0}}$ denoting evaluation of the positions $\bm{R}_{i} = \bar{\bm{R}}_{i} + \bm{Q}_{i}$ at the equilibrium positions $\bar{\bm{R}}_{i}$ where $\bm{Q}_{i}$ is the associated displacement and $\boldsymbol{\nabla}_{i}$ the derivative with respect to the position $\bm{R}_{i}$.
The initial condition is given by the zeroth order term,
\begin{equation}
    V_{\mathrm{ex}}^{(0)} |_{\bm{0}} = \bigg[ \frac{M \omega^{2}}{2} \sum_{\mathclap{i = 1}}^{N} \sum_{\mathclap{u}} \gamma_{u}^{2} R_{i; u}^{2} + \frac{C e^{2}}{2} \sum_{\mathclap{\substack{i, j = 1 \\ j \neq i}}}^{N} \frac{1}{\ab{\bm{R}_{ij}}} \bigg] \! \bigg|_{\bm{0}} \! = \frac{M \omega^{2}}{2} \sum_{\mathclap{i = 1}}^{N} \bar{R}_{i; z}^{2} + \frac{C e^{2}}{2} \sum_{\mathclap{\substack{i, j = 1 \\ j \neq i}}}^{N} \frac{1}{\ab{\bar{R}_{ij; z}}},
\end{equation}
where in order to obtain the latter expression, we have evaluated the positions $\bm{R}_{i}$ at their equilibrium $\bar{\bm{R}}_{i} = (0, 0, \bar{R}_{i; z})$.
Remarking that this term is constant, since it is independent of both the center of mass displacements $\bm{Q}_{i}$ and relative positions $\bm{r}_{i}$, it can be neglected.
The first order term then follows as
\begin{equation}
    V_{\mathrm{ex}}^{(1)} |_{\bm{0}} = \bigg[ M \omega^{2} \sum_{\mathclap{i = 1}}^{N} \sum_{\mathclap{u}} \gamma_{u}^{2} R_{i; u} Q_{i; u} - \frac{C e^{2}}{2} \sum_{\mathclap{\substack{i, j = 1 \\ j \neq i}}}^{N} \frac{\bm{R}_{ij} \cdot \bm{Q}_{ij}}{\ab{\bm{R}_{ij}}^{3}} \bigg] \! \bigg|_{\bm{0}} \! = \bm{0},
\end{equation}
which, we note, after evaluating at the equilibrium positions vanishes.
This follows from the equilibrium positions condition in Eq.~\eqref{eq:equilibrium-position-condition}.
For the second order term we then find
\begin{equation}
    V_{\mathrm{ex}}^{(2)} |_{\bm{0}} = \bigg[ \frac{M \omega^{2}}{2} \sum_{\mathclap{i = 1}}^{N} \sum_{u} \gamma_{u}^{2} Q_{i; u}^{2} + \frac{C e^{2}}{2} \sum_{\mathclap{\substack{i, j = 1 \\ j \neq i}}}^{N} \frac{3 [\bm{R}_{ij} \cdot \bm{Q}_{ij}]^{2} - \bm{R}_{ij}^{2} \bm{Q}_{ij}^{2}}{\ab{\bm{R}_{ij}}^{5}} \bigg] \! \bigg|_{\bm{0}} \! = \frac{M \omega^{2}}{2} \sum_{\mathclap{i, j = 1}}^{N} \sum_{u} K_{ij; u} Q_{i; u} Q_{j; u}.
\end{equation}
where to obtain the latter identity, we expressed the equilibrium positions in terms of the dimensionless equilibrium positions $Z_{i}$ and characteristic length $L$ and introduced the Hessian matrix components,
\begin{equation}
    K_{ij; x} = K_{ij; y} = \delta_{ij} \gamma^{2} - K_{ij} \qquad
    \text{and} \qquad
    K_{ij; z} = 2 K_{ij} + \delta_{ij},
\end{equation}
with $K_{ij}$ the generalized Hessian matrix components defined by
\begin{equation}
    K_{ij} = \delta_{ij} \sum_{\mathclap{k = 1}}^{N} \frac{1}{\ab{Z_{ik}}^{3}} - \frac{1}{\ab{Z_{ij}}^{3}}.
\end{equation}
Given that we neglected the constant zeroth order term and identically zero first order term, it follows that the potential term $V_{\mathrm{ex}} \approx V_{\mathrm{ex}}^{\smash{(2)}} |_{\bm{0}}$.
Taking together with the kinetic term in Eq.~\eqref{app:eq:external-hamiltonian} then yields the approximate expression in Eq.~\eqref{eq:hamiltonian-external-expanded} for the external vibrational dynamics (i.e., up to order $1 / \ab{Z_{ij}}^{3}$).

\subsection{Coupled vibronic dynamics}

Let us now consider the Hamiltonian in Eq.~\eqref{eq:hamiltonian-coupled} accounting for the coupled vibronic dynamics,
\begin{equation}
    H_{\mathrm{co}} = - 2 e \alpha \cos(\nu t) \sum_{\mathclap{i = 1}}^{N} [R_{i; x} r_{i; x} - R_{i; y} r_{i; y}] - 2 e \beta \sum_{\mathclap{i = 1}}^{N} [3 R_{i; z} r_{i; z} - \bm{R}_{i} \cdot \bm{r}_{i}] + V_{\mathrm{co}},
\end{equation}
where $V_{\mathrm{co}}$ denotes the contributions to the coupled dynamics arising from the multipole expansion of the interaction potential (cf. App.~\ref{app:multipole-expansion}) given by
\begin{equation}
    V_{\mathrm{co}} = \frac{C e^{2}}{2} \sum_{\mathclap{\substack{i, j = 1 \\ j \neq i}}}^{N} \bigg[ \frac{\bm{n}_{ij} \cdot \bm{r}_{i}}{\ab{\bm{R}_{ij}}^{2}} - \frac{\bm{n}_{ij} \cdot \bm{r}_{j}}{\ab{\bm{R}_{ij}}^{2}} \bigg].
\end{equation}
As before, we utilize the recursive notation used for the expansion of the external potential in Eq.~\eqref{eq:app:eq:potential-recursion}.
The initial condition given by the zeroth order term is then\footnote{Note that here and throughout this appendix, we have relabeled the index of the summation over the second term and used the fact that $\bm{R}_{ji} = - \bm{R}_{ij}$ to combine the terms.}
\begin{equation}
    V_{\mathrm{co}}^{(0)} |_{\bm{0}} = C e^{2} \sum_{\mathclap{\substack{i, j = 1 \\ j \neq i}}}^{N} \frac{\bm{R}_{ij} \cdot \bm{r}_{i}}{\ab{\bm{R}_{ij}}^{3}} \bigg|_{\bm{0}} = M \omega^{2} \sum_{\mathclap{i = 1}}^{N} \bar{R}_{i; z} r_{i; z},
\end{equation}
where to simplify this expression we have used the definition of the characteristic length $L$, dimensionless equilibrium positions $Z_{i}$, and in particular the dimensionless equilibrium position condition in Eq.~\eqref{eq:equilibrium-position-condition}.
For the first order term we get
\begin{equation}
\begin{aligned}
    V_{\mathrm{co}}^{(1)} |_{\bm{0}} & = - C e^{2} \sum_{\mathclap{\substack{i, j = 1 \\ j \neq i}}}^{N} \frac{3 [\bm{R}_{ij} \cdot \bm{Q}_{ij}] [\bm{R}_{ij} \cdot \bm{r}_{i}] - \bm{R}_{ij}^{2} [\bm{Q}_{ij} \cdot \bm{r}_{i}]}{\ab{\bm{R}_{ij}}^{5}} \bigg|_{\bm{0}} \\
    & = - C e^{2} \sum_{\mathclap{\substack{i, j = 1 \\ j \neq i}}}^{N} \bigg[ \frac{3 Q_{i; z} r_{i; z} - \bm{Q}_{i} \cdot \bm{r}_{i}}{\ab{\bar{R}_{ij; z}}^{3}} - \frac{3 Q_{i; z} r_{j; z} - \bm{Q}_{i} \cdot \bm{r}_{j}}{\ab{\bar{R}_{ij; z}}^{3}} \bigg].
\end{aligned}
\end{equation}
Taking this together with the remaining terms of the coupled Hamiltonian, we find that the Hamiltonian of the coupled vibronic dynamics is approximated up to order $1 / \ab{Z_{ij}}^{3}$ by
\begin{equation}
    H_{\mathrm{co}} = - 2 e \alpha \cos(\nu t) \sum_{\mathclap{i = 1}}^{N} [Q_{i; x} r_{i; x} - Q_{i; y} r_{i; y}] - 2 e \beta \sum_{\mathclap{i = 1}}^{N} [3 Q_{i; z} r_{i; z} - \bm{Q}_{i} \cdot \bm{r}_{i}] - 4 e \beta \sum_{\mathclap{i = 1}}^{N} \bar{R}_{i; z} r_{i; z} + V_{\mathrm{co}},
\end{equation}
where the associated coupled potential is given by
\begin{equation}
    V_{\mathrm{co}} = 4 e \beta \sum_{\mathclap{i = 1}}^{N} \bar{R}_{i; z} r_{i; z} - 4 e \beta \sum_{\mathclap{\substack{i, j = 1 \\ j \neq i}}}^{N} \bigg[ \frac{3 Q_{i; z} r_{i; z} - \bm{Q}_{i} \cdot \bm{r}_{i}}{\ab{Z_{ij}}^{3}} - \frac{3 Q_{i; z} r_{j; z} - \bm{Q}_{i} \cdot \bm{r}_{j}}{\ab{Z_{ij}}^{3}} \bigg].
\end{equation}
Noticing the cancellation and identification of similar terms then yields the expression in Eq.~\eqref{eq:hamiltonian-coupled-expanded}, where again we have made use of the Hessian matrix components $K_{ij; z}$.

\subsection{Internal electronic dynamics}

Now, we consider the Hamiltonian describing the internal electronic dynamics in Eq.~\eqref{eq:hamiltonian-internal}, which is
\begin{equation}
    H_{\mathrm{in}} = \sum_{\mathclap{i = 1}}^{N} \bigg[ \frac{\bm{p}_{i}^{2}}{2 m} + V(\ab{\bm{r}_{i}}) \bigg] - e \alpha \cos(\nu t) \sum_{\mathclap{i = 1}}^{N} [r_{i; x}^{2} - r_{i; y}^{2}] - e \beta \sum_{\mathclap{i = 1}}^{N} [3 r_{i; z}^{2} - \bm{r}_{i}^{2}] - e \sum_{\mathclap{i = 1}}^{N} \bm{r}_{i} \cdot \bm{E}(t) + V_{\mathrm{in}}.
\end{equation}
Here, we have identified by $V_{\mathrm{in}}$ the terms of the internal dynamics coming from the multipole expansion of the interaction potential (cf. App.~\ref{app:multipole-expansion}) that read
\begin{equation}
    V_{\mathrm{in}} = - \frac{C e^{2}}{2} \sum_{\mathclap{\substack{i, j = 1 \\ j \neq i}}}^{N} \bigg[ \frac{3 [\bm{n}_{ij} \cdot \bm{r}_{i}]^{2} - \bm{r}_{i}^{2}}{2 \ab{\bm{R}_{ij}}^{3}} + \frac{3 [\bm{n}_{ij} \cdot \bm{r}_{i}][\bm{n}_{ij} \cdot \bm{r}_{j}] - \bm{r}_{i} \cdot \bm{r}_{j}}{\ab{\bm{R}_{ij}}^{3}} + \frac{3 [\bm{n}_{ij} \cdot \bm{r}_{j}]^{2} - \bm{r}_{j}^{2}}{2 \ab{\bm{R}_{ij}}^{3}} \bigg].
\end{equation}
Using the same notation as for the external and coupled dynamics [cf. Eq.~\eqref{eq:app:eq:potential-recursion}], the zeroth order term,
\begin{equation}
\begin{aligned}
    V_{\mathrm{in}}^{(0)} |_{\bm{0}} & = - \frac{C e^{2}}{2} \sum_{\mathclap{\substack{i, j = 1 \\ j \neq i}}}^{N} \bigg[ \frac{3 [\bm{R}_{ij} \cdot \bm{r}_{i}]^{2} - \bm{R}_{ij}^{2} \bm{r}_{i}^{2}}{\ab{\bm{R}_{ij}}^{5}} + \frac{3 [\bm{R}_{ij} \cdot \bm{r}_{i}][\bm{R}_{ij} \cdot \bm{r}_{j}] - \bm{R}_{ij}^{2} [\bm{r}_{i} \cdot \bm{r}_{j}]}{\ab{\bm{R}_{ij}}^{5}} \bigg] \! \bigg|_{\bm{0}} \\
    & = - \frac{M \omega^{2}}{2} \sum_{\mathclap{\substack{i, j = 1 \\ j \neq i}}}^{N} \bigg[ \frac{3 r_{i; z}^{2} - \bm{r}_{i}^{2}}{\ab{\bar{R}_{ij; z}}^{3}} + \frac{3 r_{i; z} r_{j; z} - \bm{r}_{i} \cdot \bm{r}_{j}}{\ab{\bar{R}_{ij; z}}^{3}} \bigg].
\end{aligned}
\end{equation}
Given that this is of order $1 / \ab{Z_{ij}}^{3}$, it follows that after identifying similar terms associated to the static electric-field mode, the Hamiltonian approximating the internal electronic dynamics is given by the expression in Eq.~\eqref{eq:hamiltonian-internal-expanded}, where we have similarly used the Hessian matrix components $K_{ij; z}$.

\section{Diagonalization of the external dynamics via phonon modes}\label{app:phonon-modes}

In this appendix, we diagonalize the Hamiltonian governing the external vibrational dynamics, which is
\begin{equation}
    H_{\mathrm{ex}} = \frac{1}{2 M} \sum_{\mathclap{i = 1}}^{N} \bm{P}_{i}^{2} + \frac{M \omega^{2}}{2} \sum_{\mathclap{i, j = 1}}^{N} \sum_{u} K_{ij; u} Q_{i; u} Q_{j; u},
\end{equation}
where $Q_{i; u}$ and $P_{i; u}$ are the center of mass displacements and momenta.
Recall that the Hessian matrix components $K_{ij; u}$ are defined in terms of the generalized Hessian matrix components $K_{ij}$ by
\begin{equation}
    K_{ij} = \delta_{ij} \gamma^{2} - K_{ij; x} = \delta_{ij} \gamma^{2} - K_{ij; y} = \frac{K_{ij; z} - \delta_{ij}}{2} \qquad
    \text{with} \qquad
    K_{ij} = \delta_{ij} \sum_{\mathclap{k = 1}}^{N} \frac{1}{\ab{Z_{ik}}^{3}} - \frac{1}{\ab{Z_{ij}}^{3}}.
\end{equation}
The generalized Hessian matrix is diagonalized by the generalized orthogonal matrix, the components of which are $\Gamma_{ip} = \Gamma_{ip; x} = \Gamma_{ip; y} = \Gamma_{ip; z}$ and satisfy
\begin{equation}
    \sum_{\mathclap{i = 1}}^{N} \Gamma_{ip} \Gamma_{iq} = \delta_{pq} \qquad
    \text{and} \qquad
    \sum_{\mathclap{j = 1}}^{N} K_{ij} \Gamma_{jp} = \gamma_{p}^{2} \Gamma_{ip}.
\end{equation}
The generalized dimensionless phonon mode frequencies $\gamma_{\smash{p}}$ are related to the dimensionless phonon mode frequencies $\gamma_{p; u}$ via the relations
\begin{equation}
    \gamma_{p}^{2} = \gamma^{2} - \gamma_{p; x}^{2} = \gamma^{2} - \gamma_{p; y}^{2} = \frac{\gamma_{p; z}^{2} - 1}{2}.
\end{equation}

For $N \leq 3$, the generalized dimensionless phonon mode frequencies $\gamma_{p}$ and the associated generalized orthogonal matrix components $\Gamma_{ip}$ of the generalized Hessian matrix components $K_{ij}$ can be determined algebraically and are~\cite{james1998}:
\begin{equation}
\begin{aligned}
    N & = 1: \qquad
    & \gamma_{1}^{2} & = 0 \qquad
    & \boldsymbol{\Gamma}_{1} & = \begin{bmatrix}
        1 \\
    \end{bmatrix} \!, \\
    N & = 2: \qquad
    & \gamma_{1}^{2} & = 0, \qquad
    & \boldsymbol{\Gamma}_{1} & = \frac{1}{\sqrt{2}}
    \! \begin{bmatrix}
        1, & 1 \\
    \end{bmatrix} \!, \\
    & & \gamma_{2}^{2} & = 1, \qquad 
    & \boldsymbol{\Gamma}_{2} & = \frac{1}{\sqrt{2}}
    \! \begin{bmatrix}
        1, & -1 \\
    \end{bmatrix} \!, \\
    N & = 3: \qquad
    & \gamma_{1}^{2} & = 0, \qquad
    & \boldsymbol{\Gamma}_{1} & = \frac{1}{\sqrt{3}}
    \! \begin{bmatrix}
        1, & 1, & 1 \\
    \end{bmatrix} \!, \\
    & & \gamma_{2}^{2} & = 1, \qquad
    & \boldsymbol{\Gamma}_{2} & = \frac{1}{\sqrt{2}}
    \! \begin{bmatrix}
        1, & 0, & -1 \\
    \end{bmatrix} \!, \\
    & & \gamma_{3}^{2} & = \frac{12}{5}, \qquad
    & \boldsymbol{\Gamma}_{3} & = \frac{1}{\sqrt{6}}
    \! \begin{bmatrix}
        1, & -2, & 1 \\
    \end{bmatrix} \!.
\end{aligned}
\end{equation}
Here, we have employed the notation $\boldsymbol{\Gamma}_{p} = [\Gamma_{1p}, \quad \ldots, \quad \Gamma_{Np}]$ for the orthogonal vectors.
For $N > 3$ we must, however, obtain these numerically and from this make the following observations.
First, we note that the (square of the) generalized dimensionless phonon mode frequencies are strictly monotonically increasing with the mode index $p$.
As such, for any finite number of ions $N$, we have
\begin{equation}
    \gamma_{p; x}^{2} > \gamma_{p + 1; x}^{2}, \qquad
    \gamma_{p; y}^{2} > \gamma_{p + 1; y}^{2}, \qquad
    \text{and} \qquad
    \gamma_{p; z}^{2} < \gamma_{p + 1; z}^{2}.
\end{equation}
Moreover, we find that the generalized dimensionless phonon mode frequencies associated to the first and second modes, namely, the center of mass and breathing modes for which $p = 1$ and $2$, respectively, are the only modes that are independent of the number of ions $N$.
Together, this immediately implies that for the axial modes that $\gamma_{p; z}^{2} > 0$ since $\gamma_{1: z}^{2} = 1$.
However, in contrast, given that $\gamma_{1; x}^{2} = \gamma_{1; y}^{2} = \gamma^{2}$, this further suggests that, for a given number of ions~$N$ and anisotropy $\gamma$, there is a critical radial mode index $p^{*}$ for which $\gamma_{p^{*}; x}^{2} = \gamma_{p^{*}; y}^{2} < 0$ (i.e., the frequencies of the radial modes are imaginary).
Physically, this implies that there is a critical anisotropy $\gamma_{*} = \gamma_{N}$ at which a structural phase transition occurs whereby the ions reconfigure from a one- to a two-dimensional Coulomb crystal~\cite{fishman2008, li2012}.
Note that the deviation from the values given in Ref.~\cite{enzer2000} is due to the omission of the micromotion and coupling of the electronic and vibrational degrees of freedom.

Expressed in terms of the phonon mode creation and annihilation operators $a_{p; u}^{\dagger}$ and~$a_{p; u}$, the position (i.e., displacement) and momentum coordinates $Q_{i; u}$ and $P_{i; u}$ are
\begin{equation}
    Q_{i; u} = l \sum_{p = 1}^{N} \frac{1}{\sqrt{\gamma_{\smash{p; u}}}} \Gamma_{i p; u} [a_{p; u}^{\dagger} + a_{p; u}] \qquad
    \text{and} \qquad
    P_{i; u} = \mathrm{i} M \omega l \sum_{p = 1}^{N} \sqrt{\gamma_{\smash{p; u}}} \Gamma_{i p; u} [a_{p; u}^{\dagger} - a_{p; u}],
\end{equation}
where the characteristic length $l$ associated to the ions oscillations about their equilibria is given by
\begin{equation}
    l = \sqrt{\frac{\hbar}{2 M \omega}}.
\end{equation}
Substituting these into the expression for the external vibrational Hamiltonian and utilizing the defining identities for the orthogonal matrix components $\Gamma_{i p; u}$, we then get
\begin{equation}
    H_{\mathrm{ex}} = \frac{\hbar \omega}{2} \sum_{\mathclap{p = 1}}^{N} \sum_{u} \gamma_{p; u} [a_{p; u}^{\dagger} a_{p; u} + a_{p; u} a_{p; u}^{\dagger}].
\end{equation}
Exploiting the commutation relations of the bosonic creation and annihilation operators,
\begin{equation}
    [a_{p; u}, a_{q; v}^{\dagger}] = \delta_{p, q} \delta_{u, v} \qquad
    \text{and} \qquad
    [a_{p; u}, a_{q; v}] = [a_{p; u}^{\dagger}, a_{q; v}^{\dagger}] = 0,
\end{equation}
and neglecting the resultant constant term, which simply amounts to shifting the zero point energy of the oscillators, yields the expression for the Hamiltonian of the external vibrational dynamics in Eq.~\eqref{eq:hamiltonian-external-phonon}.

\section{Determination of the electric multipole matrix elements}\label{app:matrix-elements}

In this appendix, we explicitly detail the analytic calculation of the angular matrix elements of spherical harmonics using angular momentum algebra.
Additionally, we briefly discuss the numeric computation of the radial matrix elements.
Since we treat the trapped Rydberg ions as distinguishable particles within the effective two-body approximation, determining electric multipole matrix elements simply amounts to calculating single electron position-spin wavefunctions~\cite{drake2023}.
As noted in App.~\ref{app:electronic-states}, the bound electronic quantum states can be represented in terms of the basis $\ket{n, l, s, j, m_{j}}$ with $n$, $l$, $s$, $j$, and $m_{j}$ the principal, orbital angular momentum, spin angular momentum, total angular momentum, and total magnetic quantum numbers.
This facilitates the decomposition of the electric multipole matrix elements into the product of a radial and angular matrix element\footnote{To avoid ambiguity, we use the standard quantum mechanical hat notation to denote operators in this appendix.},
\begin{equation}
    \me{n'\!, l'\!, j'\!, m_{j'}}{\h{r}^{k} Y_{k}^{\smash{m_{k}}}(\h{\theta}, \h{\varphi})}{n, l, j, m_{j}} \equiv \me{n'\!, l'\!, j'}{\h{r}^{k}}{n, l, j} \! \me{l'\!, j'\!, m_{j'}}{\h{Y}_{k}^{m_{k}}}{l, j, m_{j}} \!
\end{equation}
for integer $k$ and $m_{k}$ where $\h{r}$ is the radial position operator and $\h{Y}_{k}^{m_{k}} \equiv Y_{k}^{m_{k}}(\h{\theta}, \h{\varphi})$ the spherical harmonic operator of degree $k$ and order $m_{k}$.
Akin to App.~\ref{app:electronic-states}, we have omitted the spin angular momentum quantum number $s = 1/2$ where unnecessary.

\subsection{Numeric computation of radial matrix elements}

The radial matrix elements are computed numerically using the radial wavefunctions $\Phi_{n l j}(r)$ which are themselves computed numerically by solving the radial Schr{\" o}dinger equation, see App.~\ref{app:electronic-states}.
In particular, by inserting resolutions of the identity over the radial coordinate basis (i.e., $\mathds{1} = \int \! \d r \, r^{2} \! \op{r}{r}$), the radial matrix element can be rewritten as
\begin{equation}
    \me{n'\!, l'\!, j'}{\h{r}^{k}}{n, l, j}
    = \int \!\!\!\! \int \d r \d r' \, r^{2} r'^{2} \! \ip{n'\!, l'\!, j'}{r'} \! \ip{r}{n, l, j} \! \me{r'}{\h{r}^{k}}{r}
    = \int \d r \, r^{k} [\Phi_{n' l' j'}]^{*} \Phi_{n l j}.
\end{equation}
To obtain the latter equality, we used the radial position operator eigenvalue equation $\h{r} \! \ket{r} = r \! \ket{r}$, recalled the normalization of the radial coordinate basis states $\ip{r}{r'} = \delta(r - r') / r^{2}$, which allowed us to integrate over the radial position $r'$, and identified the radial wavefunctions $\Phi_{n l j} \equiv \Phi_{n l j}(r) = r \! \ip{r}{n, l, j}$.

\subsection{Analytic calculation of angular matrix elements}

In contrast to the radial matrix elements, the angular matrix elements are calculated analytically using angular momentum algebra~\cite{louck2023}.
To start, we insert resolutions of the identity over the orbital and spin angular momentum bases (i.e., $\mathds{1} = \sum_{m_{l}, m_{s}} \! \op{l, m_{l}, s, m_{s}}{l, m_{l}, s, m_{s}}$) such that we can write
\begin{equation}
    \me{l'\!, j'\!, m_{j'}}{\h{Y}_{k}^{m_{k}}}{l, j, m_{j}} = \sum_{\mathclap{m_{l}, m_{s}, m_{l'}, m_{s'}}} \ip{l'\!, m_{l'}, s'\!, m_{s'}}{j'\!, m_{j'}} \! \ip{l, m_{l}, s, m_{s}}{j, m_{j}} \! \me{l'\!, m_{l'}, s'\!, m_{s'}}{\h{Y}_{k}^{m_{k}}}{l, m_{l}, s, m_{s}} \!.
\end{equation}
Here, we have used the Condon–Shortley phase convention, which guarantees that the Clebsch-Gordon coefficients are strictly real (i.e., $\ip{j, m_{j}}{l, m_{l}, s, m_{s}} = \ip{l, m_{l}, s, m_{s}}{j, m_{j}}$)~\cite{louck2023}.
Exploiting the separation of the orbital and spin angular momentum basis states (i.e., $\ket{l, m_{l}, s, m_{s}} = \ket{l, m_{l}} \! \ket{s, m_{s}}$), the angular matrix elements follow as
\begin{equation}
    \me{l'\!, j'\!, m_{j'}}{\h{Y}_{k}^{m_{k}}}{l, j, m_{j}} = \sum_{\mathclap{m_{l}, m_{s}, m_{l'}}} \ip{l'\!, m_{l'}, s, m_{s}}{j'\!, m_{j'}} \! \ip{l, m_{l}, s, m_{s}}{j, m_{j}} \! \me{l'\!, m_{l'}}{\h{Y}_{k}^{m_{k}}}{l, m_{l}} \!.
\end{equation}
To obtain this, we noted that the inner product $\ip{s, m_{s}}{s'\!, m_{s'}} = \delta_{ss'} \delta_{m_{s} m_{s'}}$, hence, the summation over the spin magnetic quantum number $m_{s'}$ can be performed.
Furthermore, we omitted the Kronecker delta $\delta_{s s'}$, since here $s = s' = 1/2$.
We now insert resolutions of the identity over the angular coordinate bases (i.e., $\mathds{1} = \int \!\!\! \int \! \d \theta \d \varphi \sin\theta \! \op{\theta, \varphi}{\theta, \varphi}$) such that the expression for the angular matrix element becomes
\begin{equation}
\begin{aligned}
    \me{l'\!, j'\!, m_{j'}}{\h{Y}_{k}^{m_{k}}}{l, j, m_{j}} & = \sum_{\mathclap{m_{l}, m_{s}, m_{l'}}} \ip{l'\!, m_{l'}, s, m_{s}}{j'\!, m_{j'}} \! \ip{l, m_{l}, s, m_{s}}{j, m_{j}} \\
    & \qquad \times \int \!\!\!\! \int \!\!\!\! \int \!\!\!\! \int \! \d \theta \d \varphi \d \theta' \d \varphi' \sin\theta \sin\theta' \! \ip{l'\!, m_{l'}}{\theta'\!, \varphi'} \! \ip{\theta, \varphi}{l, m_{l}} \! \me{\theta'\!, \varphi'}{\h{Y}_{k}^{m_{k}}}{\theta, \varphi} \!.
\end{aligned}
\end{equation}
Since by definition the angular coordinate basis states are eigenstates of the spherical harmonic operator $\h{Y}_{k}^{m_{k}} \! \ket{\theta, \varphi} = Y_{k}^{m_{k}} \! \ket{\theta, \varphi}$ with $Y_{k}^{m_{k}}$ the spherical harmonic function, the angular matrix element reduces~to
\begin{equation}
    \me{l'\!, j'\!, m_{j'}}{\h{Y}_{k}^{m_{k}}}{l, j, m_{j}} = \sum_{\mathclap{m_{l}, m_{s}, m_{l'}}} \ip{l'\!, m_{l'}, s, m_{s}}{j'\!, m_{j'}} \! \ip{l, m_{l}, s, m_{s}}{j, m_{j}} \! \int \!\!\!\! \int \! \d \theta \d \varphi \sin\theta [Y_{l'}^{m_{l'}}]^{*} Y_{k}^{m_{k}} Y_{l}^{m_{l}}.
\end{equation}
Note that in order to obtain this equality, we recalled the normalization of the angular coordinate basis states $\ip{\theta, \varphi}{\theta'\!, \varphi'} = \delta(\theta - \theta') \delta(\varphi - \varphi') / \sin\theta$, then integrated over the angular coordinates $\theta'$ and $\varphi'$, and identified the spherical harmonics $Y_{l}^{m_{l}}(\theta, \varphi) = \ip{\theta, \varphi}{l, m_{l}}$.
Using the following integral identity, known as a Gaunt coefficient,
\begin{equation}
    \int \!\!\!\! \int \! \d \theta \d \varphi \sin\theta [Y_{l'}^{m_{l'}}]^{*} Y_{k}^{m_{k}} Y_{l}^{m_{l}} = \sqrt{\frac{2 k + 1}{4 \pi}} \sqrt{\frac{2 l + 1}{2 l' + 1}} \! \ip{l, 0, k, 0}{l', 0} \! \ip{l, m_{l}, k, m_{k}}{l', m_{l'}} \!,
\end{equation}
the angular matrix element can be explicitly written only in terms of Clebsch-Gordon coefficients as
\begin{equation}
\begin{aligned}
    \me{l'\!, j'\!, m_{j'}}{\h{Y}_{k}^{m_{k}}}{l, j, m_{j}} & = \sqrt{\frac{2k + 1}{4 \pi}} \sqrt{\frac{2 l + 1}{2 l' + 1}} \! \ip{l, 0, k, 0}{l', 0} \sum_{\mathclap{m_{l}, m_{s}, m_{l'}}} \ip{l'\!, m_{l'}, s, m_{s}}{j'\!, m_{j'}} \! \ip{l, m_{l}, s, m_{s}}{j, m_{j}} \\
    & \qquad \times \! \ip{l, m_{l}, k, m_{k}}{l', m_{l'}} \!.
\end{aligned}
\end{equation}
We can further simplify this expression by exploiting the Clebsch-Gordon coefficients selection rules~\cite{louck2023}.
In particular, the angular matrix elements are only nonzero if the magnetic quantum numbers satisfy $m_{j'} = m_{j} + m_{k}$, $m_{l'} = m_{j} + m_{k} - m_{s}$, and $m_{l} = m_{j} - m_{s}$.
This allows us to perform the sums over the orbital magnetic quantum numbers $m_{l}$ and $m_{l'}$ and impose the constraint on the total magnetic quantum number $m_{j'} = m_{j} + m_{k}$.
Accordingly, the nonzero angular matrix elements read
\begin{equation}
\begin{aligned}
    & \me{l'\!, j'\!, m_{j} + m_{k}}{\h{Y}_{k}^{m_{k}}}{l, j, m_{j}} = \sqrt{\frac{2k + 1}{4 \pi}} \sqrt{\frac{2 l + 1}{2 l' + 1}} \! \ip{l, 0, k, 0}{l', 0} \sum_{\mathclap{m_{s}}} \ip{l'\!, m_{j} + m_{k} - m_{s}, s, m_{s}}{j'\!, m_{j} + m_{k}} \\
    & \qquad \times \ip{l, m_{j} - m_{s}, s, m_{s}}{j, m_{j}} \! \ip{l, m_{j} - m_{s}, k, m_{k}}{l', m_{j} + m_{k} - m_{s}} \!.
\end{aligned}
\end{equation}
Note also that the angular momentum quantum numbers must additionally obey the triangle inequalities,
\begin{equation}
    \ab{l - k} \leq l' \leq l + k, \qquad
    \ab{l' - s} \leq j' \leq l' + s, \qquad
    \ab{l - s} \leq j \leq l + s,
\end{equation}
where as usual the spin angular momentum quantum number $s = 1/2$.

\section{Reduction of the interacting many-body Hamiltonian}\label{app:hamiltonian-reduction}

In this appendix, we outline the reasoning behind the approximations made to derive the Hamiltonian in Eq.~\eqref{eq:model-hamiltonian}, used to implement the entangling phase gate protocols.
In particular, we investigate the relative strengths of the interactions between the valence electrons and the electric potential of the linear Paul trap, excitation laser and microwave modes, and other valence electrons.

Considering first the electron-trap interaction, it follows from selection rules that the matrix elements associated to electric quadrupolar transitions are zero when the total angular momentum quantum number $j = 1/2$.
Hence, the only nonzero angular matrix elements are
\begin{equation}
\begin{gathered}
    \me{1}{\qo{Y}_{2}^{-2}}{0} = \sqrt{\frac{1}{4 \pi}}, \qquad
    \me{1}{\qo{Y}_{2}^{-2}}{3} = \sqrt{\frac{1}{4 \pi}}, \qquad
    \me{2}{\qo{Y}_{2}^{-2}}{4} = -\sqrt{\frac{1}{5 \pi}}, \\
    \me{1}{\qo{Y}_{2}^{0}}{1} = -\sqrt{\frac{5}{49 \pi}}, \qquad
    \me{2}{\qo{Y}_{2}^{0}}{2} = -\sqrt{\frac{1}{20 \pi}},
\end{gathered}
\end{equation}
where $\me{\bm{n}'}{\qo{Y}_{2}^{2}}{\bm{n}} = \me{\bm{n}}{\qo{Y}_{2}^{-2}}{\bm{n}'}$, since the angular matrix elements are strictly real by convention~\cite{louck2023}.
For a strontium $^{88}\mathrm{Sr}^{+}$ ion with principal quantum number $n = 46$ (cf. Ref.~\cite{zhang2020}), the associated radial matrix elements are
\begin{equation}
\begin{gathered}
    \me{1}{\qo{r}^{2}}{0} = 3.58 \times 10^{-20} \, \unit{\metre}^{2}, \qquad
    \me{1}{\qo{r}^{2}}{3} = -6.50 \times 10^{-23} \, \unit{\metre}^{2}, \qquad
    \me{2}{\qo{r}^{2}}{4} = -6.29 \times 10^{-22} \, \unit{\metre}^{2}, \\
    \me{1}{\qo{r}^{2}}{1} = 3.03 \times 10^{-20} \, \unit{\metre}^{2}, \qquad
    \me{2}{\qo{r}^{2}}{2} = 2.94 \times 10^{-19} \, \unit{\metre}^{2}, 
\end{gathered}
\end{equation}
where similarly $\me{\bm{n}'}{\ro^{2}}{\bm{n}} = \me{\bm{n}}{\ro^{2}}{\bm{n}'}$, since the radial matrix elements are by convention strictly real. In current experiments, typical trapping parameters are $\alpha \sim 10^{9} \, \unit{\volt} \, \unit{\metre}^{-1}$ and $\beta \sim 10^{7} \, \unit{\volt} \, \unit{\metre}^{-1}$~\cite{mokhberi2020} and , therefore, the magnitudes of the on-diagonal matrix elements which cause levels to be shifted are
\begin{equation}
    \ab{e \beta \!\me{1}{[3 \qo{r}_{z}^{2} - \qo{\bm{r}}^{2}]}{1}\!} = 2.77 \times 10^{-32} \, \unit{\joule}, \qquad
    \ab{e \beta \!\me{2}{[3 \qo{r}_{z}^{2} - \qo{\bm{r}}^{2}]}{2}\!} = 1.89 \times 10^{-31} \, \unit{\joule}.
\end{equation}
These are, however, several orders of magnitude smaller compared to the energies of the $j > 1/2$ states (i.e., $\ket{1}$ and $\ket{2}$),
\begin{equation}
    E_{1} \equiv E_{4, 2, 5/2} \equiv E_{4 \mathrm{D}_{5/2}} = 2.95 \times 10^{-19} \, \unit{\joule}, \qquad
    E_{2} \equiv E_{6, 1, 3/2} \equiv E_{6 \mathrm{P}_{3/2}} = 1.11 \times 10^{-18} \, \unit{\joule},
\end{equation}
and so can be neglected, henceforth. Similarly, the magnitudes of the off-diagonal matrix elements which cause levels to be coupled are
\begin{equation}
\begin{gathered}
    \ab{e \alpha \!\me{1}{[\qo{r}_{x}^{2} - \qo{r}_{y}^{2}]}{0}\!} = 2.09 \times 10^{-30} \, \unit{\joule}, \qquad
    \ab{e \alpha \!\me{1}{[\qo{r}_{x}^{2} - \qo{r}_{y}^{2}]}{3}\!} = 3.80 \times 10^{-33} \, \unit{\joule}, \\
    \ab{e \alpha \!\me{2}{[\qo{r}_{x}^{2} - \qo{r}_{y}^{2}]}{4}\!} = 3.29 \times 10^{-32} \, \unit{\joule},
\end{gathered}
\end{equation}
which, likewise, can be ignored, hereafter. Notice that the magnitudes of the off-diagonal matrix elements within fine structure manifolds (i.e., between states with the same principal $n$, orbital angular momentum $l$, and total angular momentum $j$ quantum numbers, but different total magnetic $m_{j}$ quantum numbers) are still insignificant compared to the energies,
\begin{equation}
\begin{gathered}
    \ab{e \alpha \!\me{4 \mathrm{D}_{\smash{5/2}}^{\smash{-5/2}}}{[\qo{r}_{x}^{2} - \qo{r}_{y}^{2}]}{4 \mathrm{D}_{\smash{5/2}}^{\smash{-1/2}}}\!} = 8.76 \times 10^{-31} \, \unit{\joule}, \\
    \ab{e \alpha \!\me{6 \mathrm{P}_{\smash{3/2}}^{\smash{-3/2}}}{[\qo{r}_{x}^{2} - \qo{r}_{y}^{2}]}{6 \mathrm{P}_{\smash{3/2}}^{\smash{1/2}}}\!} = 1.09 \times 10^{-29} \, \unit{\joule}.
\end{gathered}
\end{equation}
Note that the position-dependent modification to the static electric-field gradient $\beta$ (i.e., the coefficient $2 K_{ii}$) contributes, however, not enough (e.g., for $N = 120$ we find that $K_{ii} < 10^{\smash{3}}$).

Let us now consider the electron-laser and electron-electron interactions, the matrix elements of which are electric dipolar transitions. For the angular matrix elements, the only nonzero terms are
\begin{equation}
\begin{gathered}
    \me{2}{\qo{Y}_{1}^{-1}}{0} = \sqrt{\frac{1}{4 \pi}}, \qquad
    \me{0}{\qo{Y}_{1}^{-1}}{4} = -\sqrt{\frac{1}{6 \pi}}, \\
    \me{1}{\qo{Y}_{1}^{-1}}{2} = \sqrt{\frac{3}{10 \pi}}, \qquad
    \me{2}{\qo{Y}_{1}^{-1}}{3} = \sqrt{\frac{1}{4 \pi}}, \qquad
    \me{3}{\qo{Y}_{1}^{-1}}{4} = -\sqrt{\frac{1}{6 \pi}},
\end{gathered}
\end{equation}
where, in contrast, $\me{\bm{n}'}{\qo{Y}_{1}^{1}}{\bm{n}} = -\me{\bm{n}}{\qo{Y}_{1}^{-1}}{\bm{n}'}$. The associated radial matrix elements are then
\begin{equation}
\begin{gathered}
    \me{2}{\qo{r}}{0} = -7.18 \times 10^{-12} \, \unit{\metre}, \qquad
    \me{0}{\qo{r}}{4} = 7.96 \times 10^{-14} \, \unit{\metre}, \\
    \me{1}{\qo{r}}{2} = -6.87 \times 10^{-12} \, \unit{\metre}, \qquad
    \me{2}{\qo{r}}{3} = 1.13 \times 10^{-12} \, \unit{\metre}, \qquad
    \me{3}{\qo{r}}{4} = -6.37 \times 10^{-8} \, \unit{\metre}
\end{gathered}
\end{equation}
with $\me{\bm{n}'}{\ro}{\bm{n}} = \me{\bm{n}}{\ro}{\bm{n}'}$. While the relative magnitude of the electric dipole transitions of the electron-laser interactions are compensated for by the electric-field gradients $E_{j; x}$, those of the electron-electron interactions are not. Therefore, in contrast to the former, we can neglect all contributions to the electric dipole-dipole interaction, except for those between the Rydberg states $\ket{3}$, $\ket{4}$ since the associated radial matrix elements are negligible in comparison.


\bibliography{main.bib}

\end{document}